%% file: protonid08.tex
\documentclass[aps,prd,twocolumn,superscriptaddress,showpacs]{revtex4}
\pdfoutput=1

\usepackage{graphicx}
\usepackage{xspace}
\usepackage{multirow}

\graphicspath{{./figures/}}

\newcommand{\superk}      {Super-Kamiokande\xspace}       

\newcommand{\nue}         {$\nu_{e}$\xspace}
\newcommand{\numu}        {$\nu_{\mu}$\xspace}

\newcommand{\mutau}       {$\nu_\mu \rightarrow \nu_{\tau}$\xspace}
\newcommand{\musterile}   {$\nu_\mu \rightarrow \nu_{sterile}$\xspace}
\newcommand{\dms}         {$\Delta m^2$\xspace}
\newcommand{\sstt}        {$\sin^2 2 \theta$\xspace}

\newcommand{\degree}      {$^\circ$\xspace}

\newcommand{\pizero}      {$\pi^{0} $}

\newcommand{\etal}        {{\em et al.}}

\newcommand{\asymerr}[2]{\ooalign{{\scriptsize \raisebox{4pt}{+~#1}}\crcr
{\scriptsize \raisebox{-4pt}{--~#2}}}}

\begin{document}

\title{Kinematic reconstruction of atmospheric neutrino events in a large
  water Cherenkov detector with proton identification}

\input{authors}
\date{\today}

\begin{abstract}
  We report the development of a proton identification method
  for the \superk detector. This new tool is applied to the search for
  events with a single proton track, a high purity neutral current
  sample of interest for sterile neutrino searches. After selection
  using a neural network, we observe 38 events in the combined SK-I
  and SK-II data corresponding to 2285.1 days of exposure, with an
  estimated signal-to-background ratio of 1.6 to 1.  Proton
  identification was also applied to a direct search for
  charged-current quasi-elastic (CCQE) events, obtaining a high
  precision sample of fully kinematically reconstructed atmospheric
  neutrinos, which has not been previously reported in water Cherenkov
  detectors.  The CCQE fraction of this sample is 55\%, and its
  neutrino (as opposed to anti-neutrino) fraction is $91.7\pm3$\%. We
  selected 78 $\mu$-like and 47 e-like events in the SK-I and SK-II
  data set.  With this data, a clear zenith angle distortion of
  the neutrino direction itself is reported in a sub-GeV sample of
  $\mu$ neutrinos where the lepton angular correlation to the incoming
  neutrino is weak.  Our fit to \mutau oscillations using the neutrino
  ${L\over E }$ distribution of the CCQE sample alone yields a wide
  acceptance region compatible with our previous results and excludes
  the no-oscillation hypothesis at 3 sigma.
\end{abstract}


\maketitle

\section{Introduction}

The \superk detector is well known for its discovery of atmospheric
neutrino oscillations. It is a large, 50-kton, cylindrical water
Cherenkov detector located under Mt. Ikenoyama near Kamioka, Japan
\cite{fukuda:2002uc}. It is separated in two concentric regions: a
2~m-thick outer veto (OD), instrumented with 1885 8 inch
photo-multiplier tubes (PMTs) facing outward, used as a cosmic ray
veto; and an inner detector (ID) instrumented with 11,146 inward
facing 20 inch PMTs. The Cherenkov light emitted by charged particles
traveling in the water travels to the walls and the resulting
ring-shaped patterns are then analyzed.

\superk has very good particle identification (PID) capabilities,
being able to separate muon tracks from electron tracks with better
than 1.8\% mis-identification \cite{ashie:2005ik}.  Until now, no
attempt has been made to extend its PID algorithm to other types of
particles produced in neutrino interactions. In this article we
demonstrate that \superk can also be used to search for proton
tracks. Other authors have published estimates predicting what would
be found in Super-K and other experiments, along with explorations of
the physics consequences~\cite{beacom:2003prd}.

This improved PID technique will be described in section
\ref{sec:fitter}. In section~\ref{sec:dataset} we will describe
atmospheric neutrino data selection and reconstruction at \superk.  In
section \ref{sec:ncelastic} we will apply this new proton selection
algorithm to a search for neutral-current (NC) elastic
$\nu+p\rightarrow\nu+p$ events. This sample has high neutral current
purity and angular correlation with the incoming neutrino, and is
potentially sensitive to \musterile
oscillations~\cite{beacom:2003prd}.

In section \ref{sec:ccqe} we will extend the technique to
charged-current quasi-elastic (CCQE) neutrino interactions,
reconstructing both the outgoing lepton and the proton.  This allows
full kinematic reconstruction of the atmospheric neutrino track,
improving neutrino zenith angle, flight length and energy
reconstruction, which is useful for oscillation studies.  This is the
first time such a reconstruction has been performed in a water
Cherenkov detector.

\section{A proton track fitter for \superk}
\label{sec:protonfit}

\subsection{Characteristics of proton tracks in \superk}
Proton tracks in water exhibit certain features that make it possible
to identify them.

The main parameter in Cherenkov light production is the velocity
$\beta$ of the particle, related to the Cherenkov cone opening angle
by $$\cos\theta = \frac{1}{\beta n},$$ where $n$ is the refraction
index of the medium.  In water this leads to a threshold of
$\beta\approx0.74$. Because of the large proton mass this corresponds
to a momentum threshold of $1.07\ \mathrm{GeV}/c$.  Due to the falling
energy spectrum of atmospheric neutrinos, we do not expect to see many
proton tracks above $3\ \mathrm{GeV}/c$, corresponding to $\beta <
0.96$.  Consequently for most of the protons seen in \superk the
half-opening angle of the Cherenkov cone will be relatively small,
below about $38$\degree. This contrasts with electrons and muons which
have lower mass, and so reach $\beta\approx 1$ and Cherenkov angles of
$\approx42$\degree at much lower momenta than protons.

\begin{figure*}[!htb]
  \begin{minipage}{3in}
    \includegraphics[width=2.0in]{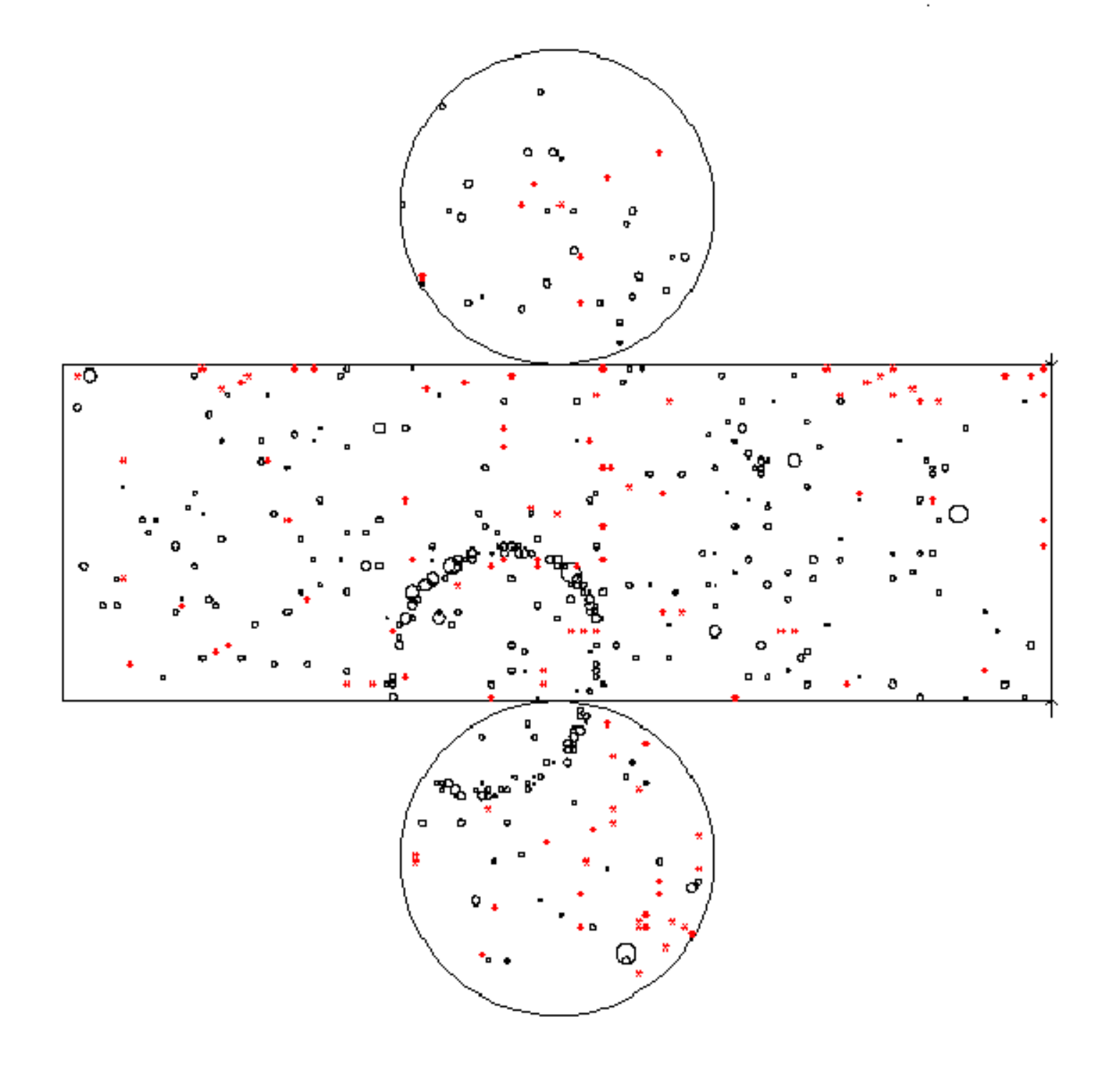}
  \end{minipage}
  \hfill
  \begin{minipage}{3in}
    \includegraphics[width=2.0in]{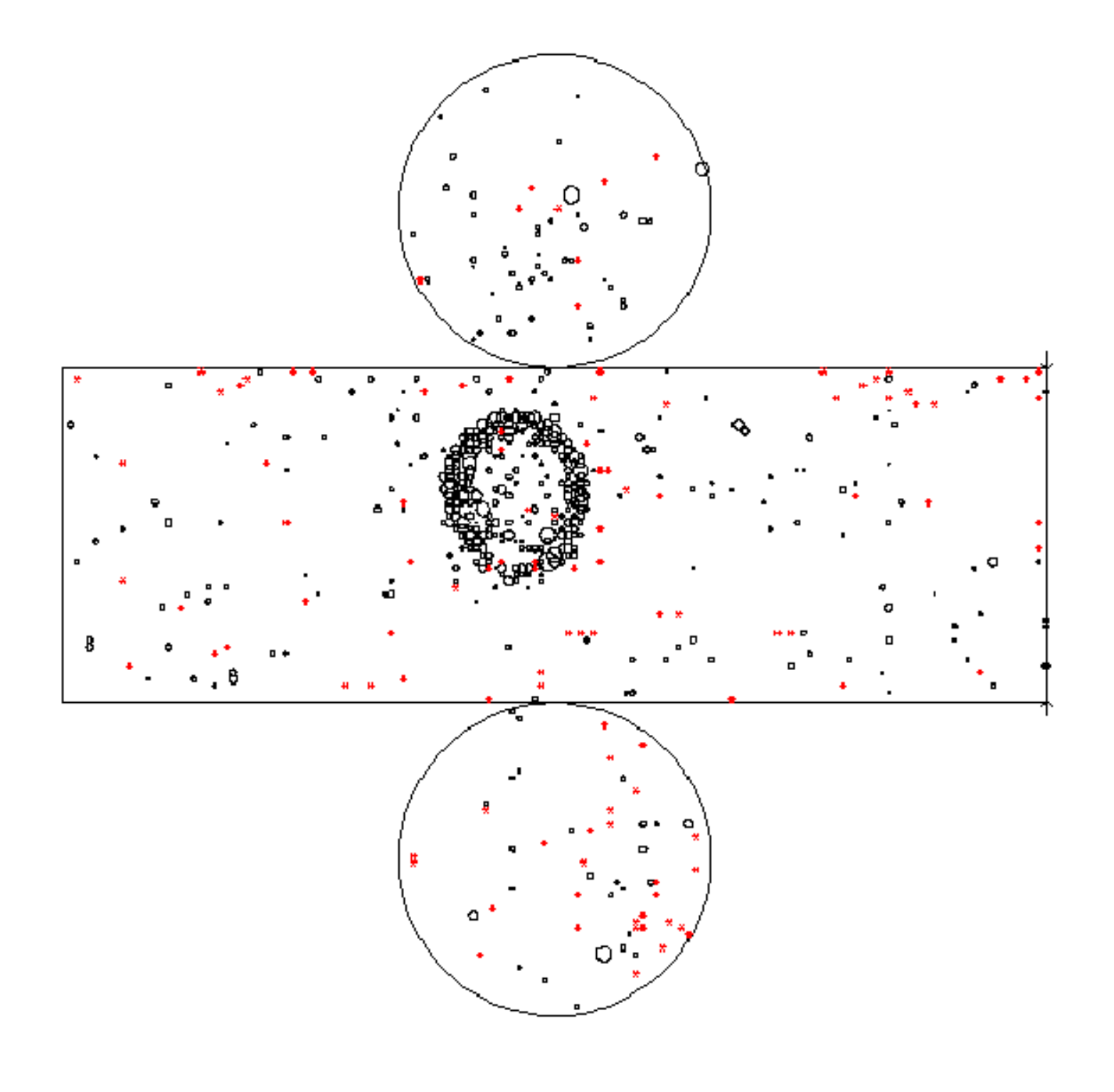}
  \end{minipage}
    \caption{Event displays of a Monte-Carlo NC elastic event, with
      proton momentum of 1490 MeV/c (left), and a Monte-Carlo 300 MeV/c muon. 
      The proton stopped early, causing a thin ring pattern on the wall. The muon
      ring is thicker than most proton rings with similar opening angles.}
    \label{fig:ncandmudisplay}
\end{figure*}

\superk's PID program uses the ring topology to separate muons and
electrons: due to scattering and radiative processes creating
electromagnetic showers, the outer edge of electron rings are blurred,
while muons have rings with sharp outer edges
\cite{ashie:2005ik}. Like muons, protons will have rings with sharp
outer edges.  Since muons at these momenta lose their energy by
ionization, their ranges are long (typically several meters), leading
to thick projected rings on the detector walls.  However protons have
a high probability of undergoing nuclear interactions with $^{16}O$ or
$H$ nuclei in the water. A large fraction of the time, neither the
proton nor the secondary particles produced in these collisions are
above Cherenkov threshold: in this case the proton's Cherenkov light
emission is seen to stop suddenly, causing a short, thin ring to
appear on the detector wall. Figure~\ref{fig:ncandmudisplay}
shows typical proton and muon event displays.

\begin{figure}[!htbp]
  \begin{center}
  \includegraphics[scale=0.3]{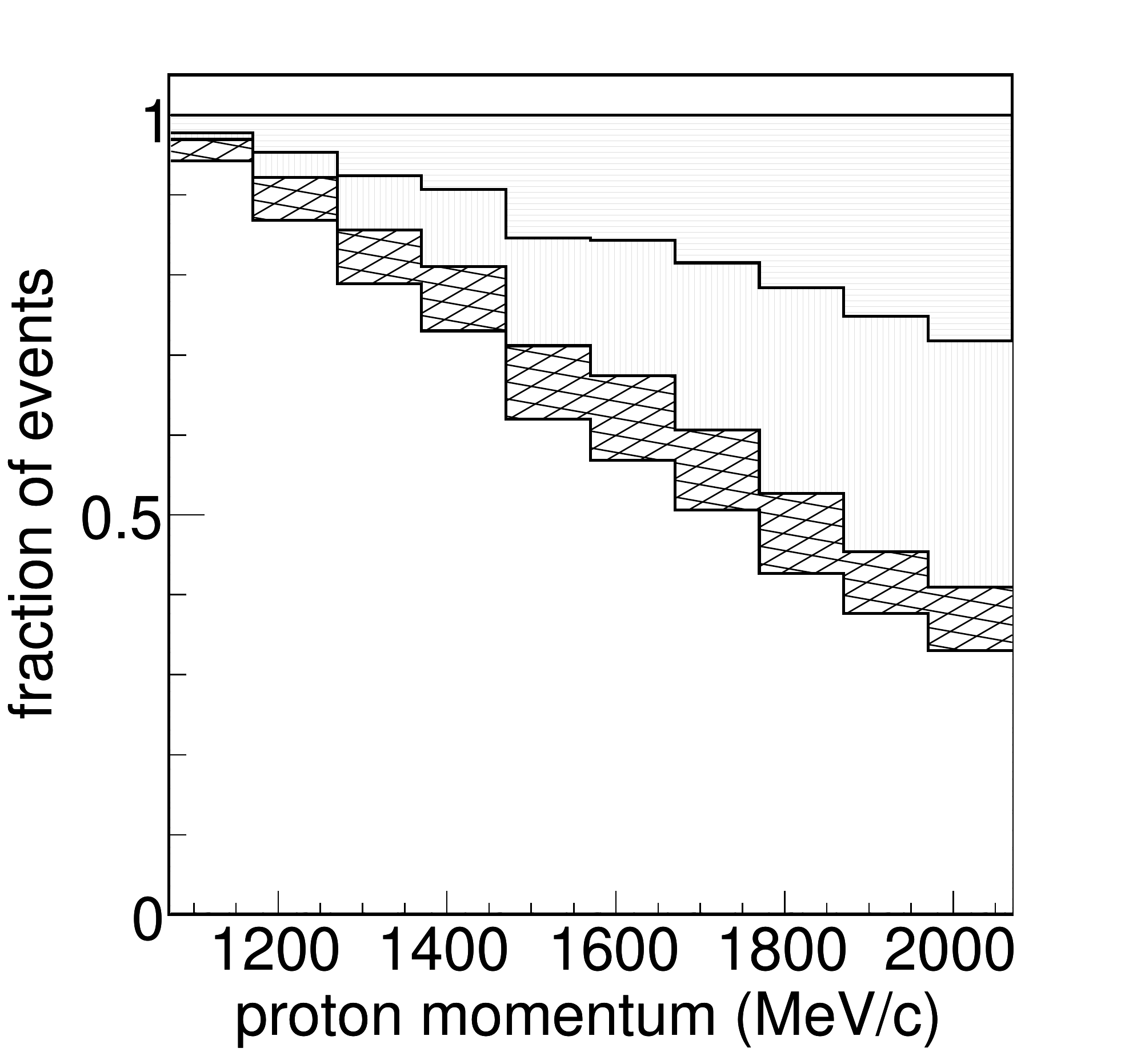}
  \caption{Probability of hadronic interactions in the water as a
    function of proton momentum. The clear region shows the fraction
    of protons that do not interact hadronically in the water.  The
    cross-hatched region shows events whose interactions produce only
    sub-threshold secondaries. The region with a
    vertical hatch pattern corresponds to production of
    above-threshold charged secondaries, with no \pizero. The
    horizontally hatched region shows the amount of
    \pizero\ production.}
  \label{fig:pizeroprod}
  \end{center}
\end{figure}

Proton tracks have an additional characteristic: as shown in
Fig.~\ref{fig:pizeroprod}, the probability for a proton to produce a
\pizero\ or a charged secondary during a hadronic interaction in the
water increases with momentum, reaching $\sim 50\%$ at around 2
GeV$/c$ based on our Monte-Carlo studies.  In particular, if a
\pizero\ is produced, at least one bright electron-like ring (from
neutral pion decay gamma showers) will be seen in \superk, greatly
reducing the chances of identifying the proton and accurately
reconstructing the event. The fraction of protons that do not produce
any visible secondary particles goes
from $\sim90\%$ at 1.2 GeV/c to $\sim 40\%$ at 2 GeV/c.\\

\subsection{Expected light pattern engine and fitter}
\label{sec:fitter} 
Particle identification is a hypothesis test, and is dealt with by
calculating a ratio of maximum likelihoods. This is attempted after a
previous fitter has already computed a vertex position, Cherenkov
opening angle and candidate ring direction. A \emph{light pattern
  engine} generates the average Cherenkov pattern corresponding to the
given vertex and track configuration inside the tank. Depending on the
hypothesis being tested, several parameters of the Cherenkov cone are
then adjusted to maximize a \emph{pattern likelihood} built from the
observed and expected charges on individual phototubes.  This
procedure is carried out twice, using the two different particle types
that are being studied to generate the light patterns.
In this analysis, a special particle identification
algorithm that takes into account proton-like features was developed.
During the first stage, we assume the ring was
created by a proton, and during the second stage it is assumed to be
a muon.
We will now briefly describe the light pattern engine as well as the
fit performed during the proton hypothesis test.
  
  \subsubsection{Calculation of expected charge patterns}
  Expected light patterns are produced using a program described in
  \cite{thesis:tomasz}. For proton identification we have extended the
  program's capabilities. Using a GEANT4 \cite{Agostinelli:2002hh}
  Monte-Carlo simulator, we obtained tables of the Cherenkov photon
  density $$\frac{d^3N}{dr\,dp\, d\cos\theta},$$ where $r$ is the
  distance from the vertex, $p$ is the particle momentum and $\theta$
  is the angle with respect to the particle track (cone axis). This
  table was built in ``pure'' water, \textit{i.e.} all scattering and
  absorption effects were turned off; moreover, proton hadronic
  interactions and $\delta$-ray production were turned off, but energy
  loss by ionization was kept. Thirty batches of monochromatic protons
  at momenta from 1.1 to 4.0 GeV/c were used, with 50 MeV/c spacing
  until 1.5 GeV/c, 100 MeV/c spacing until 3500 MeV/c and one batch at
  4 GeV/c.  The distance $r$ was sampled every 50 cm.  We used 400
  bins in $\cos\theta$ from $1$ to $0.6$ to sample the edge of the
  Cherenkov ring as finely as possible.

  For a given vertex, track direction and momentum $p$, the amount of
  Cherenkov light on a given phototube is calculated by linear
  interpolation from this table. The computed photon flux is
  azimuthally symmetric around the proton track. PMT solid angle and
  acceptance corrections, water attenuation and scattering effects are
  then applied to the light pattern. Since the photon flux is
  tabulated at fixed momenta, the Cherenkov opening angle of the cone
  is quantized in relatively large steps; consequently before any
  interpolation is done, each $\frac{d^2N}{dr\,d\cos\theta}$
  distribution is reweighted so that its peak angle matches the
  Cherenkov angle at input momentum $p$ to avoid any numerical
  problems.

  As explained earlier, one of the main features of protons is that
  their Cherenkov light emission can be stopped suddenly in the
  water. The pattern engine was modified to produce light patterns for
  any path length: stopping the path length is simulated by masking
  the PMTs inside the inner edge of the cone. A smooth cubic spline is
  used to describe the effects of finite PMT size for PMTs just on the
  edge of the lit region.

  A MINUIT-based fit \cite{minuit} was built to fit the observed light
  pattern with an expected light pattern. The free parameters in the
  fit are the proton momentum and proton track length.

\begin{figure}[!htb]
  \includegraphics[width=2.in]{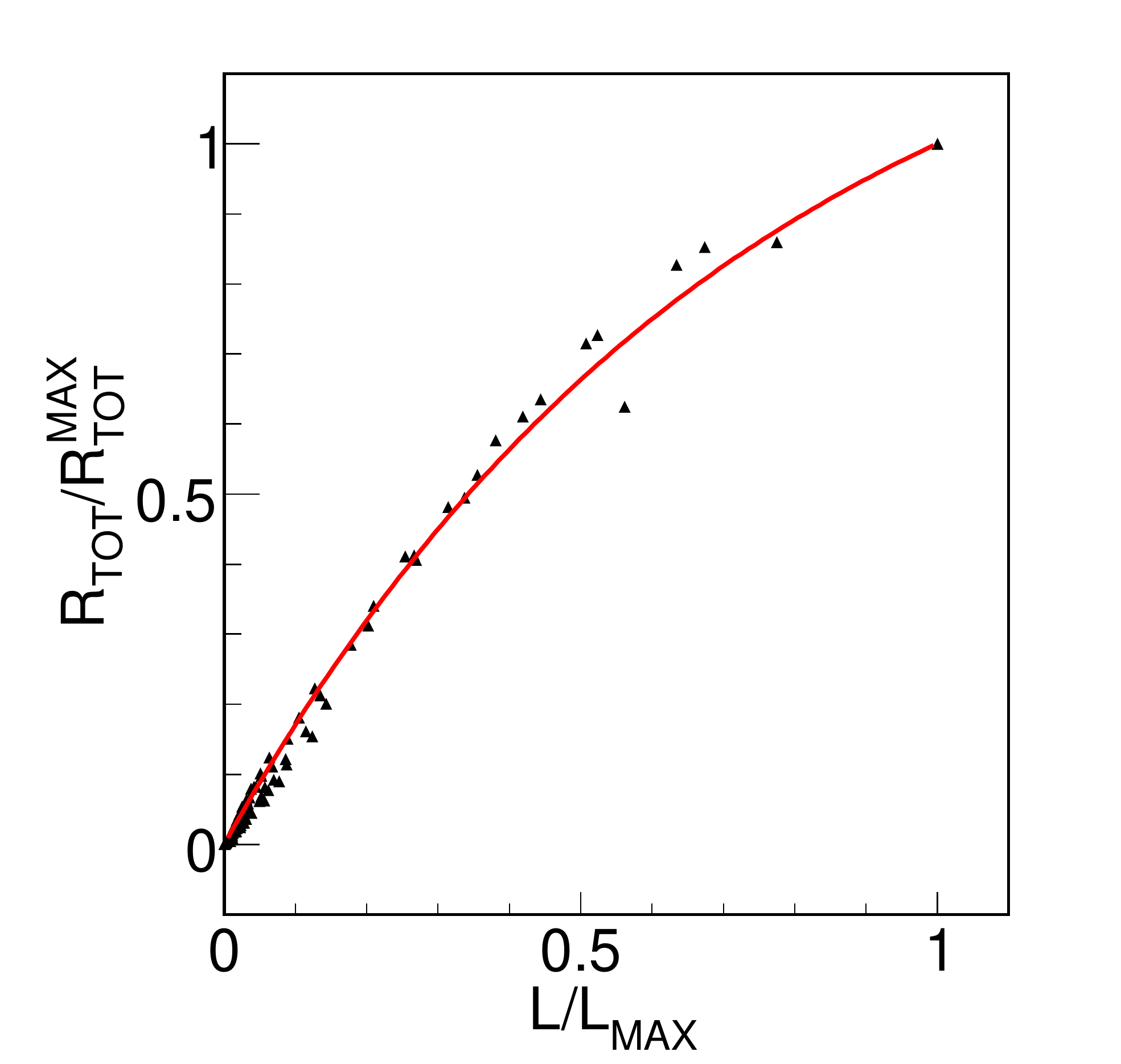}
  \caption{$\frac{R_{tot}}{R_{tot}^{max}}$ as a function of 
    $\frac{L}{L_{max}}$ for Monte-Carlo batches at different
    momenta and fixed path-lengths in the detector.
    The red line shows the best fit to the functional form
    $a (\exp (-b\frac{L}{L_{max}}) -1)$, with $a=-1.29$ and $b=1.5$.
    The function allows to calculate the expected amount of light
    for any proton pattern.}
  \label{fig:universalfunction}
\end{figure}

  The expected light pattern then needs to be normalized.  The
  relevant variable in \superk is $R_{tot}$, the amount of
  photo-electrons collected in a 70\degree cone around the track,
  correcting for water attenuation and scattering as well as PMT
  acceptance \cite{skphd}. For electrons and muons, tables linking the
  particle momentum to $R_{tot}$ are used for momentum determination
  throughout the reconstruction. For protons, a one-to-one conversion
  function does not exist since protons may stop abruptly in the
  water, yielding the same amount of light for different initial
  momenta.  Using \superk's Monte-Carlo simulation with hadronic
  interactions turned off, conversion tables from proton momentum to
  $R_{tot}^{max}$, the maximum achievable value, were calculated. The
  proton average track length without hadronic interactions,
  $L_{max}$, was also obtained in this way. Then, by simulating
  batches of protons at fixed path lengths, it was found that $
  \frac{R_{tot}}{R_{tot}^{max}}$ is only a function of
  $\frac{L}{L_{max}}$ and not of the initial momentum of the
  proton. Figure~\ref{fig:universalfunction} shows this function as
  well as a fit. These calculations allow us to renormalize the
  expected light pattern to its expected $R_{tot}(l,p)$ value for any
  given input length $l$ and momentum $p$, ensuring that the maximum
  likelihood fit is well behaved.
 
The fitter returns the maximum likelihood $L_{\mathrm{proton}}$, as
well as the proton-like momentum and proton-like length of the event.
In a separate stage, $L_{\mathrm{muon}}$, the
likelihood that the pattern is muon-like, is calculated.

 \subsection{Performance of the proton fitter}

\begin{figure}[!htb]
  \includegraphics[width=2.in]{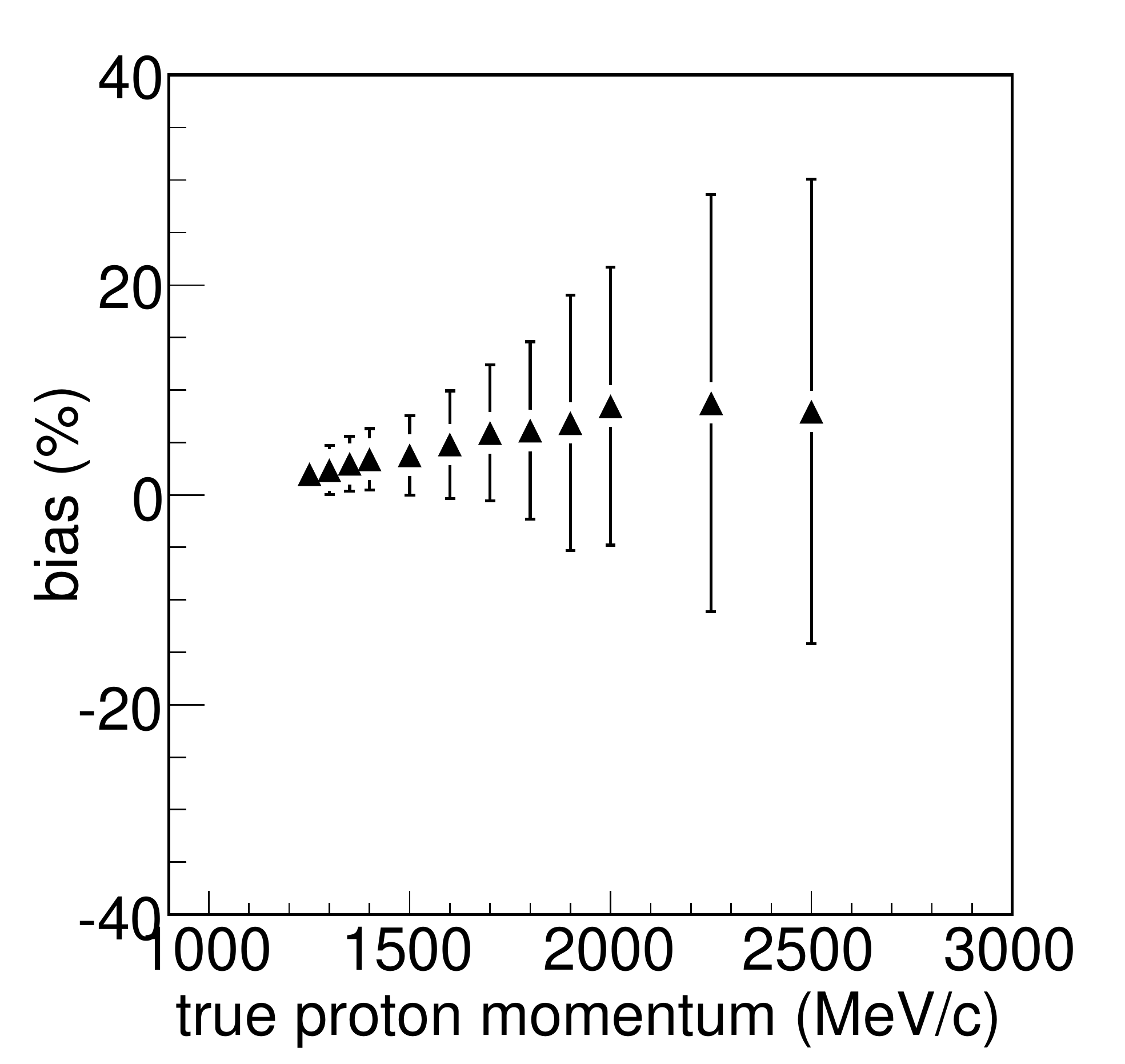}
  \includegraphics[width=2.in]{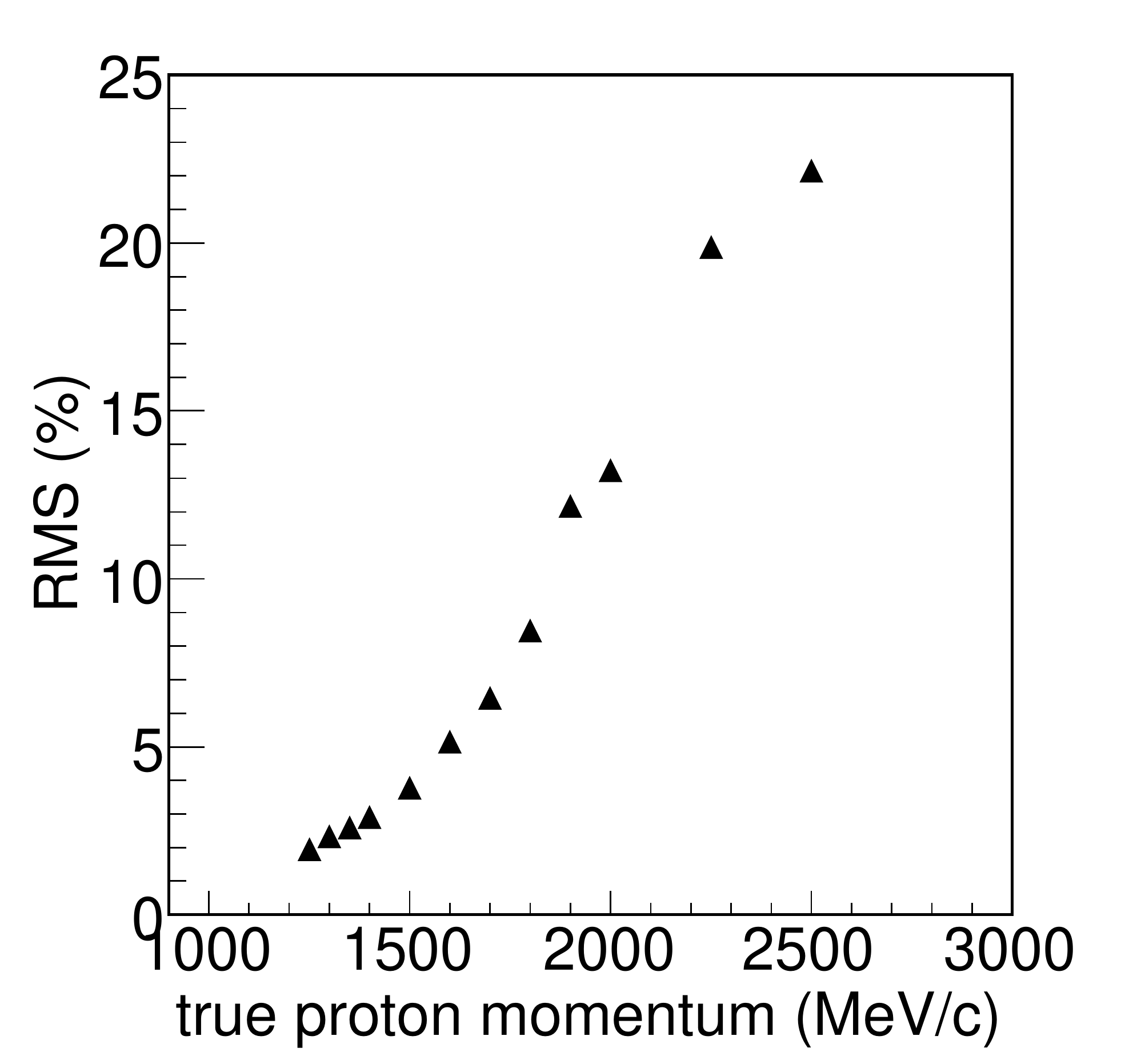}
  \caption{Bias (left) and RMS (right) of proton momentum reconstruction,
  using monochromatic Monte-Carlo.}
  \label{fig:protonmomres}
\end{figure}

 The performance of this new algorithm has been tested using
 Monte-Carlo samples of muons, protons, and charged pions. The samples
 were produced using GEANT3 and include full fluctuations of all
 processes and other detector effects.  The momentum resolution for protons
 varies from about 3\% below 1.5 GeV/c, to about 10\% at 2 GeV/c,
 reaching 30\% at 2.5 GeV/c. The mean resolution of the momentum fit
 over the NC elastic proton spectrum (for the atmospheric neutrino
 spectrum) is about 7\%, with a small positive bias in the momentum
 determination of 3\%. The resolution gets worse as the momentum
 increases because it is mostly controlled by the opening angle
 measurement of the Cherenkov ring. As the angle increases, small
 variations in the angle measurement result in larger changes in the
 fitted momentum. Figure~\ref{fig:protonmomres} summarizes the results
 of momentum reconstruction.
 
\begin{figure}[!htb]
  \includegraphics[width=3.5in]{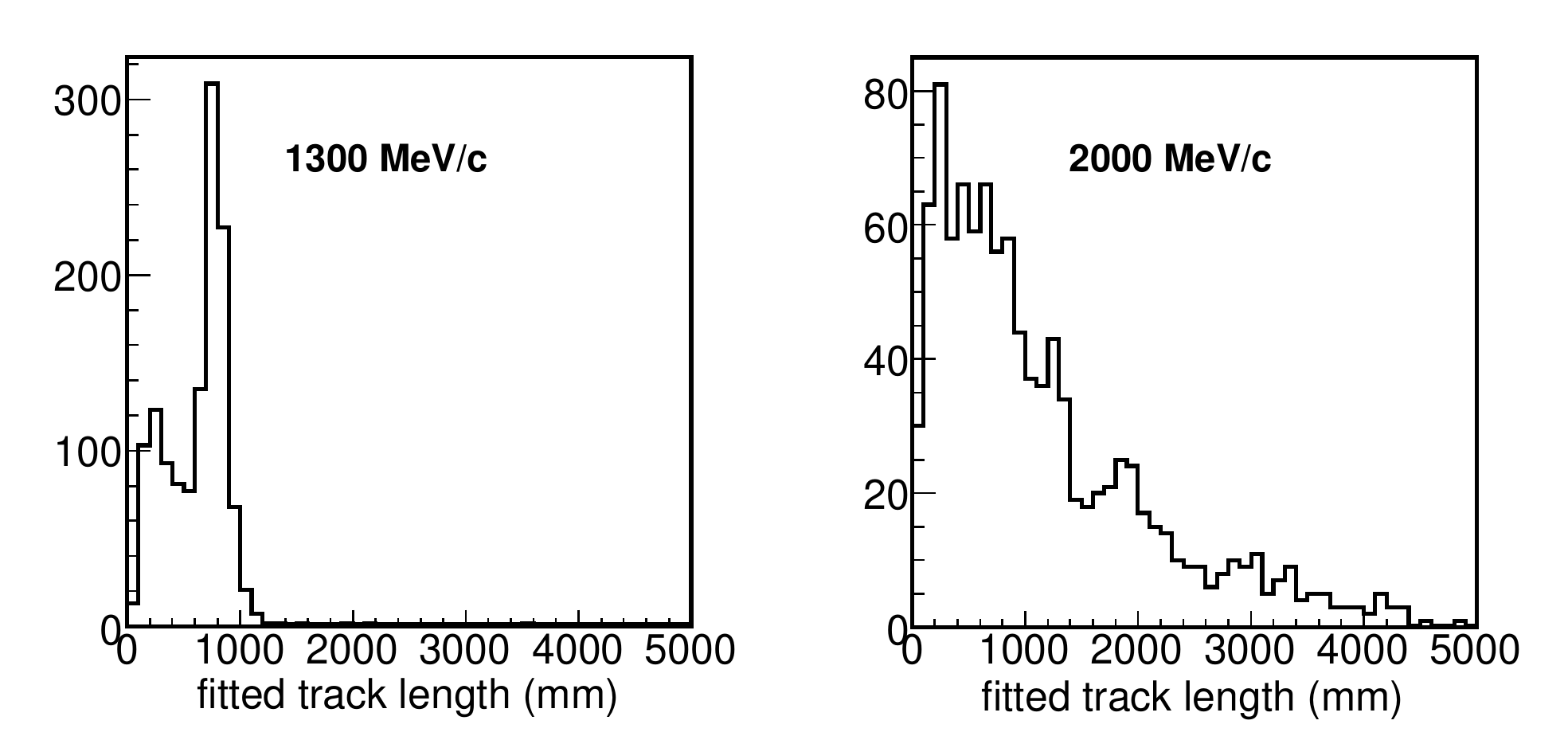}
  \caption{Fitted track length for Monte-Carlo protons 
  at 1.3 and 2.0 GeV/c. At 2.0 GeV/c the protons never reach
  their maximum path length.}
  \label{fig:protonlength}
\end{figure}

\begin{figure}[!htb]
  \includegraphics[width=2.in]{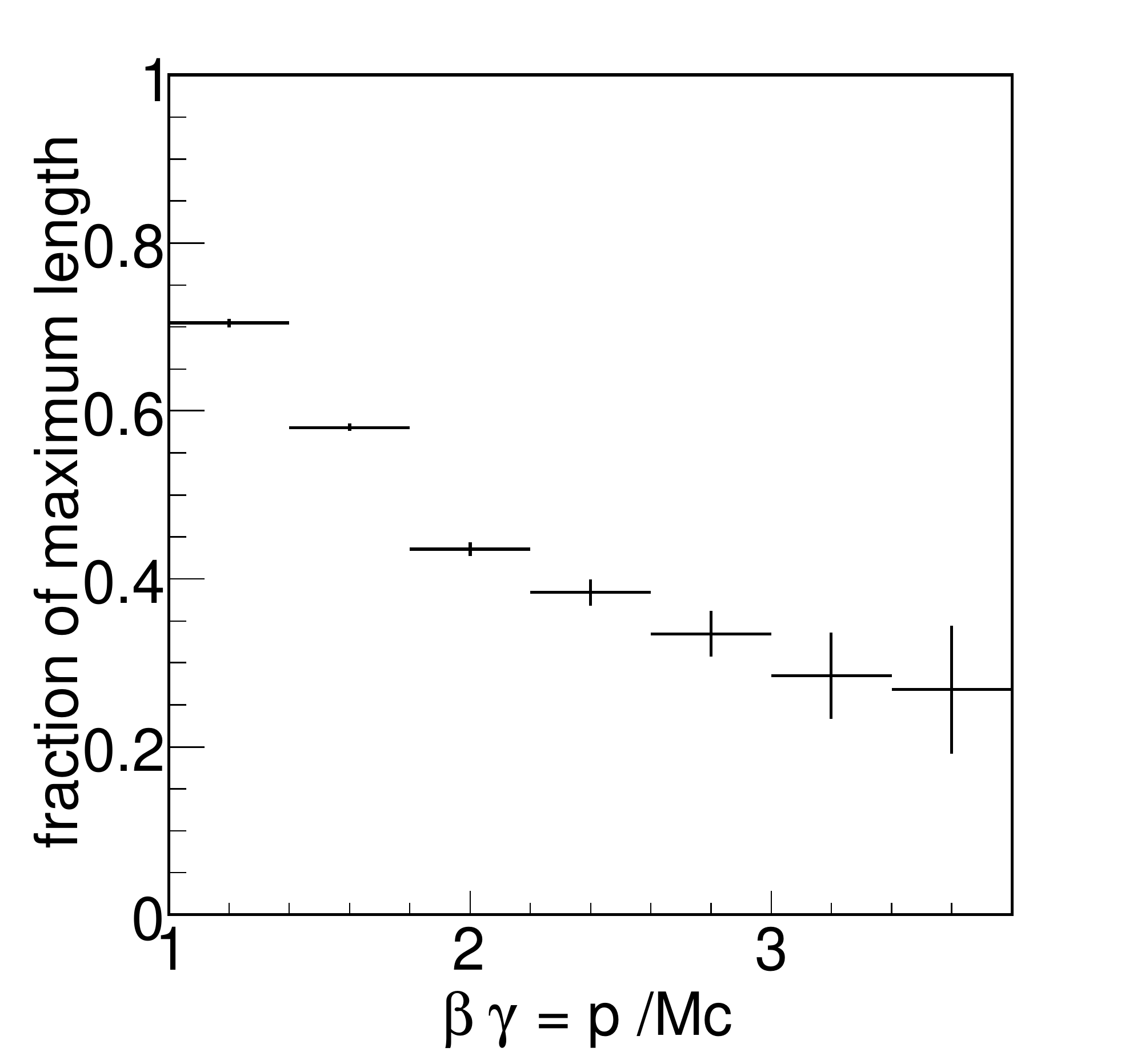}
  \caption{Average fraction of the maximum path length in water
    traveled in water as a function of proton momentum (estimated with
    NC elastic Monte-Carlo events). The effect of hadronic
    interactions is clear.}
  \label{fig:xsechadr}
\end{figure}

 Figure~\ref{fig:protonlength} shows the distributions of the fitted
 path lengths for mono-energetic simulated protons for two different
 momenta. At low energy, the distribution has a sharp peak at higher
 values, corresponding to protons that traveled their full path-length
 (no interaction in the water), and a region at lower fitted track
 lengths corresponding to proton tracks that stopped early.  At higher
 input momenta all protons traveled less than their maximum path
 length and the distribution falls off.  This observation of hadronic
 interactions in the water can be summarized as the average ratio
 $\frac{L}{L_{max}}$ as a function of the input value of
 $\beta\gamma$.  Figure~\ref{fig:xsechadr} shows that the ratio
 clearly decreases with input proton momentum, showing the increase of
 the hadronic cross-section with momentum.

\section{The Super-Kamiokande atmospheric data set}
\label{sec:dataset}
\subsection{Fully-contained atmospheric data}

The data used in this paper were collected at \superk during two
separate run periods. The first one, labeled SK-I, ran from April
1996 to July 2001, with 40\% photocathode coverage (11,146 ID
PMTs). The second one, called SK-II, ran from September 2002 to
October 2005, with 19\% photocathode coverage (5,183 ID PMTs), as a
result of partial reconstruction following an accident that occurred in
Nov 2001. In SK-II the PMTs were encased in a protective fiberglass
shell with a transparent acrylic window to protect against the risk of
tube implosion. The two data sets respectively correspond to 1489.2
and 798.6 integrated days of livetime.

\superk data first undergo a reduction process, which selects
atmospheric neutrino events and splits them into separate samples and
eliminates backgrounds. In this paper, the only relevant sample is the
fully-contained (FC) data set, since protons have relatively short
tracks in the water and do not escape the tank.  The purpose of FC
reduction is to remove events entering the detector - mainly
downward-going cosmic ray muons - as well as events caused by flashing
PMTs, and events with tracks escaping the inner detector. The optical
separation between the inner detector (ID) and the outer detector (OD)
is used during this process.  Details of the reduction process can be
found in \cite{ashie:2005ik}.

 \subsection{Event reconstruction}
 \label{sec:afit}
 Fully-contained data is then fitted by a series of programs which
 find the vertex and particle tracks.  Details about the
 reconstruction can be found in \cite{ashie:2005ik,skphd}. The initial
 step is to obtain a starting vertex and track direction using timing
 information only.  PMTs on the edge of the brightest Cherenkov ring
 are then identified. The method for tagging edge PMT, and Cherenkov
 opening angle measurement, is different in SK-I and SK-II. In SK-I,
 the distribution $Q(\theta)$ of observed charges as a function of
 opening angle is constructed. The edge of the ring is taken to be the
 first zero of the second derivative $\frac{d^2Q}{d\theta^2}$
 occurring after the maximum of $Q$. In SK-II, with half the
 photo-coverage, the sampling of $Q(\theta)$ was found to be too low
 for an accurate determination of the derivatives; therefore, a method
 based on light patterns is used. The expected light pattern of a
 Cherenkov ring (e or $\mu$) is fit to the observed pattern, allowing
 its opening angle to vary; the edge of the ring is found at the best
 fit, and edge PMTs are then tagged.
   
 Once edge PMTs are tagged, they are used to refine the Cherenkov ring
 direction. An estimated track length is calculated, and the vertex
 position is changed by maximizing an estimator based on PMT timing
 and charge, treating PMTs inside and outside of the Cherenkov edge
 differently. This procedure is iterated until a stable vertex and
 direction are found. The next step is to search for other rings in the
 event.

 A Hough-transform method~\cite{Davies:machine-vision} is used to look
 for possible ring candidates. A pattern likelihood function is built
 to determine whether adding each Hough seed matches the observed
 charge pattern better. Other estimators involving the charge density
 in the candidate ring are also considered. If a better match is found
 by adding a ring, the seed is added to the list of tracks.  This
 procedure is iterated until a maximum of 5 rings is found.

 The next step is particle identification (PID): each observed ring is
 compared to a muon pattern and an electron pattern using a likelihood
 method, and is then assigned a particle type, either e-like or
 mu-like. After PID, single ring events are processed by a refined
 track fitter that uses PID information to improve the vertex and
 track fit.

 An important variable is the \emph{visible energy}, which is the
 energy of an electron that would cause the same amount of light in
 the tank. It is estimated at the end of the reconstruction process.

 The performances of PID and the refined vertex fitter are not optimal
 for proton tracks, since by design the particle is assumed to be a
 lepton in these algorithms.  In the search for NC elastic events
 described in the present section we do not use the results of these
 two algorithms.  However, for the analysis described in section
 \ref{sec:ccqe} we do apply them.

 \subsection{Monte-Carlo simulation}
   
Monte-Carlo simulation of atmospheric neutrino events in \superk relies
on three successive steps:
\begin{enumerate}
\item \emph{Atmospheric neutrino flux determination}: the model by
  Honda \etal \cite{honda} is used in this paper.

\item \emph{Simulation of neutrino interactions in the water}: Two
  different programs were used: the NEUT simulation \cite{neut} and
  the NUANCE (version 3) \cite{nuance} simulation. Details about NEUT
  are also available in \cite{ashie:2005ik}. The versions used in
  this paper includes rescattering of the outgoing nucleons, which
  is relevant for estimating the number of protons expected above
  Cherenkov threshold.  Both NEUT and NUANCE assume the axial-vector
  form factor to have a dipole shape, with the axial mass set to 1.21
  $\mathrm{GeV}$ (for CCQE and CC single pion interactions).

\item \emph{Tracking of interaction products in \superk}: A complete
  simulation of \superk using the GEANT3 \cite{Brun93aa} package is
  used.  The Cherenkov light production and propagation models have
  been tuned to calibration data \cite{fukuda:2002uc}. In addition,
  careful studies of the reflection of light on the acrylic surfaces,
  PMT glass and plastic lining of the tank have been carried out. The
  output of the GEANT3 simulation is in the same format as the data,
  and undergoes the same reduction and fitting process.  The events
  are reweighted for livetime, solar wind activity 
  and for \mutau neutrino oscillations.

\end{enumerate}

\section{Search for NC Elastic $\nu+p\rightarrow\nu+p$ events in \superk atmospheric data}
\label{sec:ncelastic}
  The first application of the proton fitter described above is to
  search for single-ring proton events caused by the neutral current
  elastic interaction $\nu+p\rightarrow\nu+p$.

\subsection{Single proton selection}

We initially select a sample of fully-contained atmospheric events,
with their vertices (calculated by the initial fitter) more than 2~m
away from the wall, which defines a 22.5~kton fiducial volume. We also
require that there be only one ring.  The vertex resolution - defined
as the radius of the sphere centered on the true vertex that contains
68.3\% of reconstructed vertices - for single proton tracks from
simulated atmospheric NC elastic events is estimated to
be $\sim 85\ \mathrm{cm}$, with an angular resolution of $\sim
2.8$\degree on the reconstructed direction. The Cherenkov angle is
measured with an accuracy of $\sim 1.6$\degree, with a bias of
$-1.4$\degree.  

We estimate that 7.0 visible NC elastic interactions occur every year
in the fiducial volume.  These are events for which proton NC elastic
collisions produced a single outgoing proton above threshold which in turn
produced no visible secondary in hadronic collisions in the water.

However, other interaction modes create similar events with a single
visible proton, especially neutral current single-pion (NC1$\pi$)
events where the pion is absorbed; and NC elastic collisions on
neutrons where the neutron produces a proton above Cherenkov threshold
through hadronic collisions. Using Monte-Carlo information, we
determined that 5.9 such events are produced per year in the fiducial
volume, for a total of 12.9 single visible protons per year. Of those,
11.4 (88\%) events are fitted as one ring events with their vertex
inside the fiducial volume by our reconstruction. In the remainder of
this section we explain how to separate these fitted events from the
large backgrounds using the proton identification tool described in
section \ref{sec:fitter}.

\subsubsection{Removal of low energy backgrounds}

In atmospheric neutrino analyses, data with less than 30 MeV of
visible energy are discarded; they are not used for oscillation
searches. However in the present analysis such a reduction cut is too
stringent and would remove about 20\% of visible protons.  At visible
energies below 30 MeV, the FC atmospheric neutrino sample does
contain neutrino interactions, but also low energy backgrounds,
especially spallation events.  These are radioactive $\beta$ decays of
fragments left over from the collisions of cosmic ray muons with
$^{16}O$ nuclei in the water.  We use the spallation removal technique
used in solar neutrino searches, which is described in detail in
\cite{Hosaka:2005um}. Spallation events are correlated in time and
space with one of the previous cosmic ray muon tracks; there is also a
correlation with muon energy.

As in the solar analysis, a likelihood function based on these three
parameters was built. The vertex position and event time of each FC
event is checked against all the cosmic ray muons occurring during the
previous 100~s, and the highest value of the likelihood is saved. For
SK-I, two different likelihood functions are used, depending on the
success or failure of the cosmic ray muon track fit. For SK-II both
likelihoods were merged into one. Cuts on the likelihood value were
chosen to remove the events with a high correlation with a preceding
muon. This cut was only applied if $E_{vis}<50$ MeV.  Using real muon
data and uniformly distributed times and vertices, we determined that
these cuts incur a $8.2\%$ ($9.1\%$) inefficiency on non-spallation
atmospheric neutrino events in SK-I (SK-II) below 50 MeV. We also
removed 2.7 days of livetime from the SK-I sample when performing the
spallation cut because of missing cosmic ray muon data.  The value of
the cuts and induced inefficiencies differ from those in
\cite{Hosaka:2005um} because we are looking at atmospheric events
after fully-contained reduction and vertex reconstruction, and not low
energy data after solar neutrino reduction.

\begin{figure}[!htb]
  \begin{center}
\includegraphics[width=2.25in]{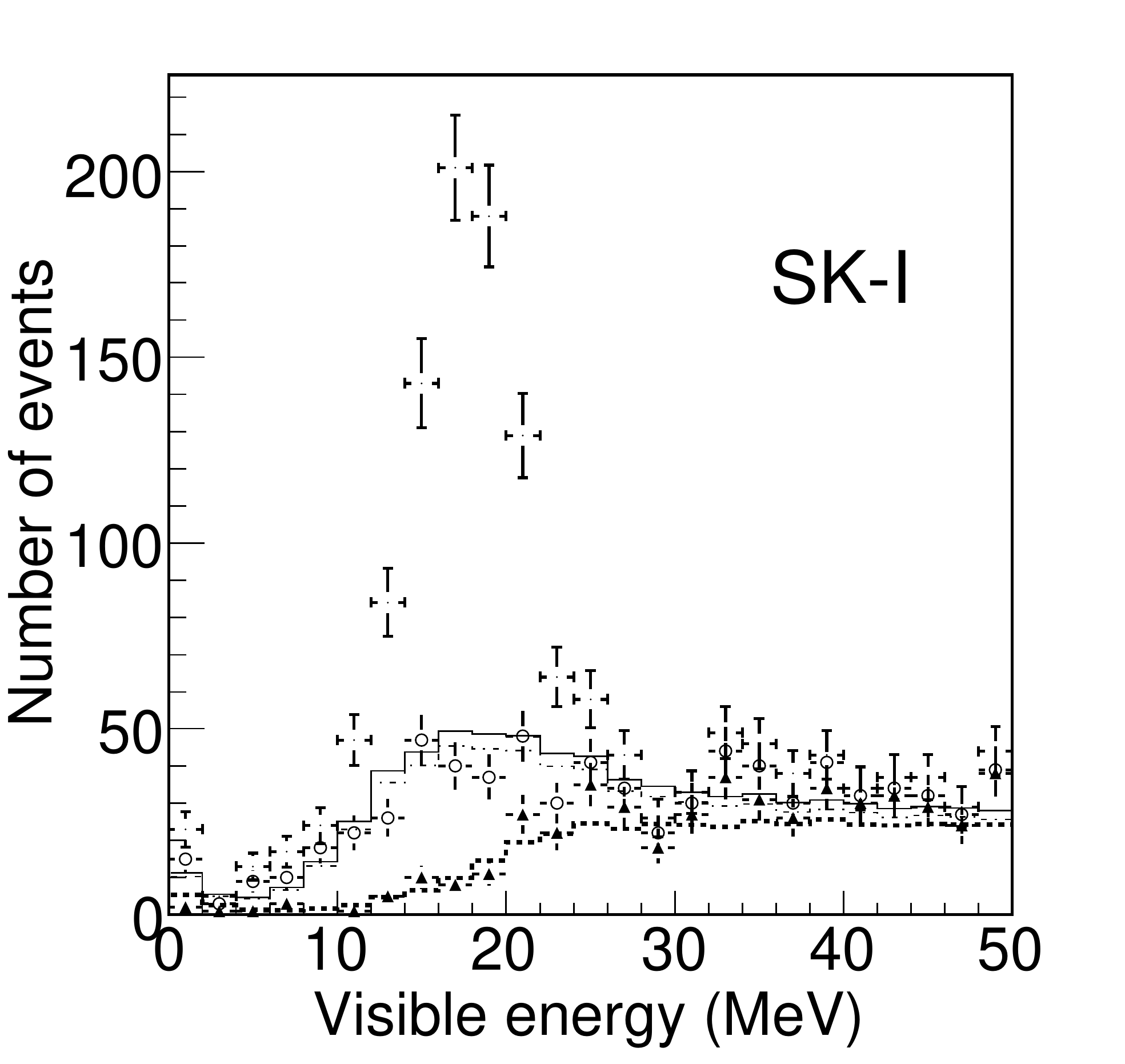}
\includegraphics[width=2.25in]{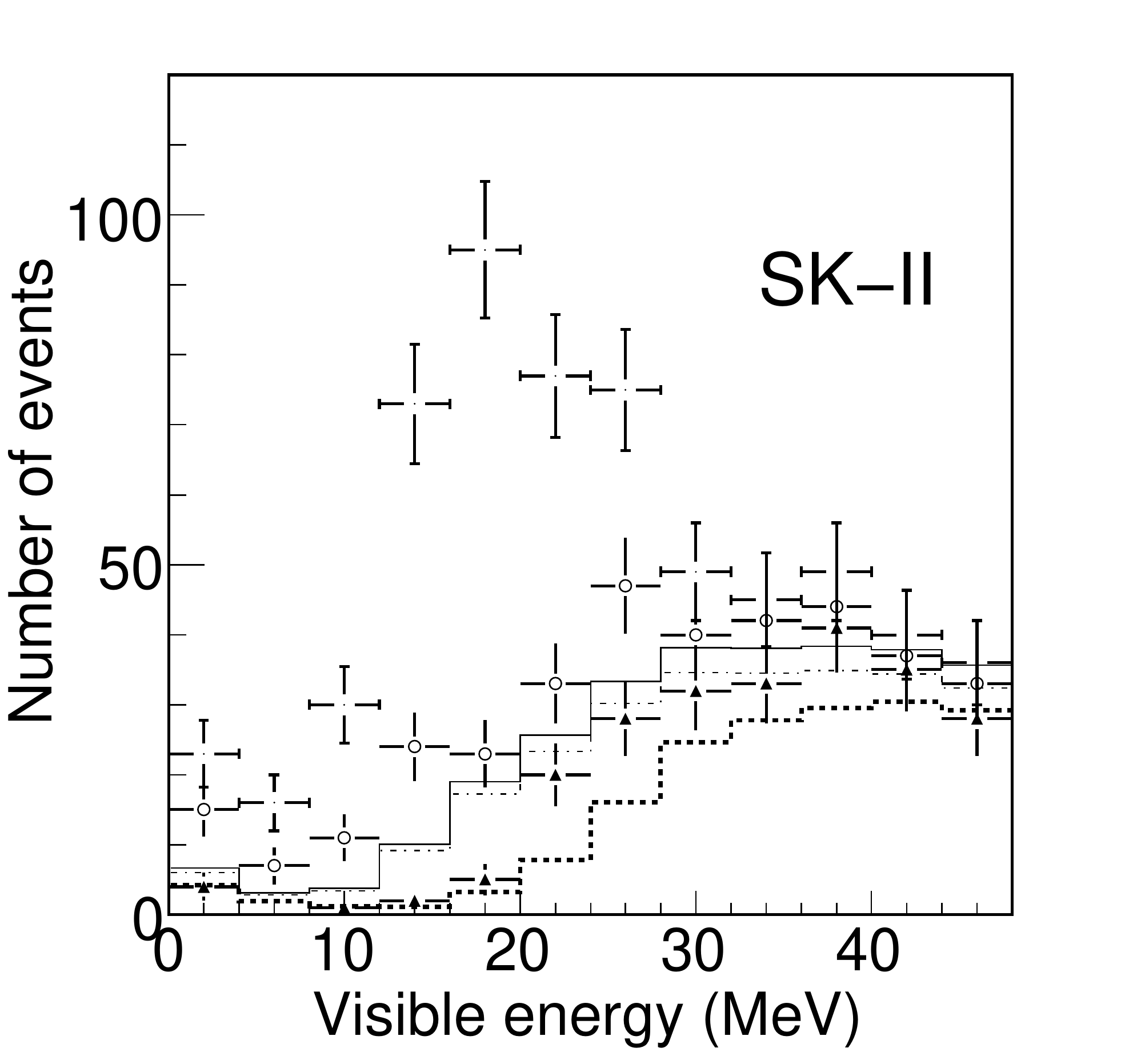}
\caption{Visible energy distribution for data and MC single-ring FC
  events for SK-I and SK-II is shown with no spallation cuts applied
  (crosses), after the spallation cut (open circles), and after the
  sparse ring cut (triangles). The corresponding MC distributions are
  shown (full line: no cuts, dash-dotted line: only the spallation
  inefficiency is applied, thick dashed line: the sparse ring cut and
  spallation inefficiency are applied). Spallation events are present
  only in the data.}
\label{fig:evissparse}
\end{center}
\end{figure}

In addition, an extra cut was developed to remove some remaining low
energy events with a sparse ring, \textit{i.e.} an incomplete Cherenkov ring
pattern. Using the fitted vertex and direction, an 6\degree-wide
annulus was constructed around the Cherenkov ring, and binned in the
azimuthal direction in 36 $\phi$-bins. A sparse ring will have empty
$\phi$-bins, and the average charge per bin will be low. A
two-dimensional cut on events with an average charge of less 1 (0.5) 
pe and more than 15 (22) empty $\phi$-bins in SK-I (SK-II) 
was applied. Studies show that this cut improves signal to
background significance at later stages.  Good agreement between data
and MC is reached once the low energy backgrounds are removed, as
can be seen in Fig.~\ref{fig:evissparse}.

\subsubsection{Proton track selection}

After the spallation and sparse ring removal cuts, three selection
criteria are applied on the sample to select proton events:

\begin{enumerate}

\item Events with more than 200 MeV of visible energy are
  discarded. This removes the bulk of the events used for the usual
  atmospheric analyses, caused by high energy muon or electron tracks.

\item Events with a Cherenkov opening angle greater than 37\degree
(35\degree in SK-II) are rejected. This effectively
removes high energy muons, as well as showers caused by
electrons or $\gamma$ rays from \pizero\  decays.

\item Finally, the pattern likelihood difference calculated by the
  proton fitter described in section \ref{sec:protonfit}, viz.
$$\Delta\log L =\log (L_{\mathrm{proton}}) - \log
  (L_{\mathrm{muon}})$$ is required to be positive. This selects
  events with a proton-like pattern.

\end{enumerate}

\begin{figure*}[!htb]
  \begin{minipage}{7in}
  \includegraphics[width=2.25in]{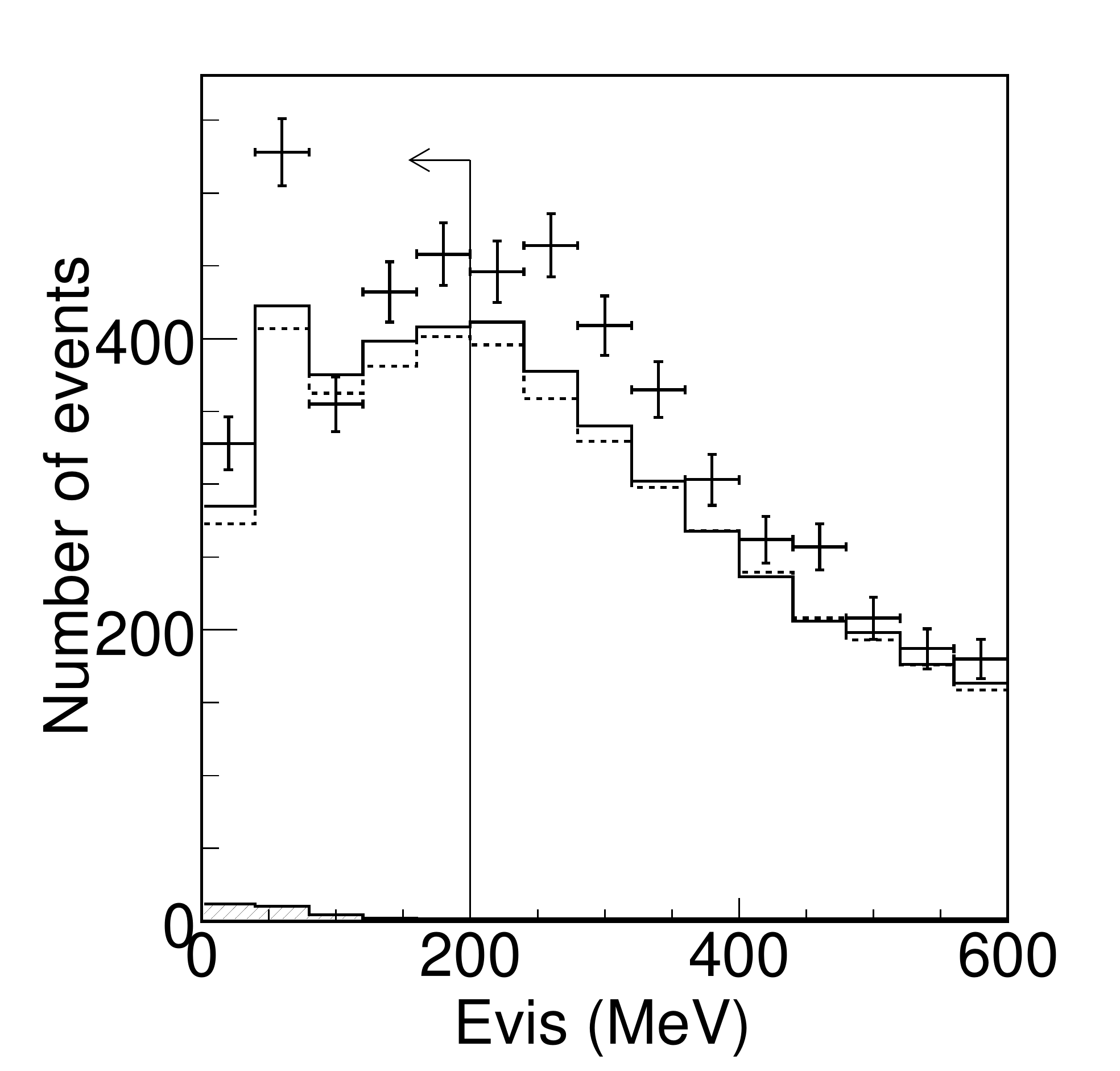}
  \includegraphics[width=2.25in]{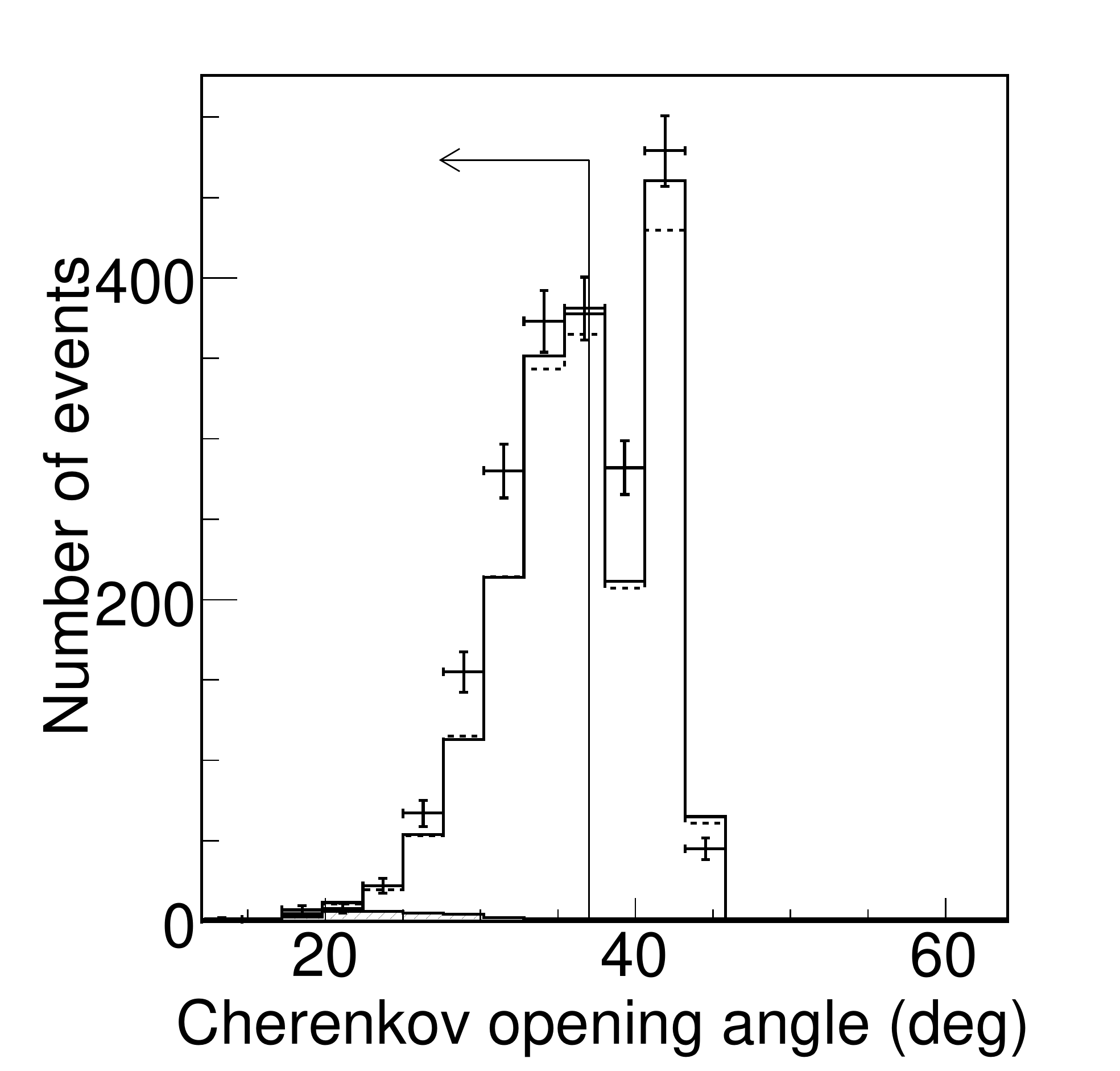}
  \includegraphics[width=2.25in]{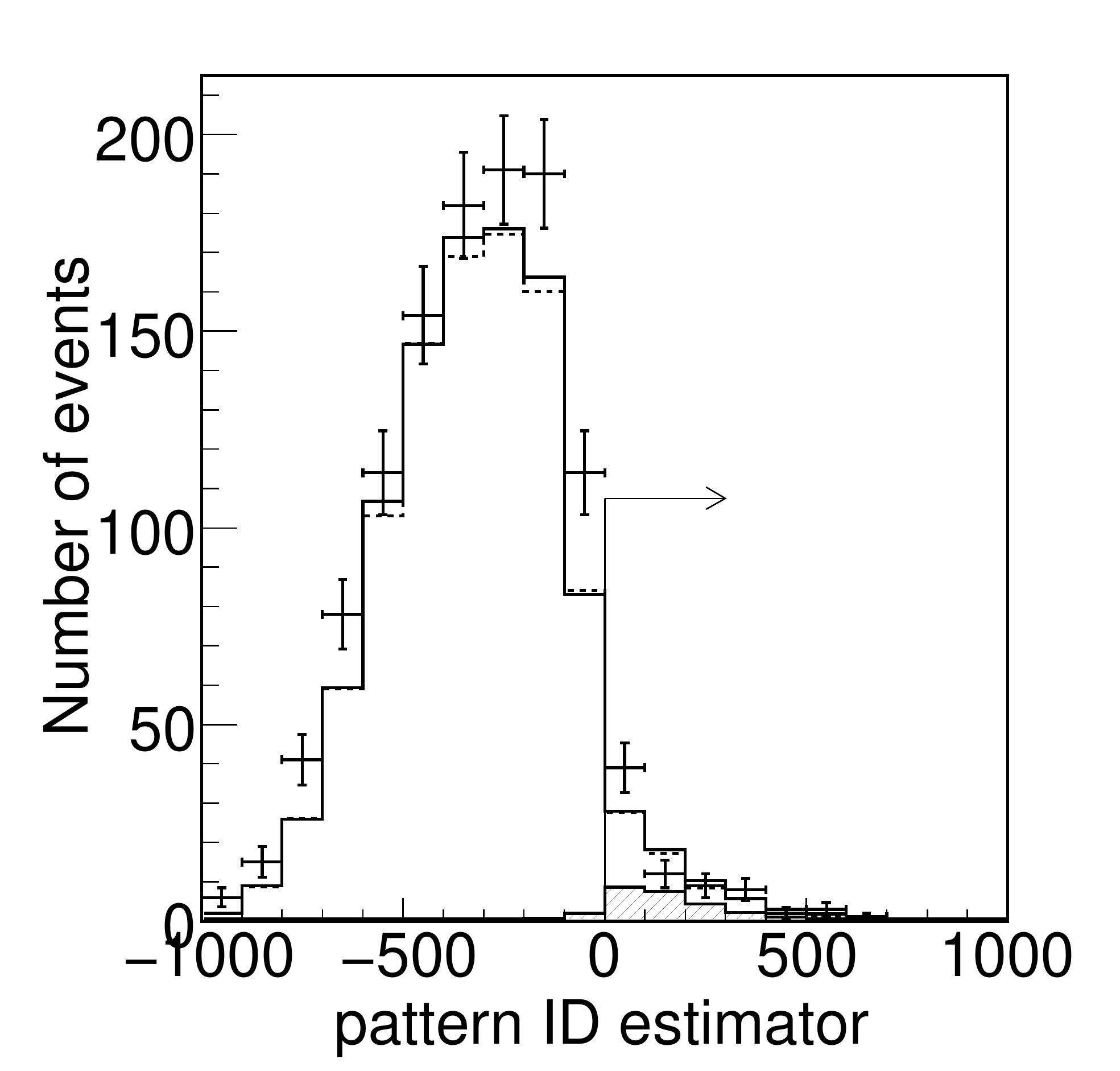}
  \end{minipage}
  \begin{minipage}{7in}
  \includegraphics[width=2.25in]{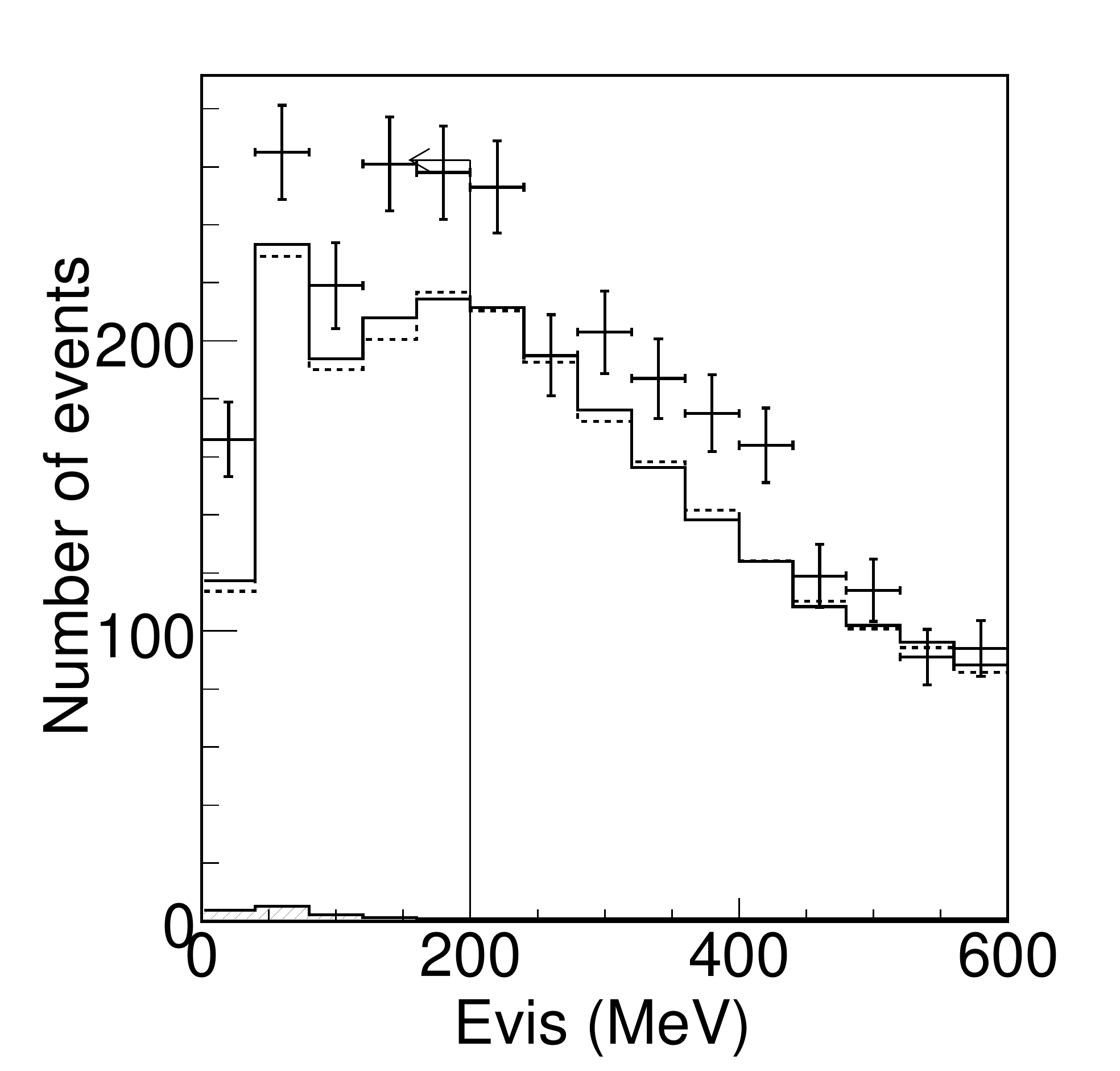}
  \includegraphics[width=2.25in]{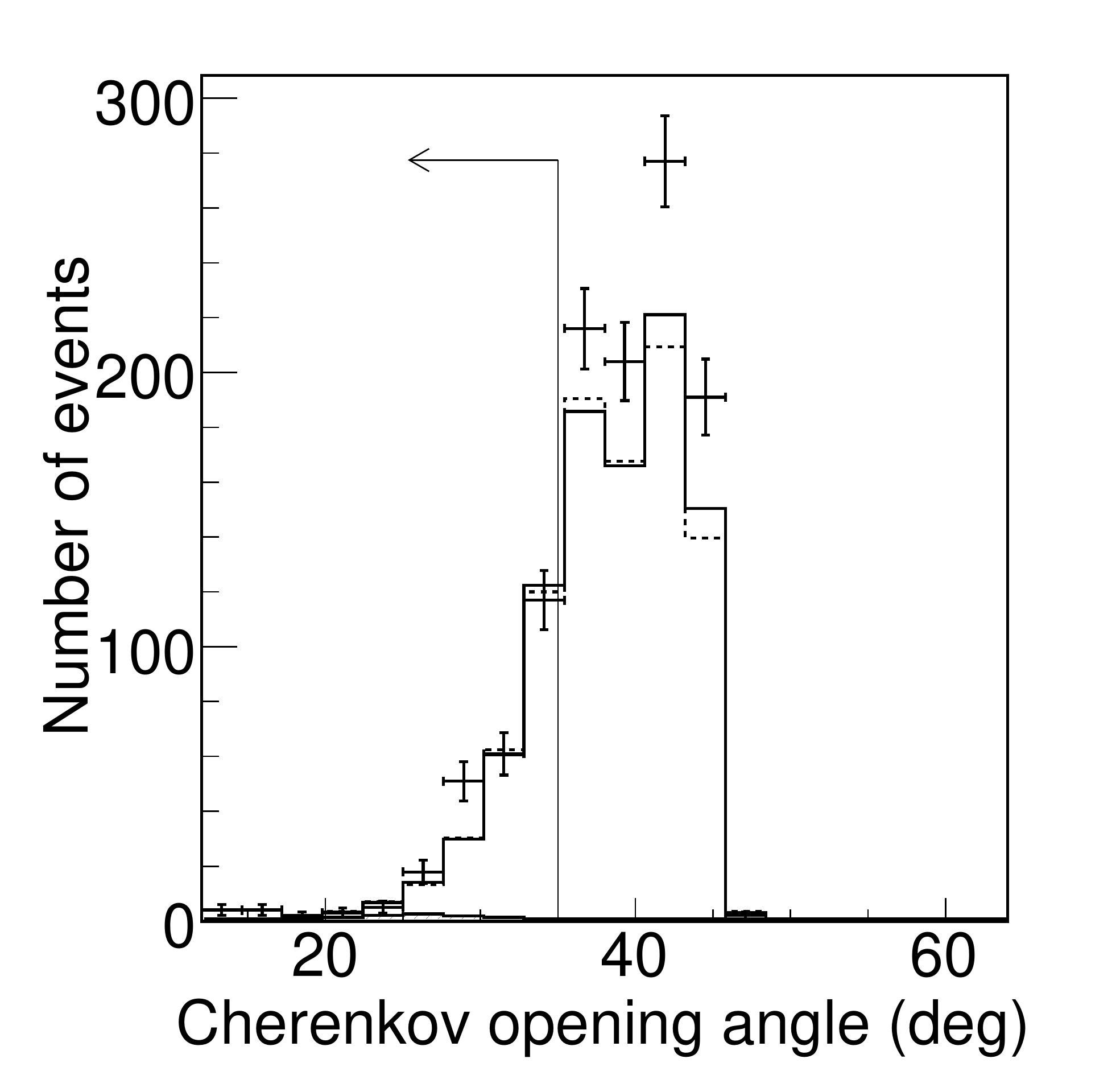}
  \includegraphics[width=2.25in]{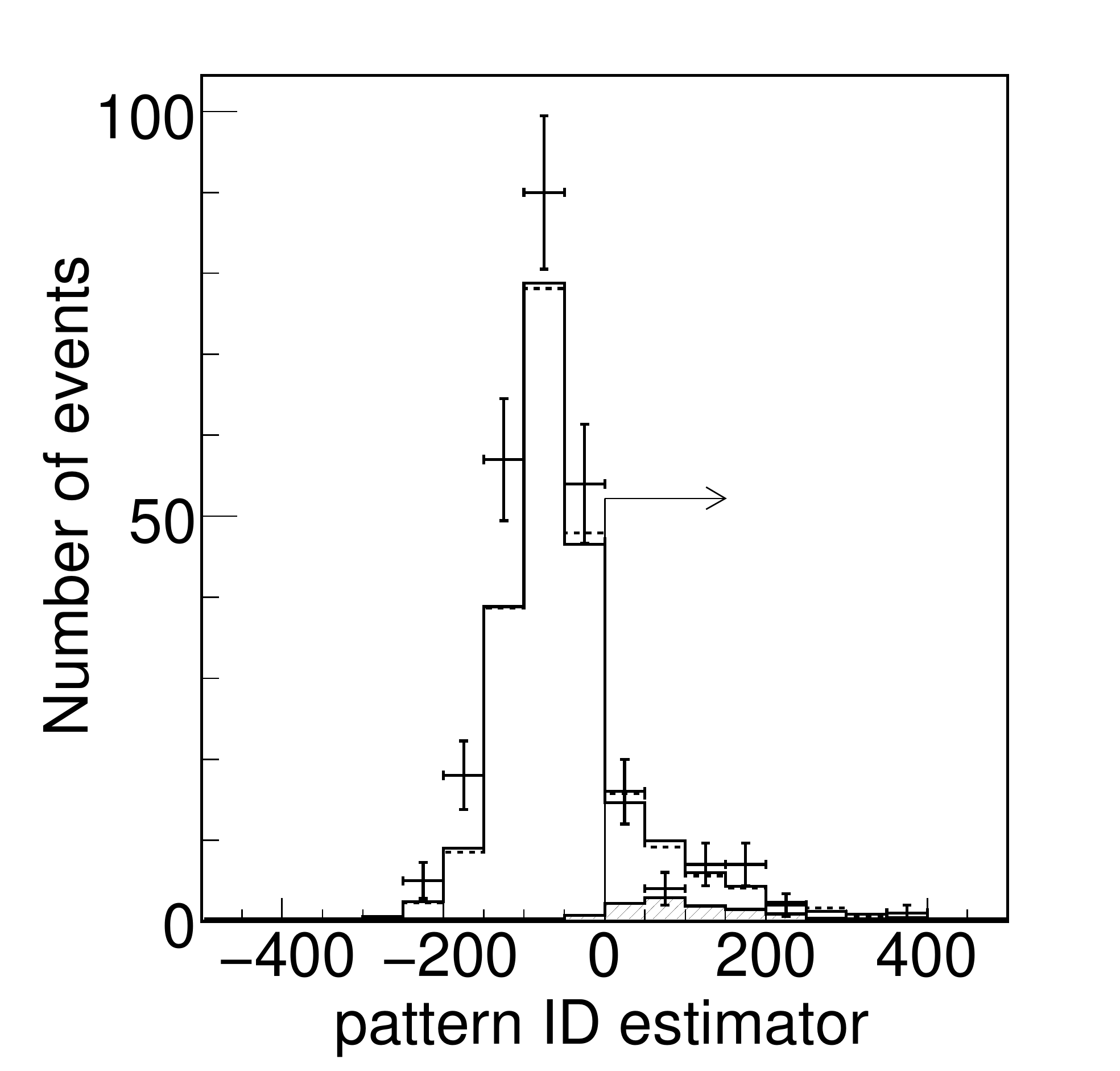}
  \end{minipage}
  \caption{Distribution of the visible energy, Cherenkov opening
    angle, and proton pattern likelihood for single ring events after
    spallation and sparse ring removal, for SK-I (top row) and SK-II
    (bottom row).  The full and dashed line show Monte-Carlo
    expectation from NEUT and NUANCE, with oscillation reweighting,
    but no correction for the absolute flux normalization. The hatched
    histogram shows the fraction of signal events according to NEUT.}
\label{fig:nccuts}
\end{figure*}

The distribution of the variables are shown in Fig.~\ref{fig:nccuts}.
The efficiencies of all cuts are summarized in
tables~\ref{table:nc-sk1} and \ref{table:nc-sk2}. In these tables,
events labeled as signal are events with a single visible track caused
by a proton above Cherenkov threshold; this is comprised of $\sim
55\%$ NC elastic events on protons, the rest being mostly neutral
current pion production collisions where the pion was absorbed. These
interactions are an irreducible background to a search for proton NC
elastic events, but we include them in the signal since this analysis
technique is designed to maximize the efficiency of proton track
selection while reducing contamination from muon and pion tracks.  At
this stage, there are $\sim 25.6$ ($\sim 10.4$) expected signal events
in SK-I (SK-II).

\begin{table*}[!htbp]
  \begin{tabular}{lccccc}
    \hline
    \hline
    \multirow{2}{100pt}{\superk -I} & ~Data~ & ~Total MC~ & ~Signal MC~ & ~Total MC~ & ~Signal MC~\\
    & & ~NEUT~ & ~NEUT~ & ~NUANCE~ & ~NUANCE~ \\
    \hline FC, FV, single-ring, spallation removed & 8946 (100\%) & 8138.1(100\%) & 45.1 (100\%) & 8031.5 (100\%) & 41.2 (100\%) \\
    Sparse ring removal cut & 8509 (95.1\%) & 7729.7 (95.0\%) & 31.7 (70.4\%) & 7673.4(95.5\%) & 29.3 (71.1\%)\\
    $E_{vis} < 200$ MeV &  2101 (23.5\%) & 1894.2 (23.3\%) & 29.7 (65.9\%) & 1843.5 (23.0\%) &  27.9 (67.7\%)\\
    Cone opening angle $<37$\degree & 1161 (13.0\%) & 1020.0 (12.5\%) & 28.9 (64.2\%) & 1009.4(12.6\%) & 26.6 (64.5\%) \\
    Pattern ID estimator cut & 74 (0.83\%) & 68.8 (0.85\%) & 25.6 (56.8\%) & 65.8 (0.82\%) & 22.7 (55.0\%) \\
    \hline \hline
  \end{tabular}
  \caption{Summary of SK-I NC elastic selection. In this table the
    Monte-Carlo has been reweighted according to live time, solar
    activity for SK-I, as well as neutrino oscillations assuming
    \dms$=2.5\times 10^{-3}~\mathrm{eV}^{2}$ and \sstt $=1$. Both NEUT and
    NUANCE simulations return similar results.}
  \label{table:nc-sk1}
\end{table*}

\begin{table*}[!htbp]
  \begin{tabular}{lccccc}
    \hline
    \hline
    \multirow{2}{100pt}{\superk -II} & ~Data~ & ~Total MC~ & ~Signal MC~ & ~Total MC~ & ~Signal MC~\\
    & & ~NEUT~ & ~NEUT~ & ~NUANCE~ & ~NUANCE~ \\
    \hline FC, FV, single-ring,spallation removed & 4700 (100\%) & 4190.1 (100\%) & 18.9 (100\%) & 4174.2 (100\%) & 17.9 (100\%) \\
    Sparse ring removal cut & 4464 (95.0\%) & 4004.6 (95.6\%) & 13.9 (73.3\%) & 3992.0(95.6\%)  & 12.6 (70.4\%) \\
    $E_{vis} < 200$ MeV &  1169 (24.9\%) & 969.4 (23.3\%) & 12.8 (67.5\%)  & 953.0 (22.8\%) & 11.8 (66.1\%)\\
    Cone opening angle $<35$\degree & 261 (5.55\%) & 217.1 (5.18\%) & 11.6 (61.1\%) & 216.7 (5.2\%) & 10.6 (59.5\%)  \\
    Pattern ID estimator cut & 37 (0.79\%) & 39.9 (0.95\%) & 10.4 (54.9\%) & 39.6 (0.95\%) & 9.7 (54.5\%) \\
    \hline \hline
  \end{tabular}
  \caption{Summary of SK-II NC elastic selection. In this table the
    Monte-Carlo has been reweighted according to live time, solar
    activity for SK-II, as well as neutrino oscillations assuming
    \dms$=2.5\times 10^{-3}~\mathrm{eV}^{2}$ and \sstt $=1$. The cut on the
    cone opening angle is tighter than in SK-I, since the fitter
    relies on a different ring edge finding method. Both NEUT and
    NUANCE simulations yield similar results.}
  \label{table:nc-sk2}
\end{table*}

\subsection{Neural network selection}

 At this stage, the signal-to-noise ratio has been decreased from
 $\sim 1/200$ to $\sim 1/3$. It is possible to improve this using
 multivariate analysis techniques. Here we chose to use a neural
 network. We used the TMVA library \cite{TMVA} for this calculation.

 \subsubsection{Neural network input variables and training}

\begin{figure*}[!htb]

  \includegraphics[width=5.5in]{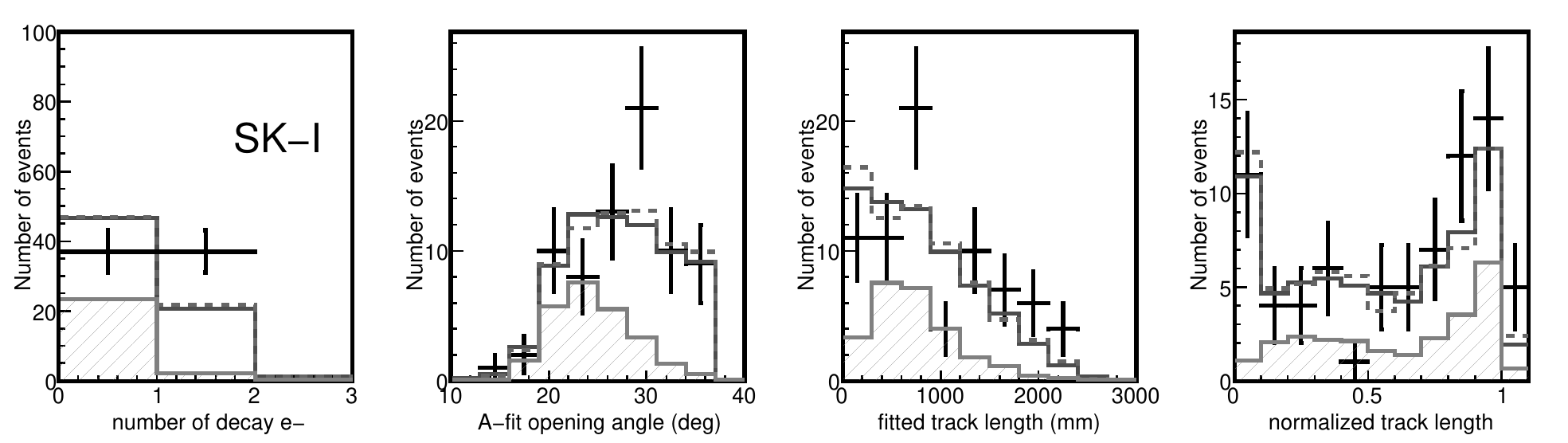}
  \vspace{0.2in}
  \includegraphics[width=5.5in]{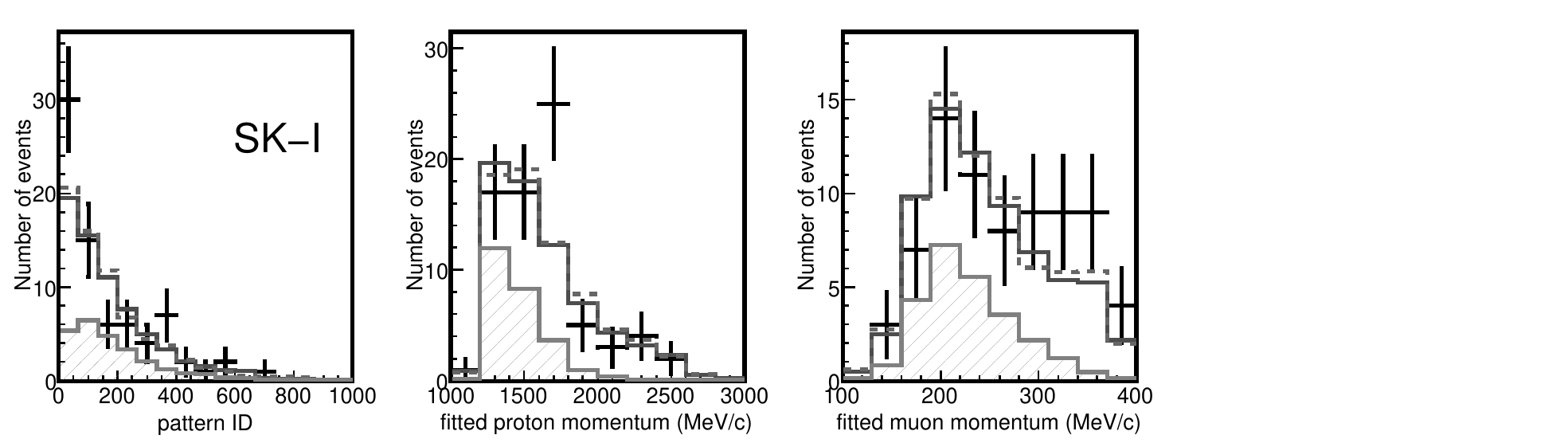}
  \vspace{0.2in}
  \includegraphics[width=5.5in]{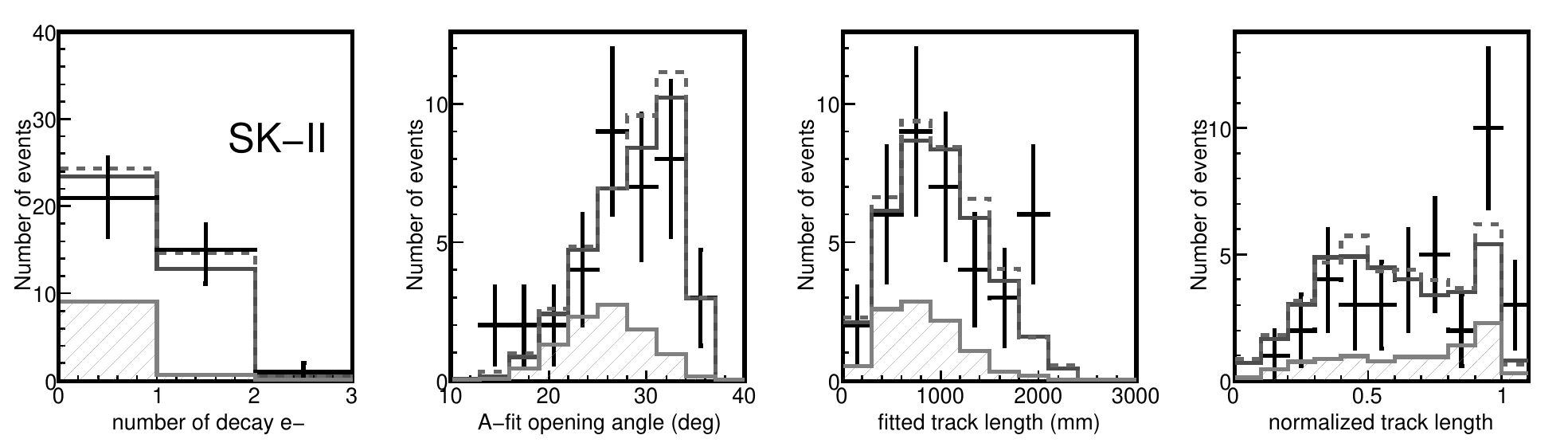}
  \vspace{0.2in}
  \includegraphics[width=5.5in]{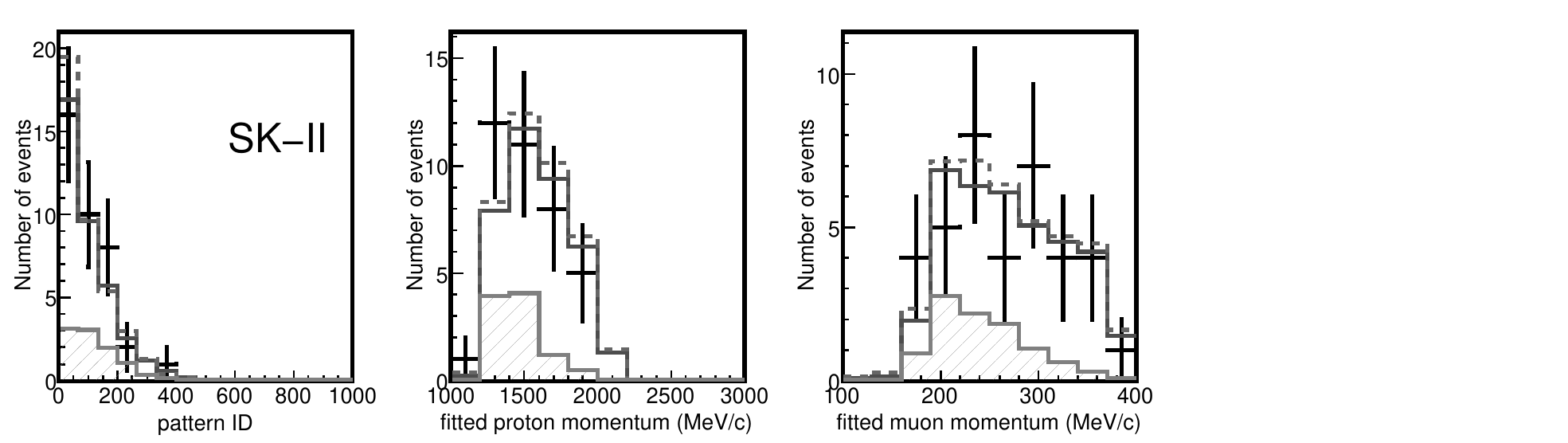}    
\caption{Discriminating variables used as input for the neural network.
  The expectations from NEUT (full line) and NUANCE (dashed line) are 
  shown. The hatched histogram shows the fraction of signal events according
  to NEUT.}
\label{fig:nninput}

\end{figure*}

 As inputs to the network, seven variables were used: the Cherenkov
 opening angle, the $\mu$-like momentum of the event, the fitted
 proton-like length, the fitted proton-like momentum, the pattern
 likelihood difference, the number of decay electrons in the event,
 and the \emph{normalized length}. The normalized proton length $L_n$
 is defined by $$ L_n = \frac{L}{L_{max}(P)},$$ where $L$ and $P$ are
 the measured proton-like track length and momentum, and $L_{max}(P)$
 is the maximum pathlength in water for a proton of momentum $P$.
 Figure~\ref{fig:nninput} shows the distributions of these variables
 for data and Monte-Carlo. Signal events are shown as hatched histograms.
 
 A large sample of NC elastic events on protons was prepared to be
 used as signal for training purposes.  Several samples of low
 momentum monochromatic charged pions (from 180 to 1000 MeV/c) and
 muons (from 140 to 500 MeV/c) were used as background training
 samples. All reconstruction algorithms and previous selection cuts
 were applied to those events, which were then fed into the neural
 network for training. Checks were conducted to make sure no
 overtraining occurred. The best architecture was found to be a single
 hidden layer with 6 neurons.

 Since the distributions of the discriminating variables are different
 for SK-I and SK-II, two separate networks with identical
 architectures were used.
 
\subsubsection{Neural network response}
\begin{figure*}[!htbp]
  \begin{center}
      \includegraphics[width=2.25in]{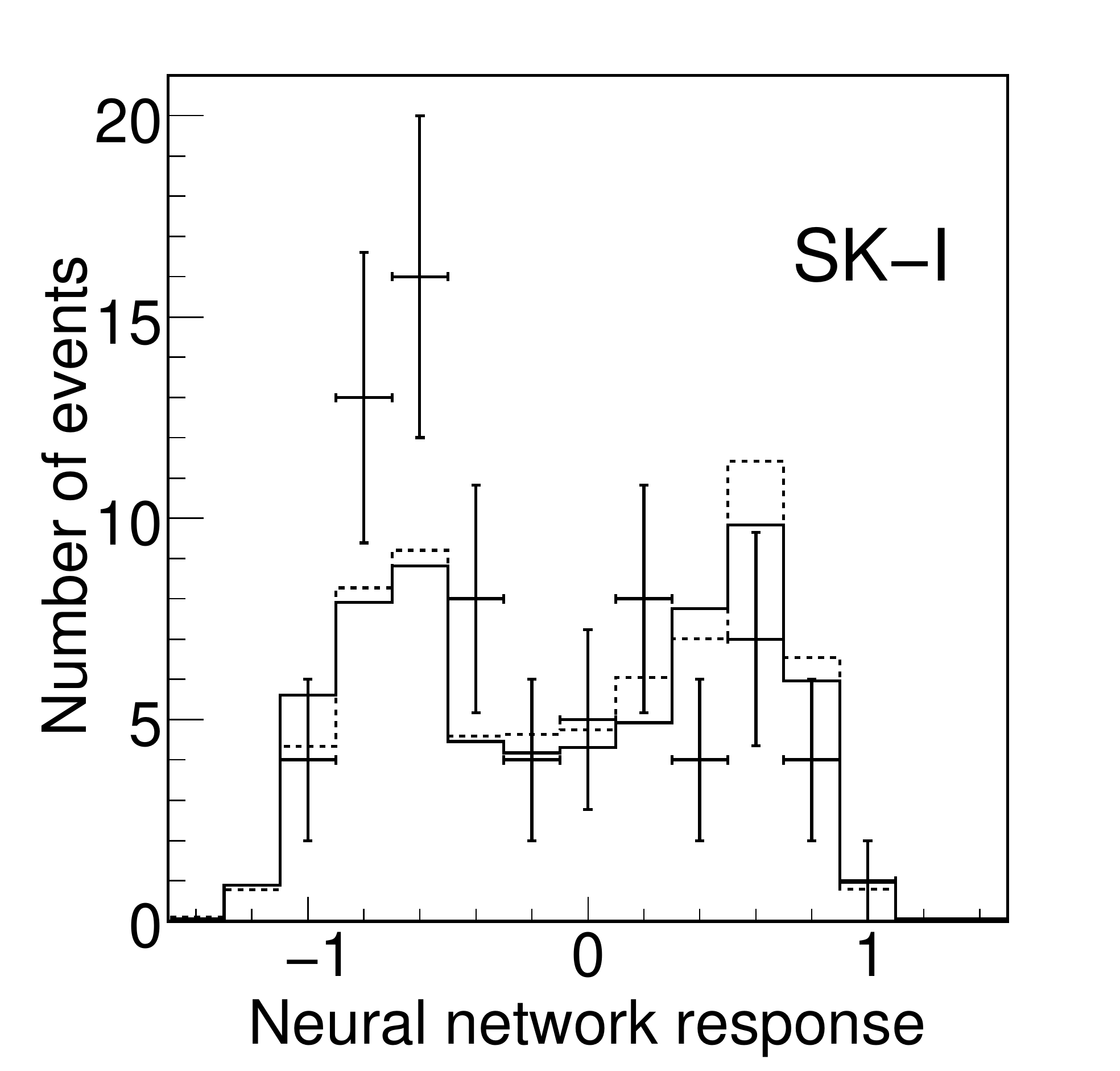}
      \includegraphics[width=2.25in]{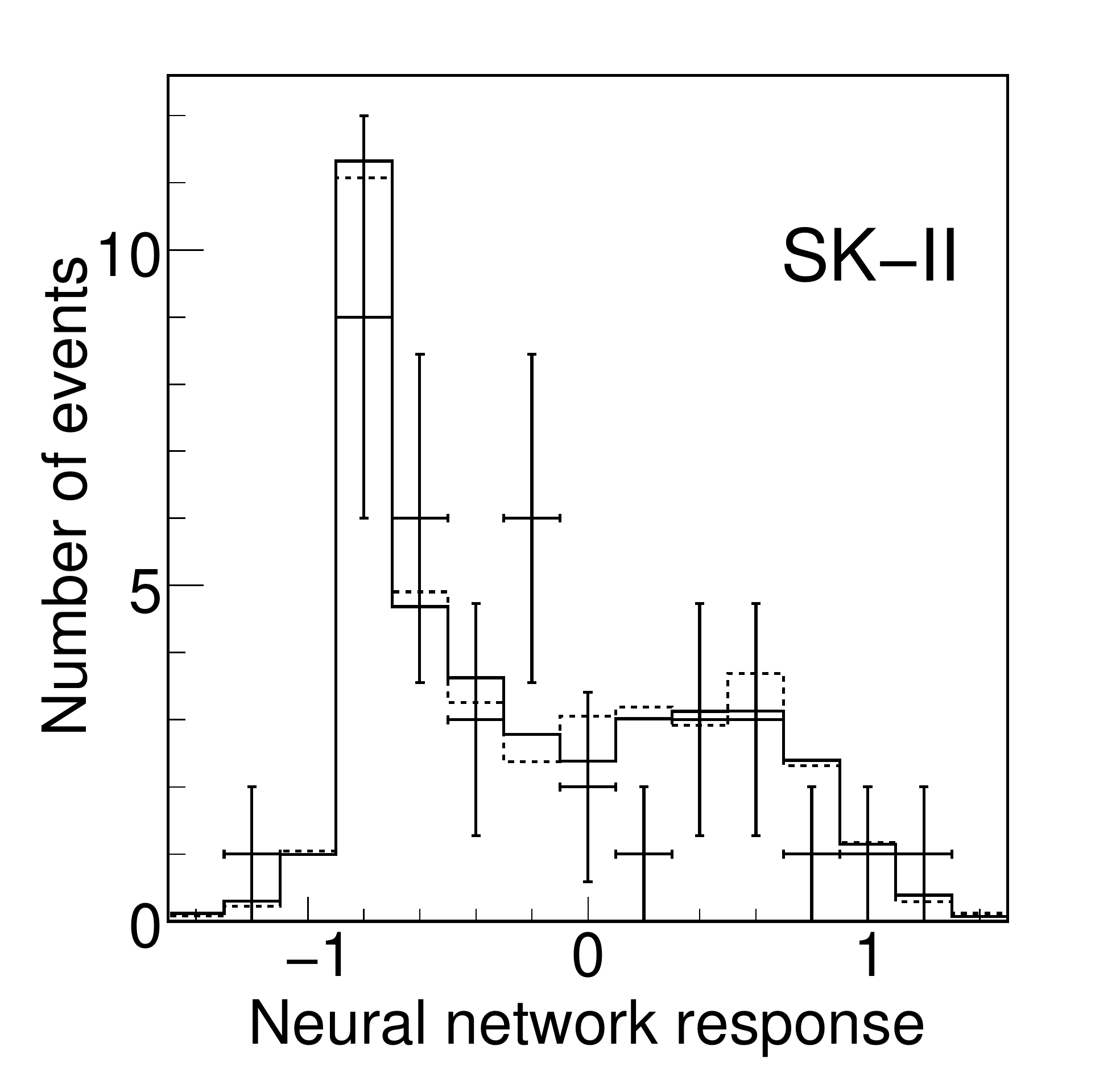}
      \includegraphics[width=2.25in]{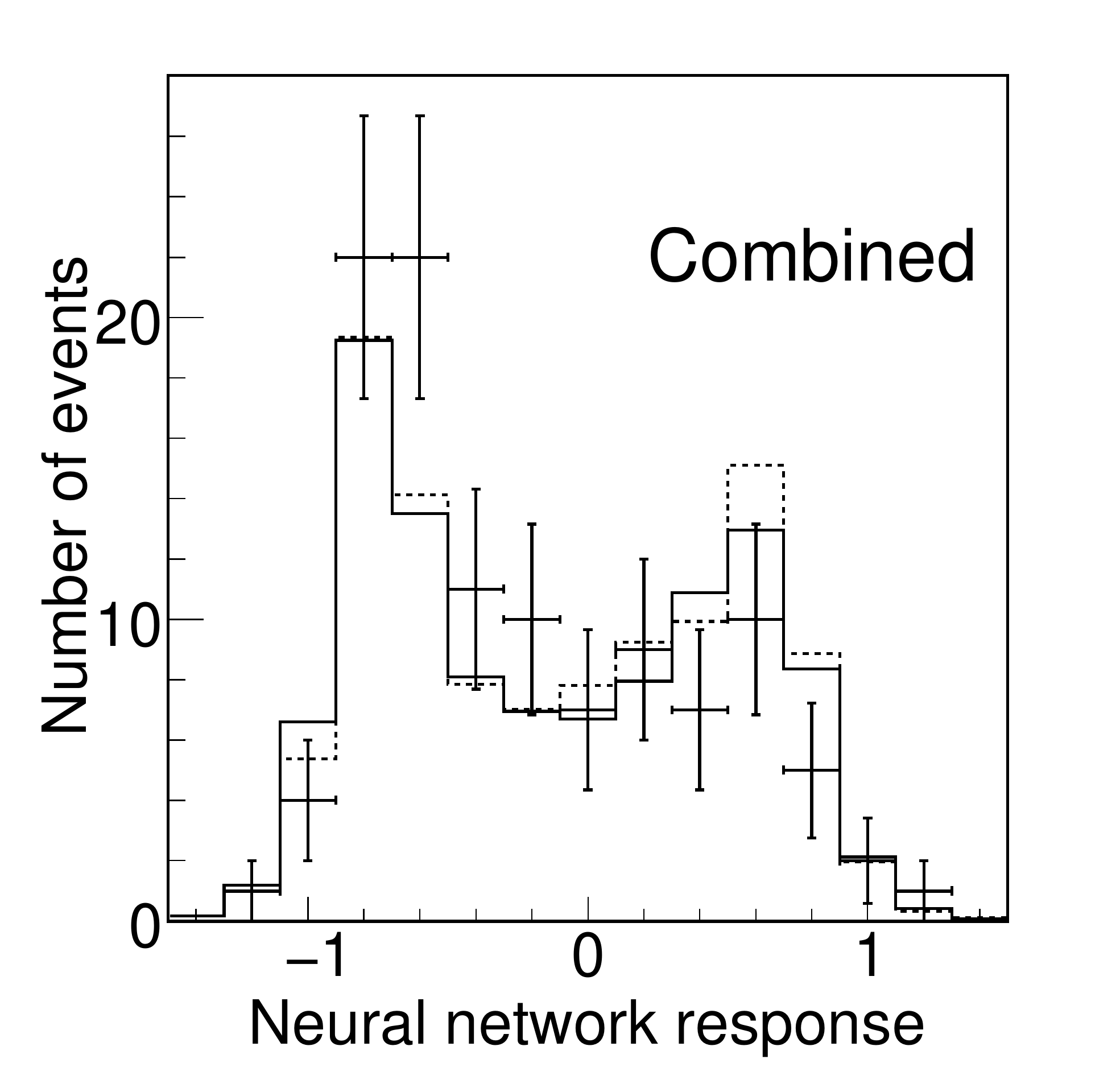}
   \end{center}
  \caption{Neural network response for data and MC (full line: NEUT,
    dashed line: NUANCE).  Left: SK-I, Middle: SK-II, Right:
    combined. The MC were oscillated with $(\Delta
    m^2=2.5\times 10^{-3}\ \mathrm{eV}^2,\sin^2 2\theta=1.0)$.}
\label{fig:nnout}
\end{figure*}

\begin{figure*}[!htb]
\begin{center}
\begin{minipage}{4.5in}
\includegraphics[width=2.in]{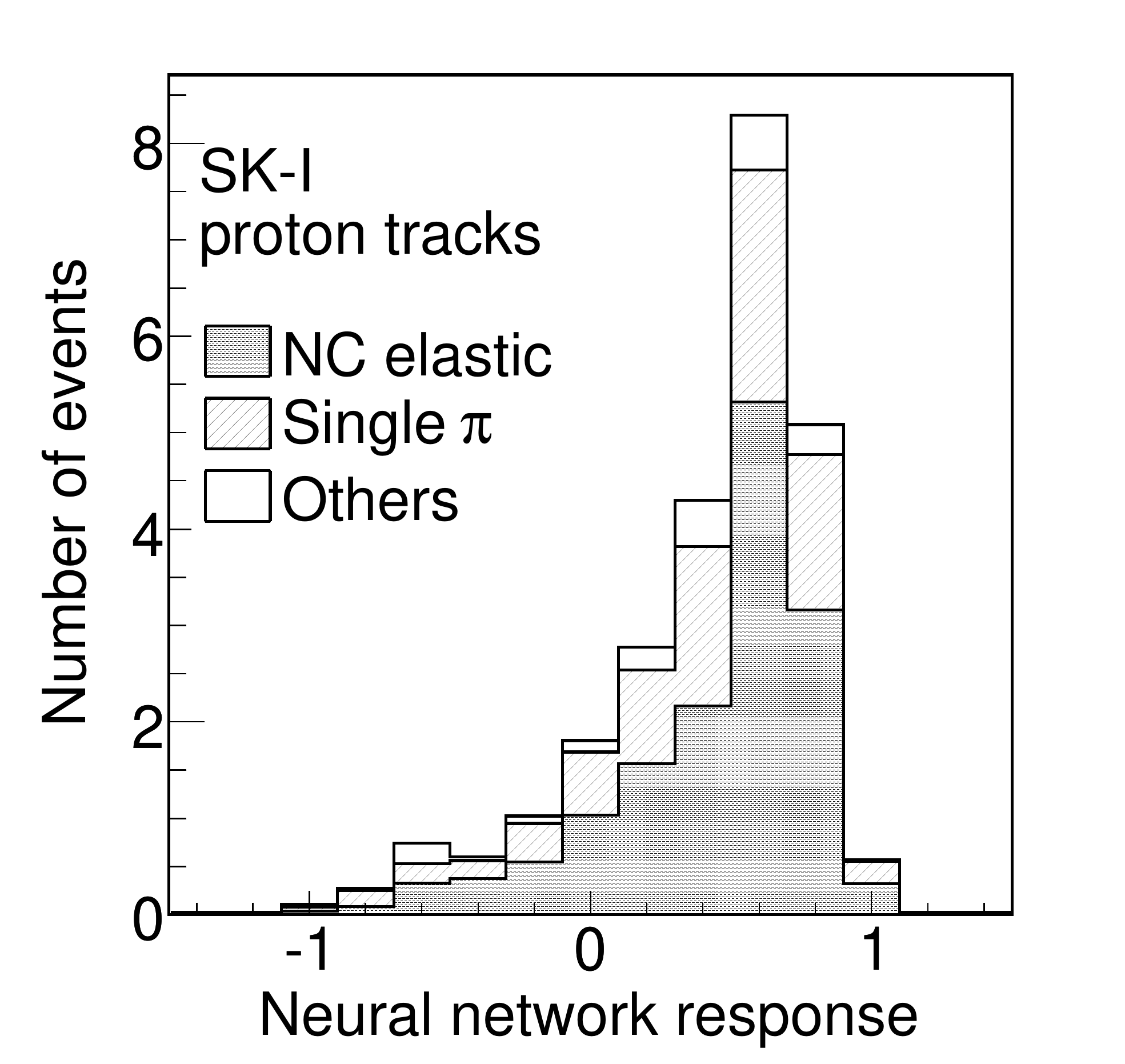}
\includegraphics[width=2.in]{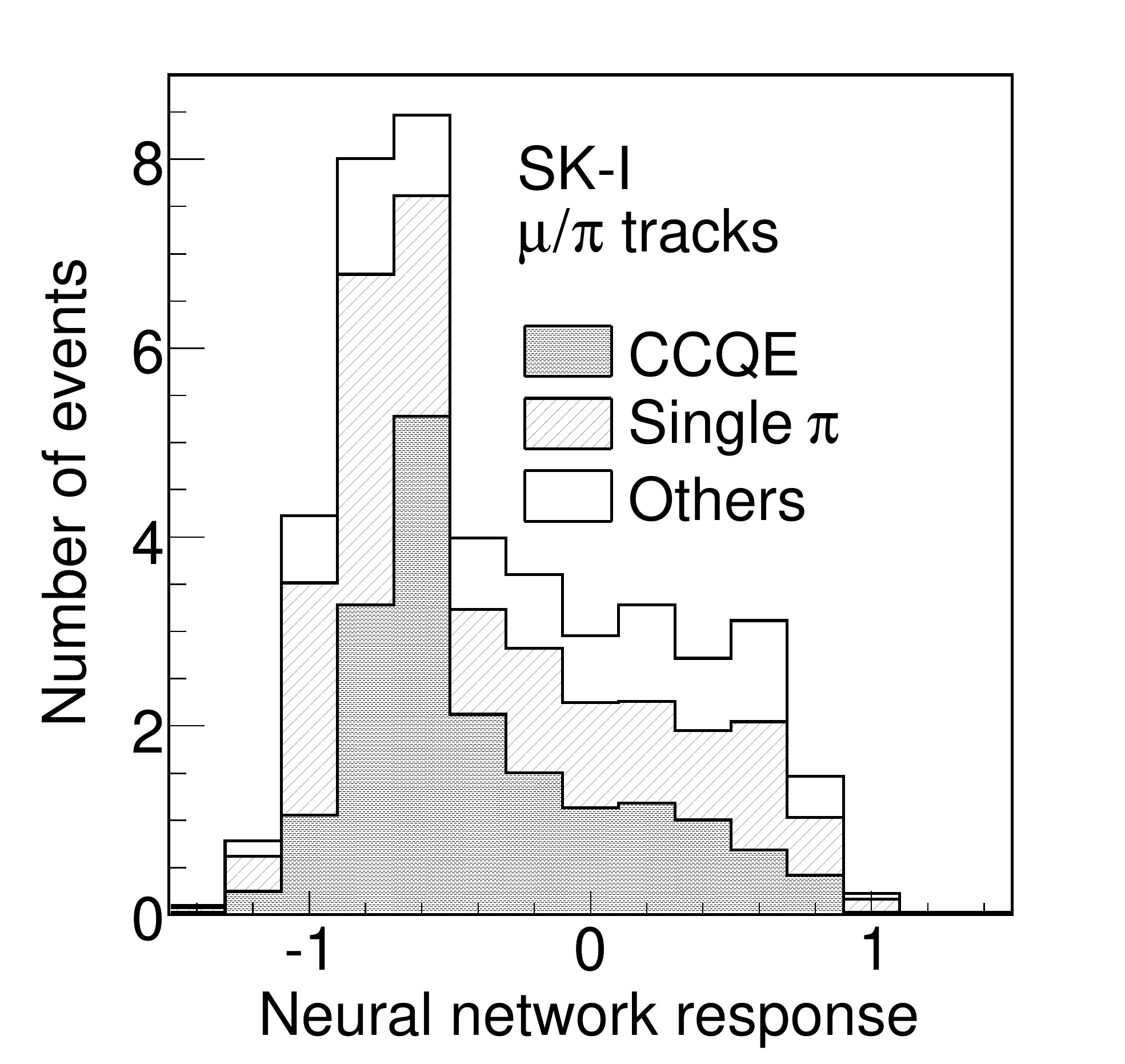}
\end{minipage}
\begin{minipage}{4.5in}
\includegraphics[width=2.in]{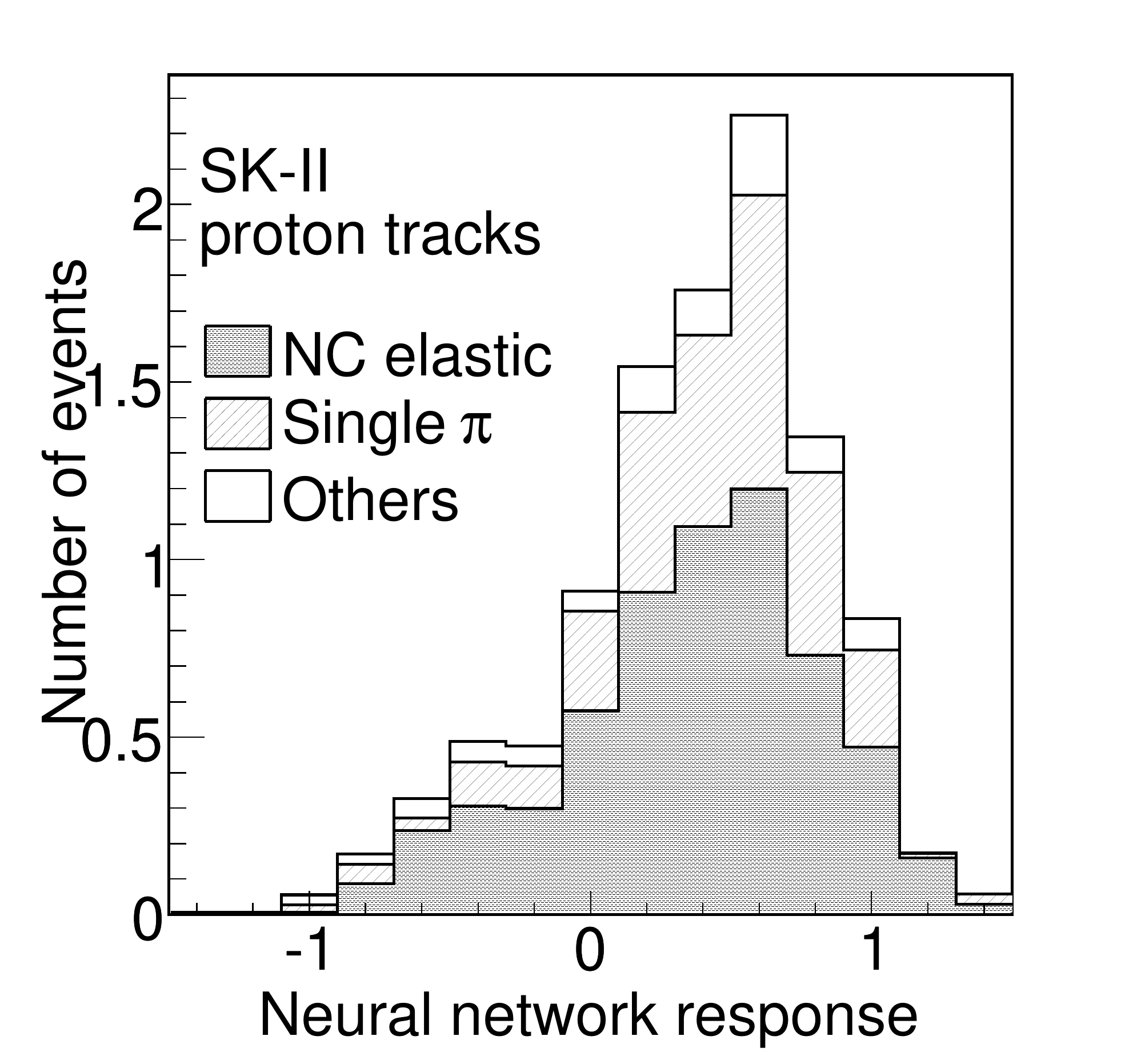}
\includegraphics[width=2.in]{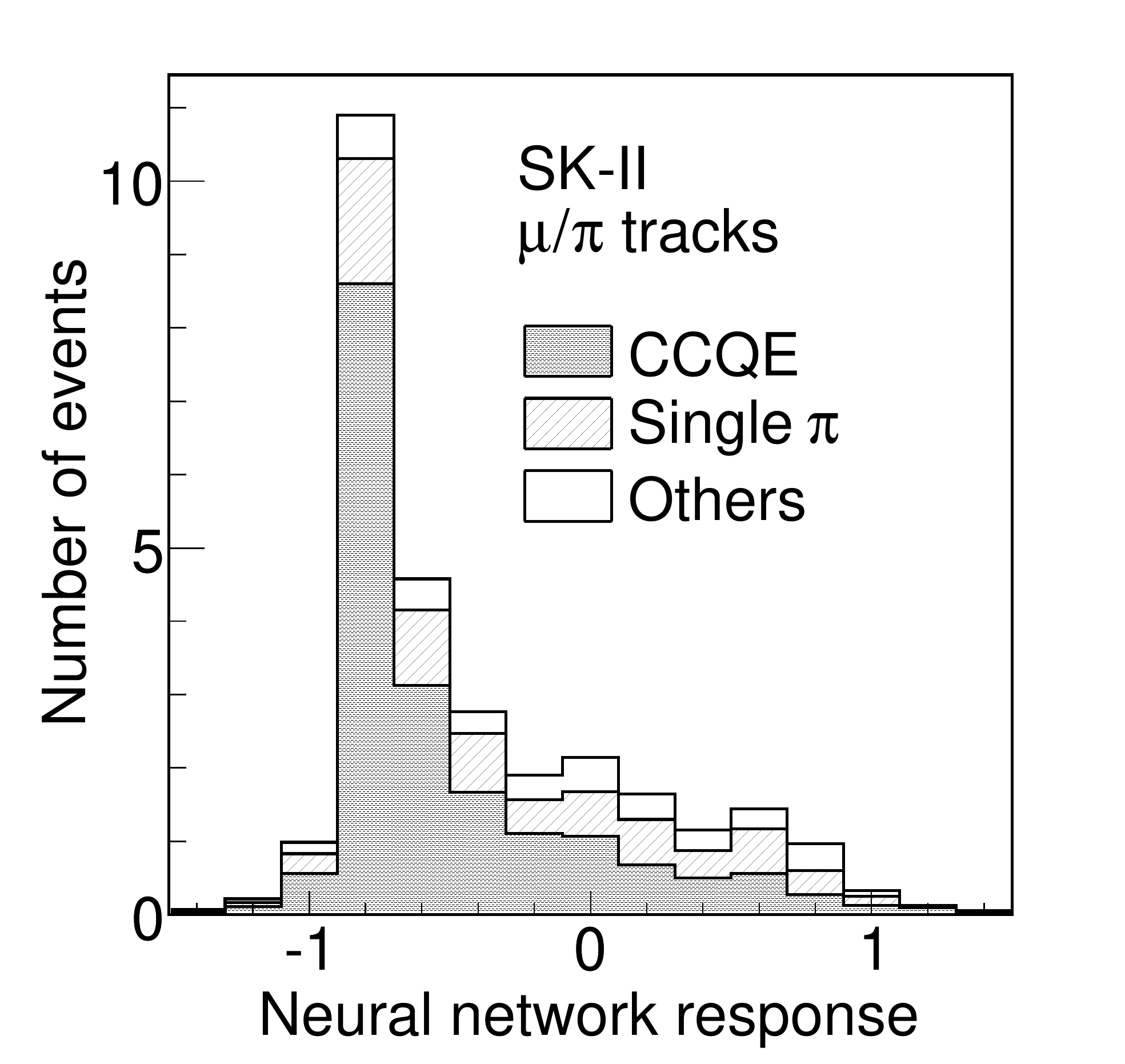}
\end{minipage}
\caption{Top: SK-I neural network response for signal and background,
separated according to the main interaction modes. For signal events,
the main mode is NC elastic on proton, followed by NC single pion
production where the pion is absorbed. The rest is mostly NC elastic
on neutrons where the neutron produced a proton. Background mostly
comes from low energy CCQE muon events, and pion tracks from single
pion interactions.}
\label{fig:modebreakdown}
\end{center}
\end{figure*}

  Figure~\ref{fig:nnout} shows the response of the neural networks,
  labelled NN in what follows.  Using the Monte-Carlo, we have studied
  the composition of the signal and background as a function of neural
  network response NN. Figure~\ref{fig:modebreakdown} shows the
  results. It should be noted that the neutral current content of the
  sample before any NN cut is 66\% (45\%) for SK-I (SK-II); when
  keeping only events with $NN>0$, this fraction reaches 85\%
  (76\%). The $NN>0$ cut was chosen because it maximizes the
  significance $\mathrm{signal}/\sqrt{\mathrm{signal}+
    \mathrm{background}}$.

\subsection{Results from the high purity neutral-current proton
  enriched sample}

  After the $NN>0$ cut we expect 34.3 events in SK-I data and 15.3 in
  SK-II data; out of those, 22.2 and 8.5 respectively are events with
  a single track caused by a visible proton; 13.1 and 5.0
  (respectively) are neutral current elastic collisions on protons
  according to the NEUT simulation.  We observe 27 events in SK-I and
  11 events in SK-II.

  \begin{table*}[!htbp]
    \begin{tabular*}{0.75\textwidth}{@{\extracolsep{\fill}}lccc}
      \hline
      \hline
      Run period & Data & Expected signal & Expected background \\
      SK-I       & $27$ & 22.1  &  12.2  \\
      \hline
      SK-II      & $11$ &  8.5  &   6.8 \\
      \hline \hline
    \end{tabular*}
    \caption{Summary of neural network selection.}
    \label{table:nnout}
  \end{table*}

  We calculated the $\chi^2$ of the data neural output distributions
  to Monte-Carlo with and without single visible proton events,
  (using 6 bins from -1 to 1 to increase the bin contents for 
  statistical purposes).
  Using the Monte-Carlo including all events the $\chi^2$ is
  9.3 for 6 bins (probability $15.7\%$); when all visible protons
  are excluded the $\chi^2$ value is 15.8 (probability 1.5\%).
  The data is clearly compatible with the observation of single
  visible protons.

\begin{figure*}[!htb]
  \includegraphics[width=2.0in]{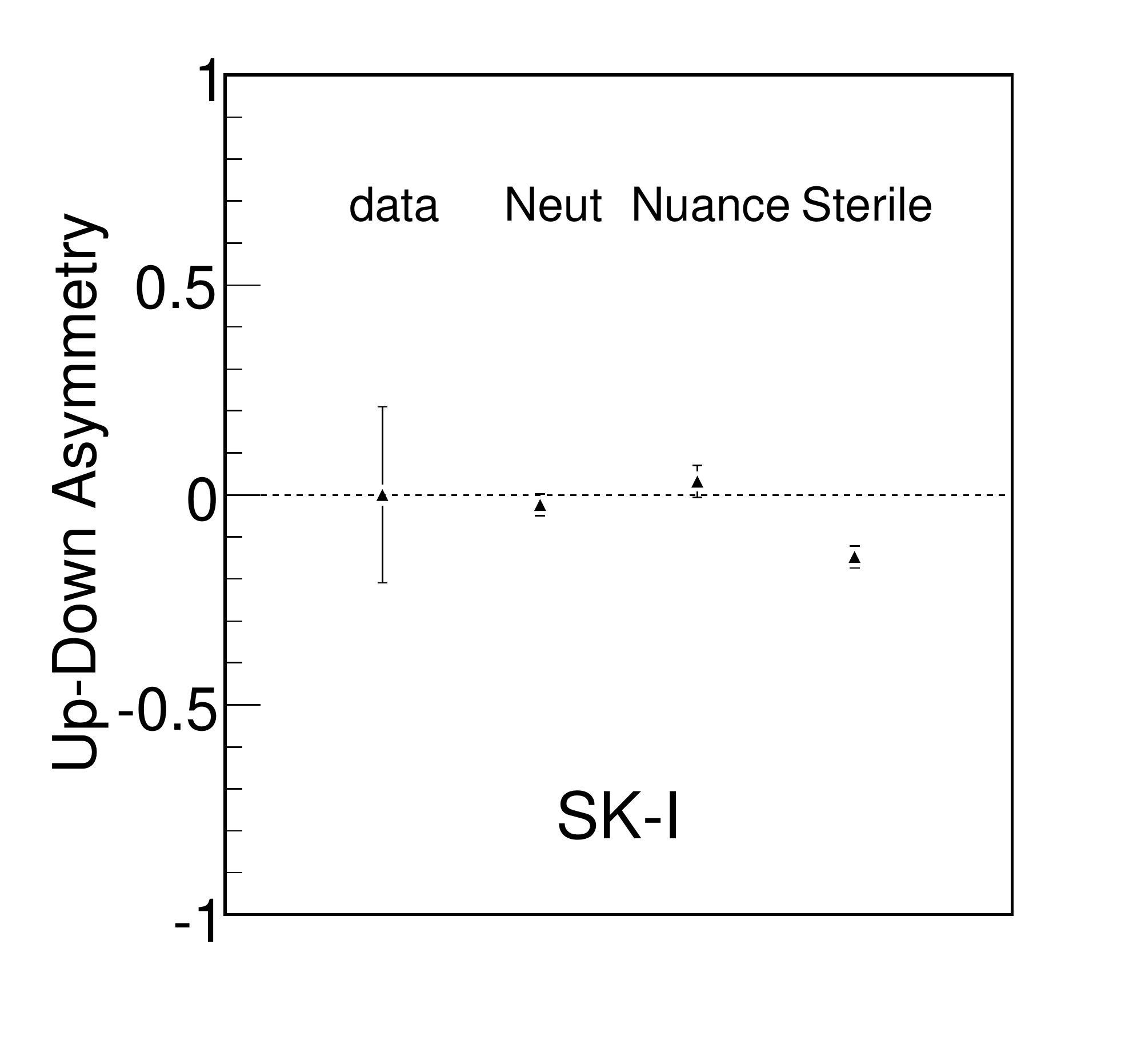}
  \includegraphics[width=2.0in]{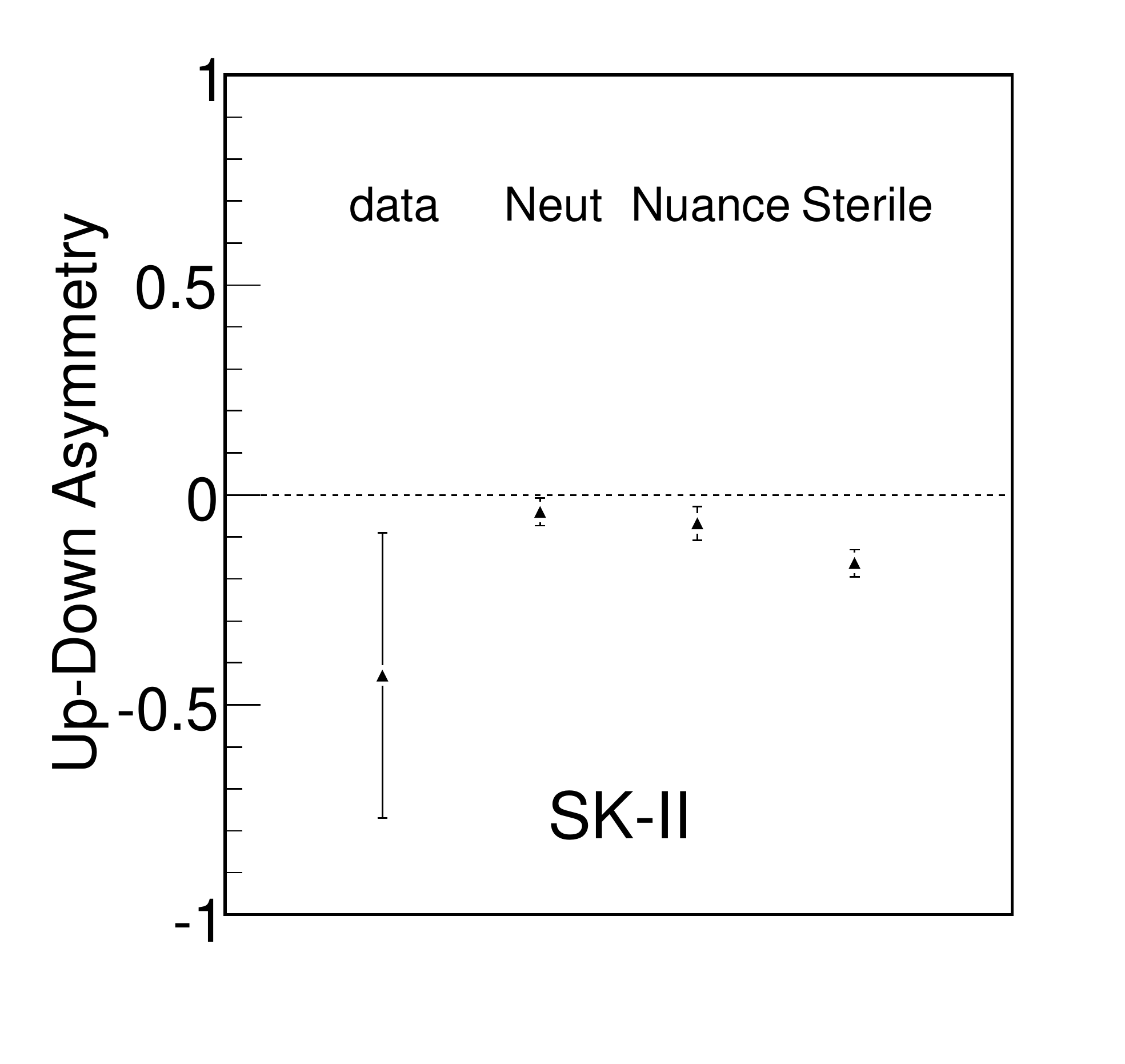}
  \includegraphics[width=2.0in]{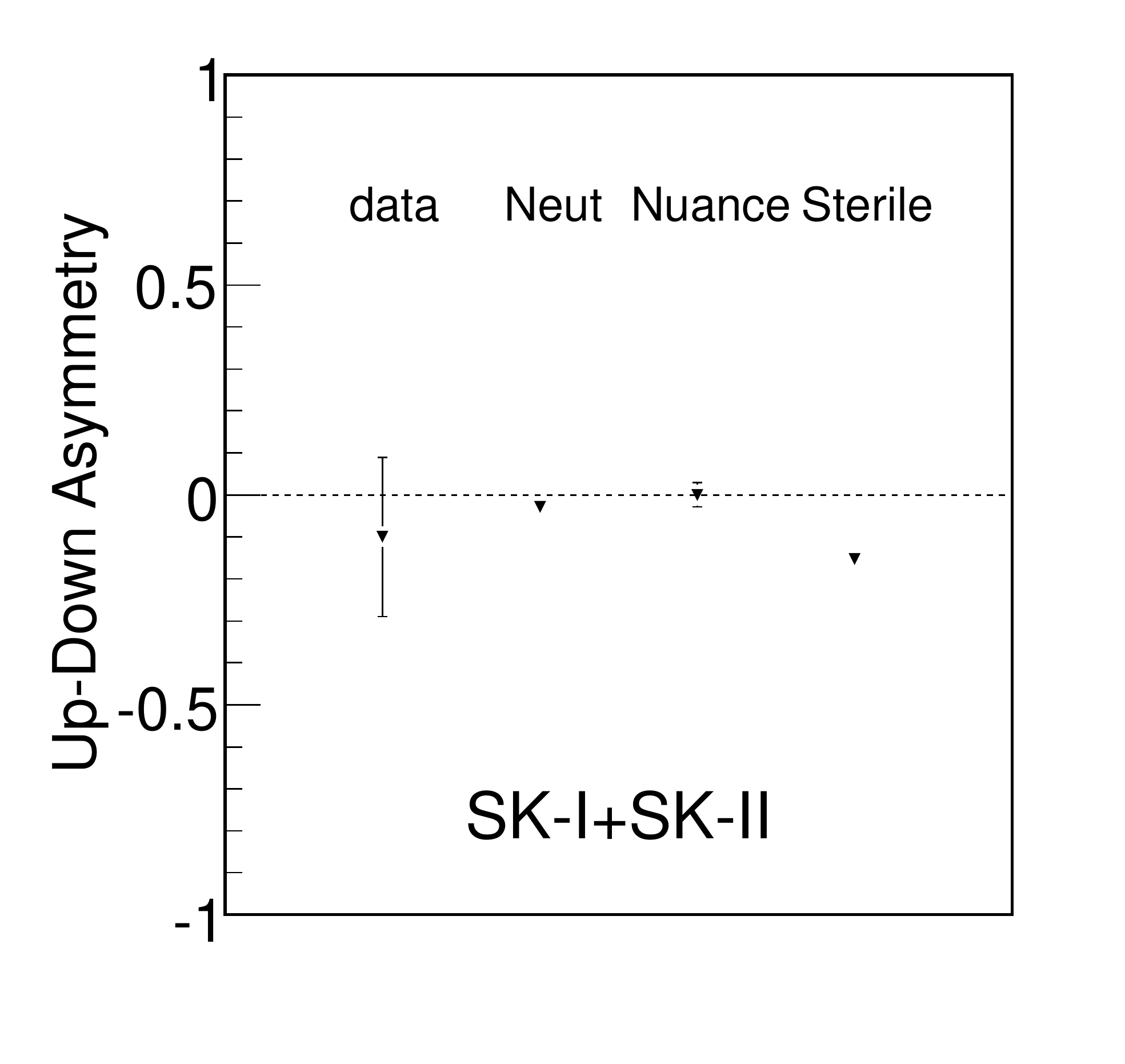}
  \caption{Up-Down asymmetries with statistical errors. In each plot
    the present measurement is the left-most point, labelled ``data''.
    The middle-points, labelled Neut and Nuance, show the MC
    expectation including 2-flavor \mutau oscillation, with MC
    statistical error only. The right-most point labelled ``sterile''
    shows the expectation of the Neut MC assuming oscillations are
    only \musterile. The error bars for simulated events correspond to
    Monte-Carlo statistics.}
\label{fig:updown}
\end{figure*}

  We have determined that in this final sample the proton-like
  particle has a mean angle of 38\degree (resp. 41\degree for SK-II)
  with the incoming neutrino track, which means that the proton still
  has some correlation with the original neutrino direction and allows
  zenith angle studies.  In order to cancel out any flux normalization
  uncertainties we calculate the up-down asymmetry $\frac{U-D}{U+D}$
  where up-going events are those with
  $-\cos\theta_{\mathrm{zenith}}<-0.2$ and down-going events have
  $-\cos\theta_{\mathrm{zenith}}>0.2$. The results are summarized in
  table~\ref{table:updown} and Fig.~\ref{fig:updown}, and are
  compatible with no sterile oscillations.  Clearly the statistics of
  our sample are too weak to extract meaningful information on
  \musterile oscillation. Full \musterile oscillations (already ruled
  out by \cite{fukuda:2000np}) would result in an asymmetry of
  $\approx -15\%$ which is comparable with our data's present
  statistical error. \superk has also already constrained the
  admixture of sterile neutrinos to be less than 26\% at 90\%
  CL\cite{Walter:2003zj} using the so-called 2+2 model parametrized
  in \cite{Fogli:2000ir}. The amount of sterile admixture is not
  measurable with the single visible proton sample alone.  Although
  such a study is beyond the scope of this paper, the single visible
  proton sample could be merged with other samples sensitive to
  sterile neutrinos and included in a larger fit.

  \begin{table}[!htbp]
    \begin{tabular}{lccc}
      \hline
      \hline
      & Up & Down & $\frac{U-D}{U+D} \pm\ \mathrm{stat}$ \\
      SK-I Data & $11$ & $11$ & $0\pm 0.21$ \\
      SK-I NEUT MC & $13.76$ & $ 14.44$ & $-0.024\pm 0.026$ \\
      SK-I NUANCE MC & $12.61$ & $11.82$ & $0.032\pm 0.038$\\
      \hline
      SK-II Data & $2$ & $5$ & $-0.43 \pm 0.34$ \\
      SK-II NEUT MC & $5.88$ & $6.38$ & $-0.041 \pm 0.033$ \\
      SK-II NUANCE MC & $5.44$ & $6.23$ & $-0.068\pm 0.040 $\\   
      \hline
      Combined Data & $13$ & $16$ & $-0.10 \pm 0.19$ \\
      Combined NEUT MC & $19.63$ & $20.81$ & $-0.029 \pm 0.021$ \\
      Combined NUANCE MC & $18.05$ & $18.05$ & $0.00\pm 0.029 $\\   
      \hline \hline
    \end{tabular}
    \caption{Up and down-going events with corresponding asymmetries.
      All NC elastic selection cuts have been applied, including the
      $NN>0$ cut.  In this table the Monte-Carlo has been reweighted
      according to live time, solar wind activity for SK-II, as well as
      neutrino oscillations assuming \dms$=2.5\times 10^{-3}~\mathrm{eV}^{2}$
      and \sstt $=1$. The error bars for simulated events correspond to
      Monte-Carlo statistics.}
    \label{table:updown}
  \end{table}

\section{Search for Charged-Current Quasi-Elastic events}
\label{sec:ccqe}
Another study made possible with proton identification is the search
for charged-current quasi-elastic (CCQE) events with both the lepton
and the proton above Cherenkov threshold. The reaction is $\nu +
n\rightarrow p + \mathrm{lepton}$.  Having a proton in the final state
can only happen for neutrinos and not anti-neutrinos, providing a
possibility for neutrino tagging, which has so far not been done in
\superk. Knowledge of both outgoing tracks allows kinematic
reconstruction of the incoming atmospheric neutrino, which has never
been reported in a large water Cherenkov detector: in the zenith-angle
\superk analyses with atmospheric fully-contained data
\cite{ashie:2005ik}, the lepton direction in the data and Monte-Carlo
are compared directly.  From our Monte-Carlo simulations, we expect
about 150 (75) events in the SK-I (SK-II) data sample with both tracks
above Cherenkov threshold, assuming 2-flavor oscillations with
$(\Delta m^2 = 2.5\times 10^{-3}\ \mathrm{eV}^2,\sin^2 2\theta = 1)$.

\begin{figure*}[!htb]
  \begin{minipage}{3.0in}
  \includegraphics[width=2.8in]{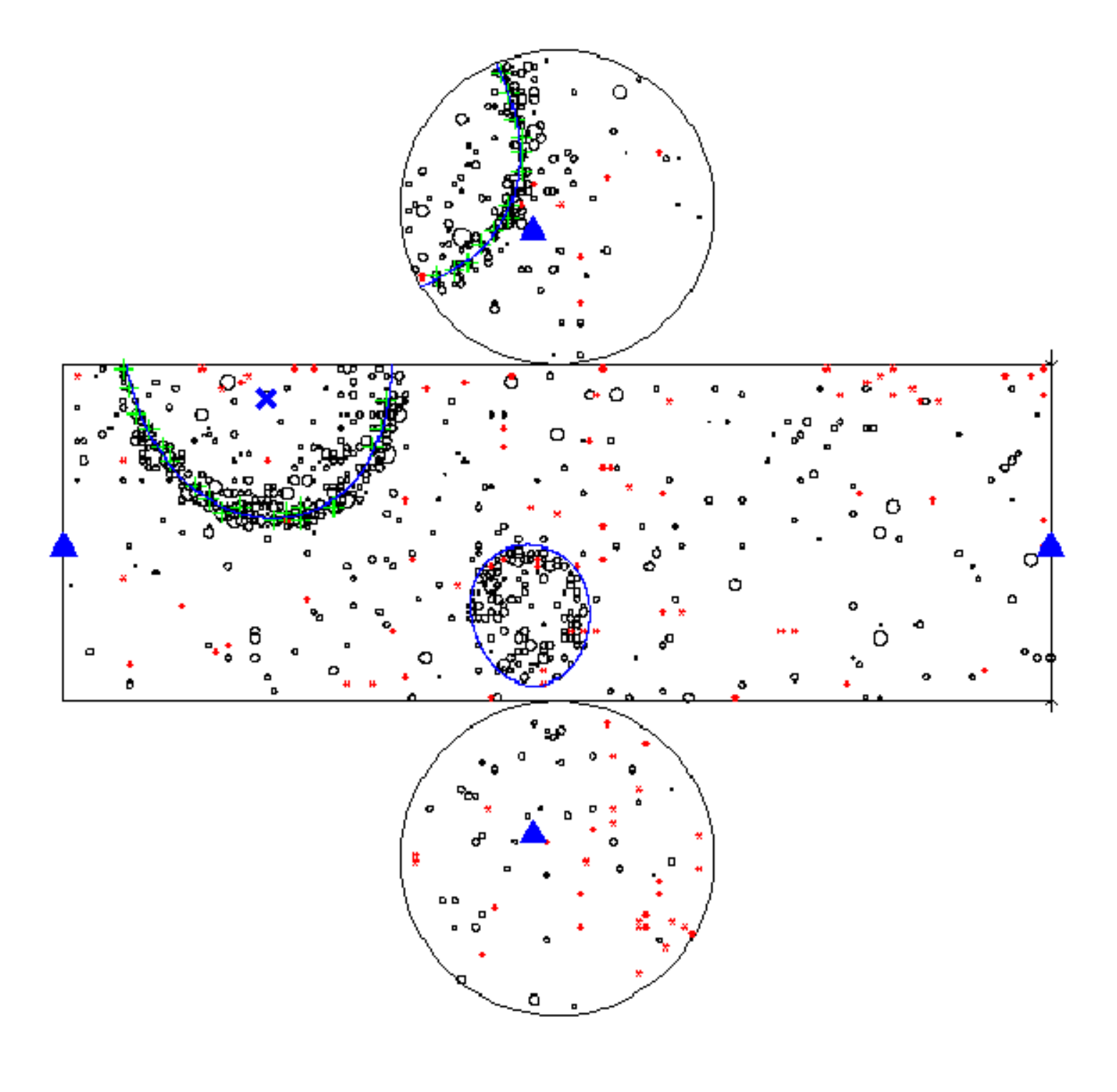}
  \end{minipage}
  \begin{minipage}{3in}
    \includegraphics[width=2.8in]{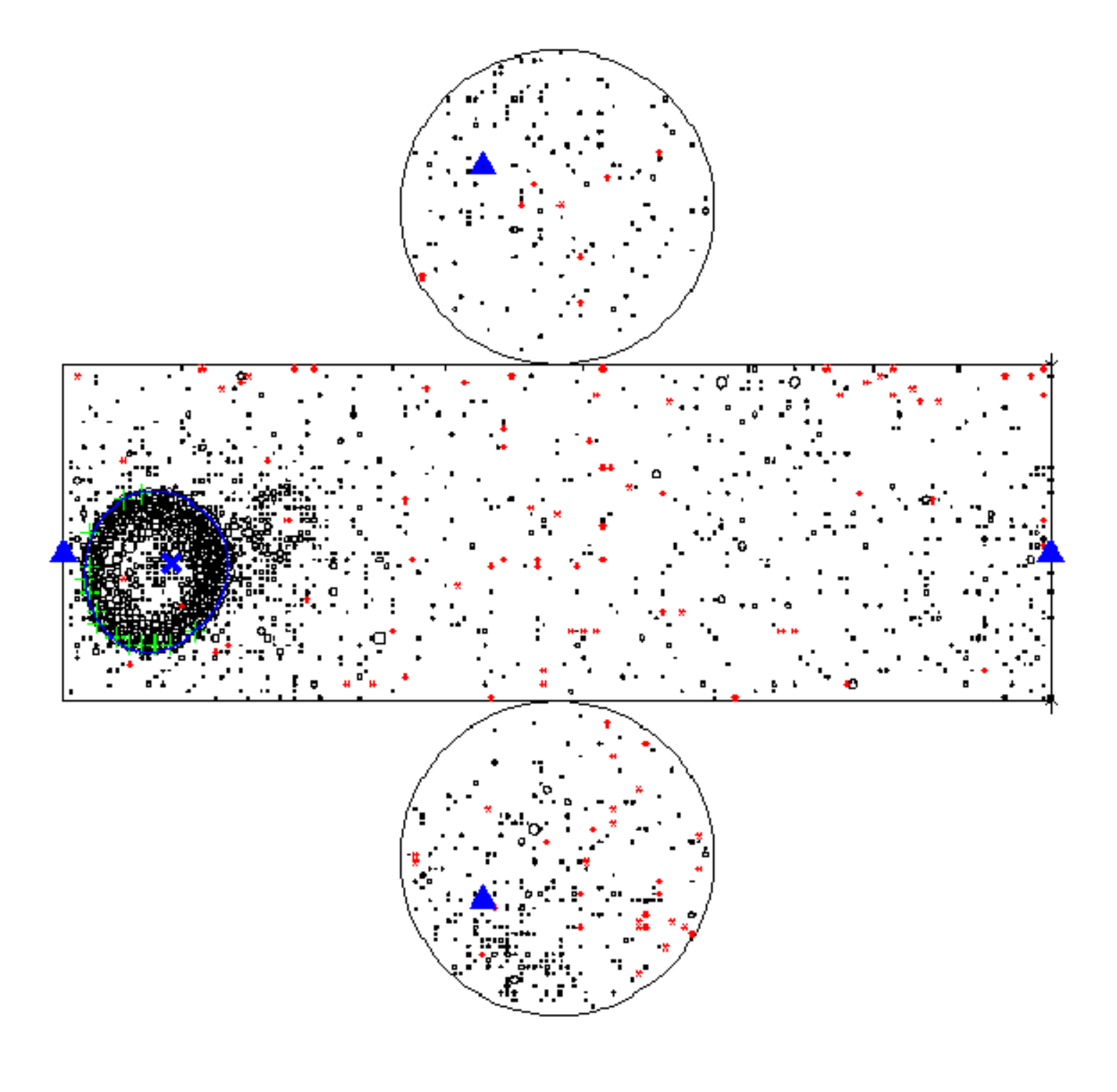}
  \end{minipage}
  \caption{Event displays of Monte-Carlo CCQE events, where both the
    muon (large ring) and the proton (smaller ring) are visible.  In
    the left figure, both tracks were found by the standard fitter
    (the thin lines show the fit results).  In the figure on the right
    only the lepton (a muon in this example) was fitted; the proton is
    visible as a smaller ring next to the muon.  }
  \label{fig:displays-ccqe1}
\end{figure*}

We have focused on events with either one or two rings reconstructed
by the standard ring finder.  Each class requires a different analysis
technique and will be studied separately.
Figure~\ref{fig:displays-ccqe1} shows two typical CCQE events, one for
each class.

\subsection{Events with two fitted rings}
\label{sec:2ring}
The first case to consider is that for which both rings were
identified by the standard ring finder described earlier. When this
happens, in more than 97\% of the cases the proton is found as the
second ring, so we make that assumption in this analysis. The vertex,
ring directions and lepton momentum are known from the reconstruction
algorithms. We have modified the light pattern engine to superimpose a
lepton and a proton pattern; the lepton ring is kept fixed, its
parameters being set to the fitted information.  The likelihood of
this combined pattern is then maximized using the same MINUIT-based
fitting, varying the proton momentum and track length. The
normalization of the light patterns is done using the same procedure
as for single ring protons, simply adding the lepton pattern to it.

Then, a second likelihood corresponding to a lepton and a muon light pattern
is calculated. The discriminating variables are the fitted proton-like
momentum and track length, as well as the log-likelihood difference
$\log L_{\mathrm{lepton+proton}}-\log L_{\mathrm{lepton+muon}}$.

\begin{figure*}[!htb]
  \begin{center}
  \begin{minipage}{7in}
  \includegraphics[width=5in]{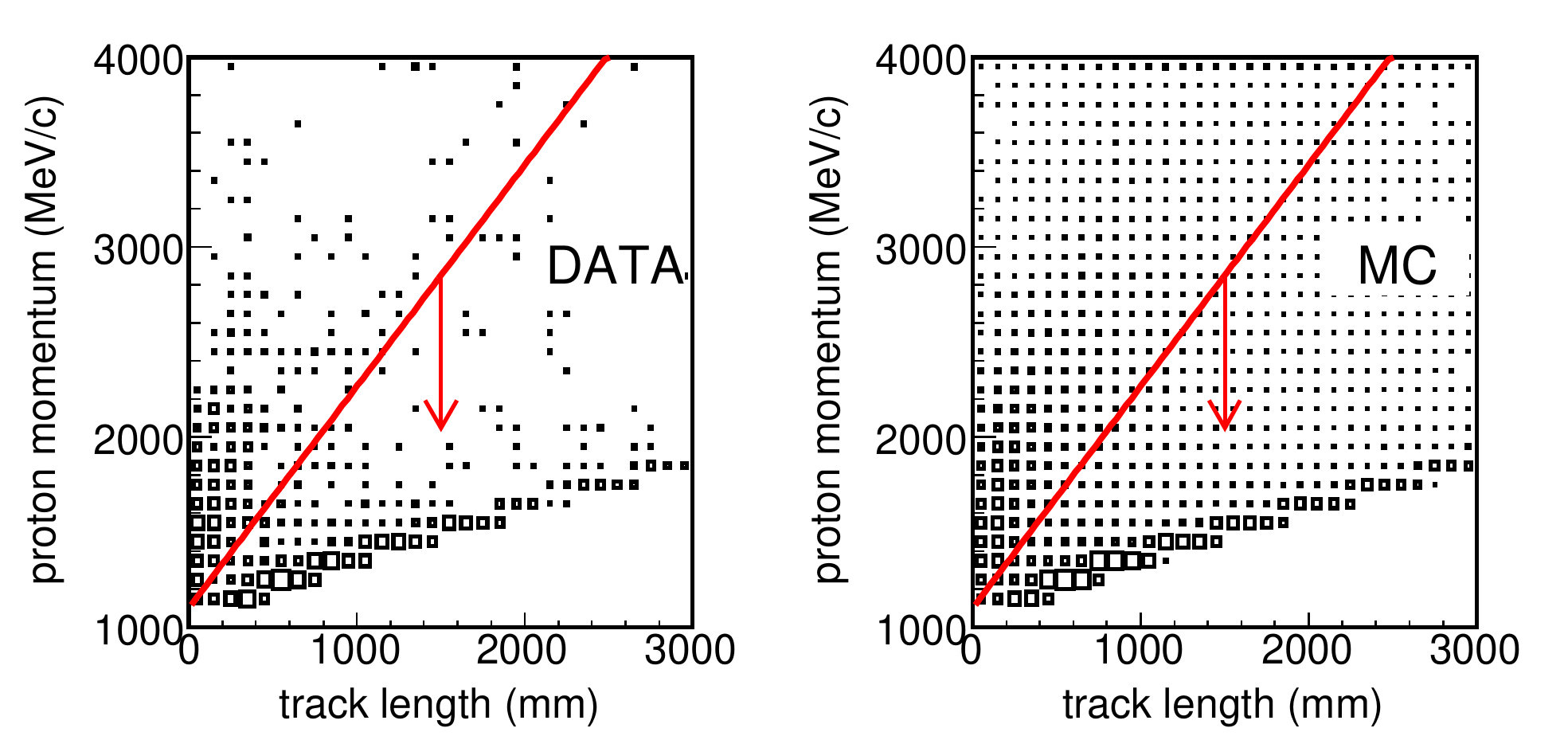}
  \end{minipage}
  \vspace{0.2in}
  \begin{minipage}{7in}
  \includegraphics[width=2.25in]{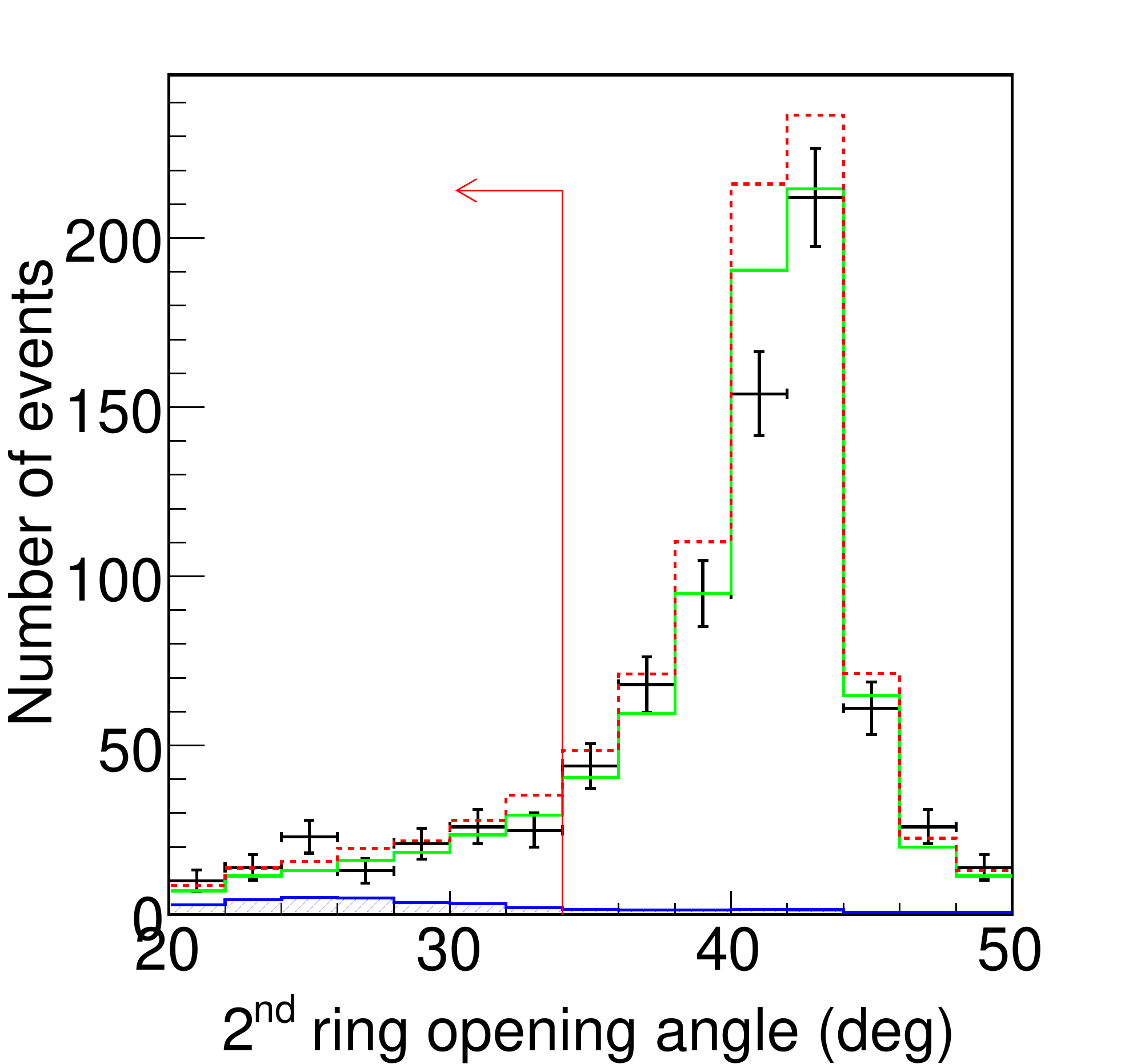}
  \includegraphics[width=2.25in]{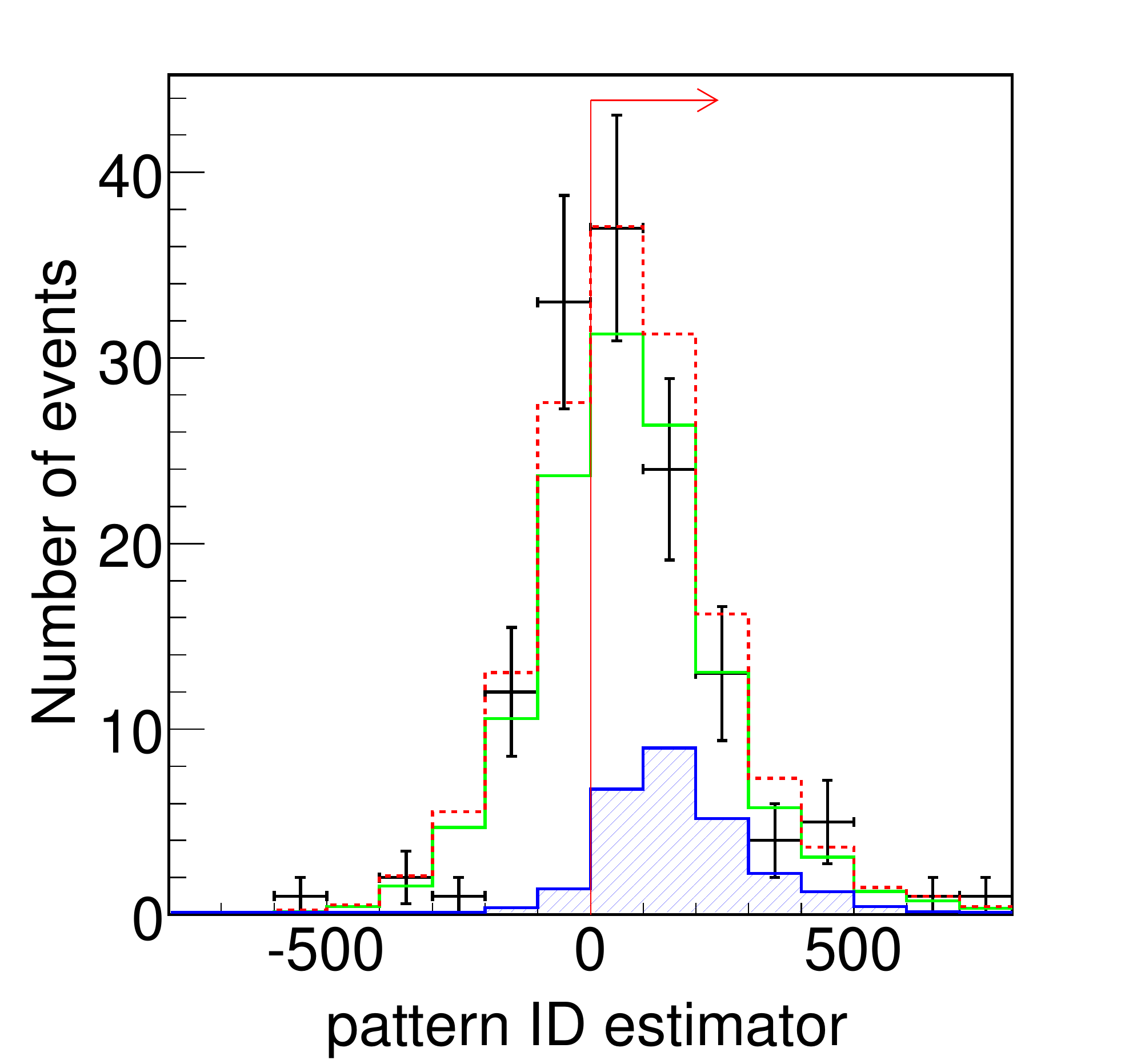}
  \includegraphics[width=2.25in]{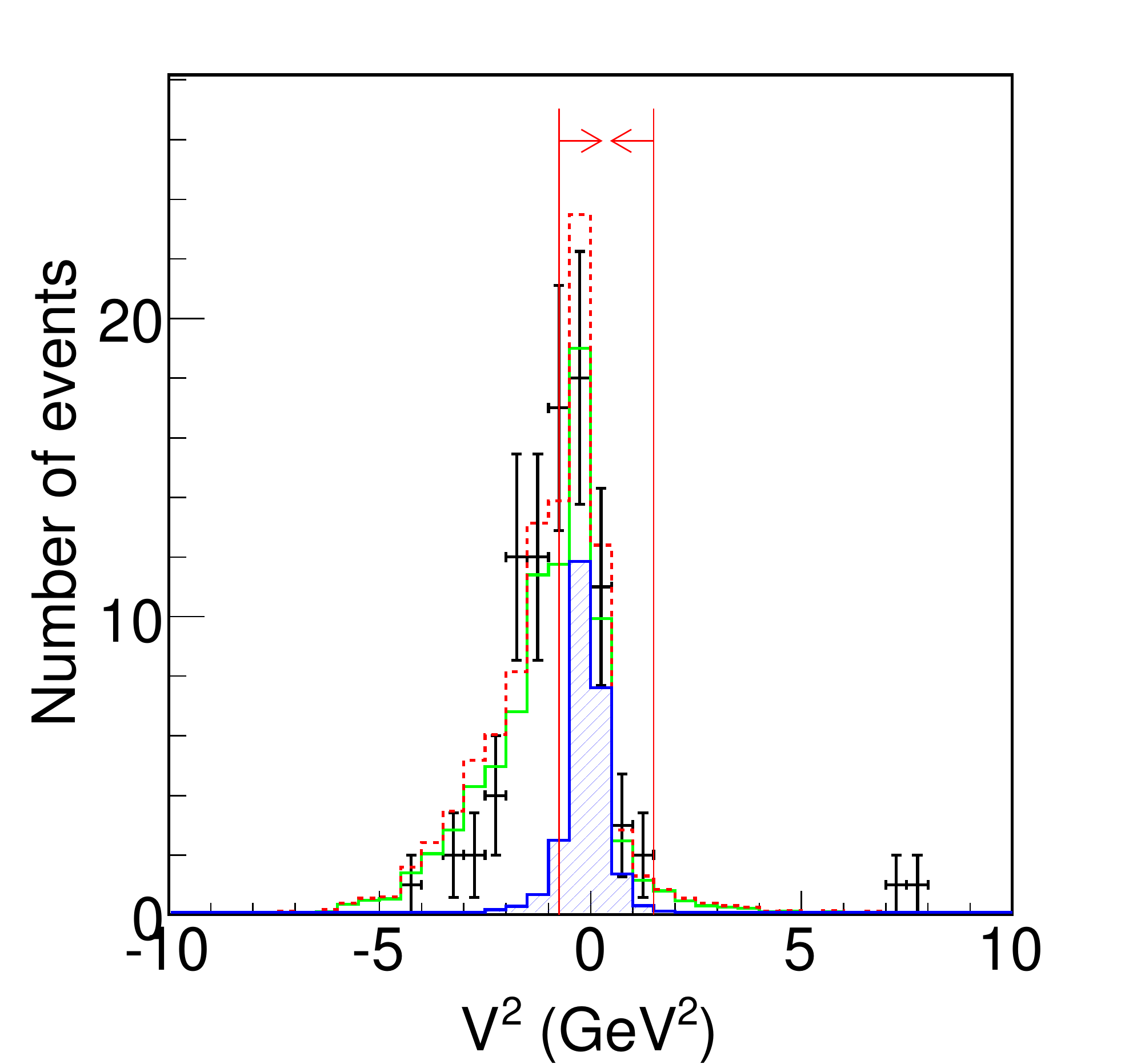}
  \end{minipage}
  \caption{Distribution of analysis variables with cuts for 2-ring
    CCQE event selection in SK-I. The top two plots are
    two-dimensional $L,P$ distributions for data and MC, shown with
    the charged pion removal cut; the arrows show the selected domain.  
    In the bottom three plots, the full
    and dashed line show Monte-Carlo expectation from NEUT and NUANCE,
    with oscillation reweighting, but no correction for the absolute
    flux normalization. The hatched regions show the contribution from
    CCQE events.}
\label{fig:ccqe2-cuts-sk1}
\end{center}
\end{figure*}

\begin{figure*}[!htb]
  \begin{center}
  \begin{minipage}{7in}
  \includegraphics[width=5in]{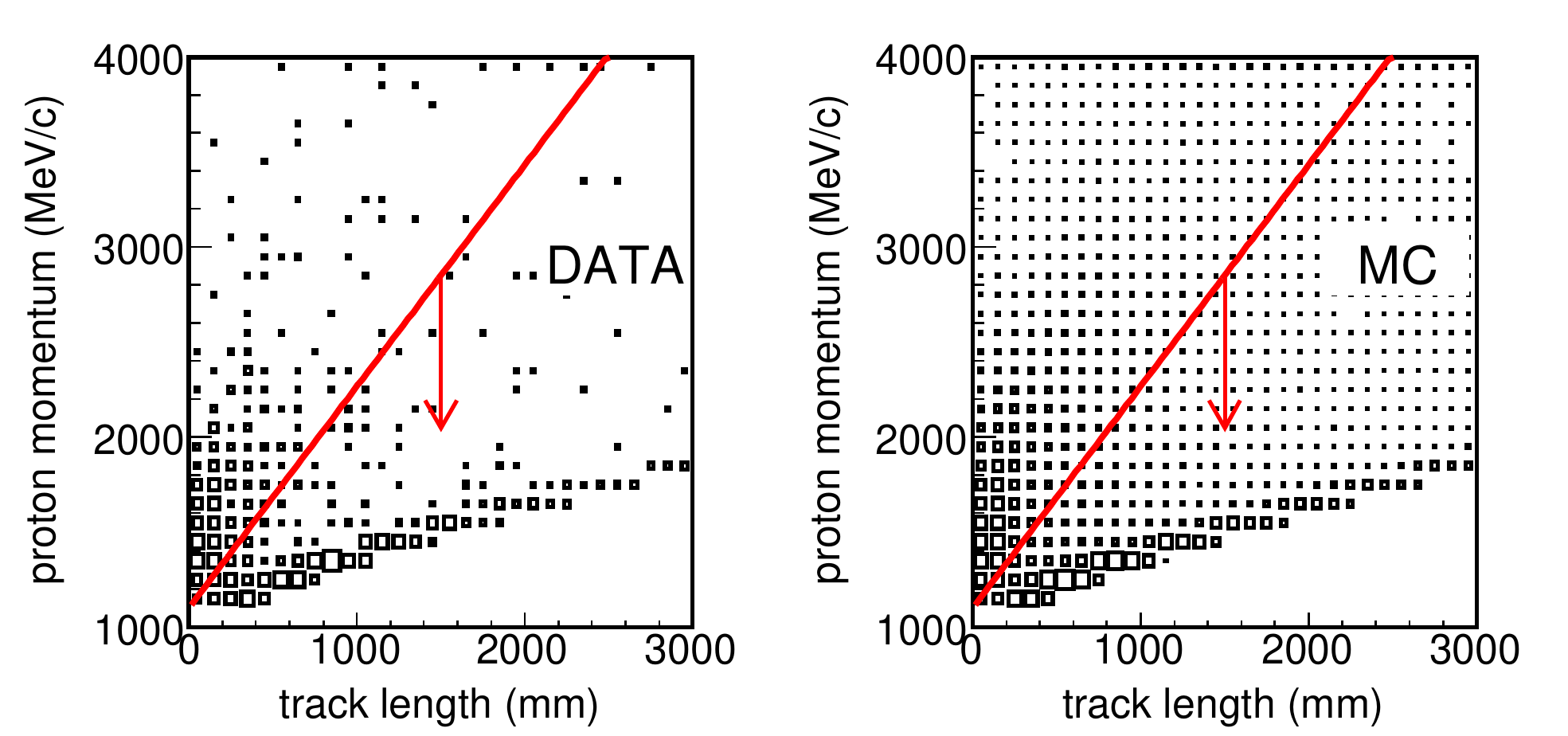}
  \end{minipage}
  \vspace{0.2in}
  \begin{minipage}{7in}
  \includegraphics[width=2.25in]{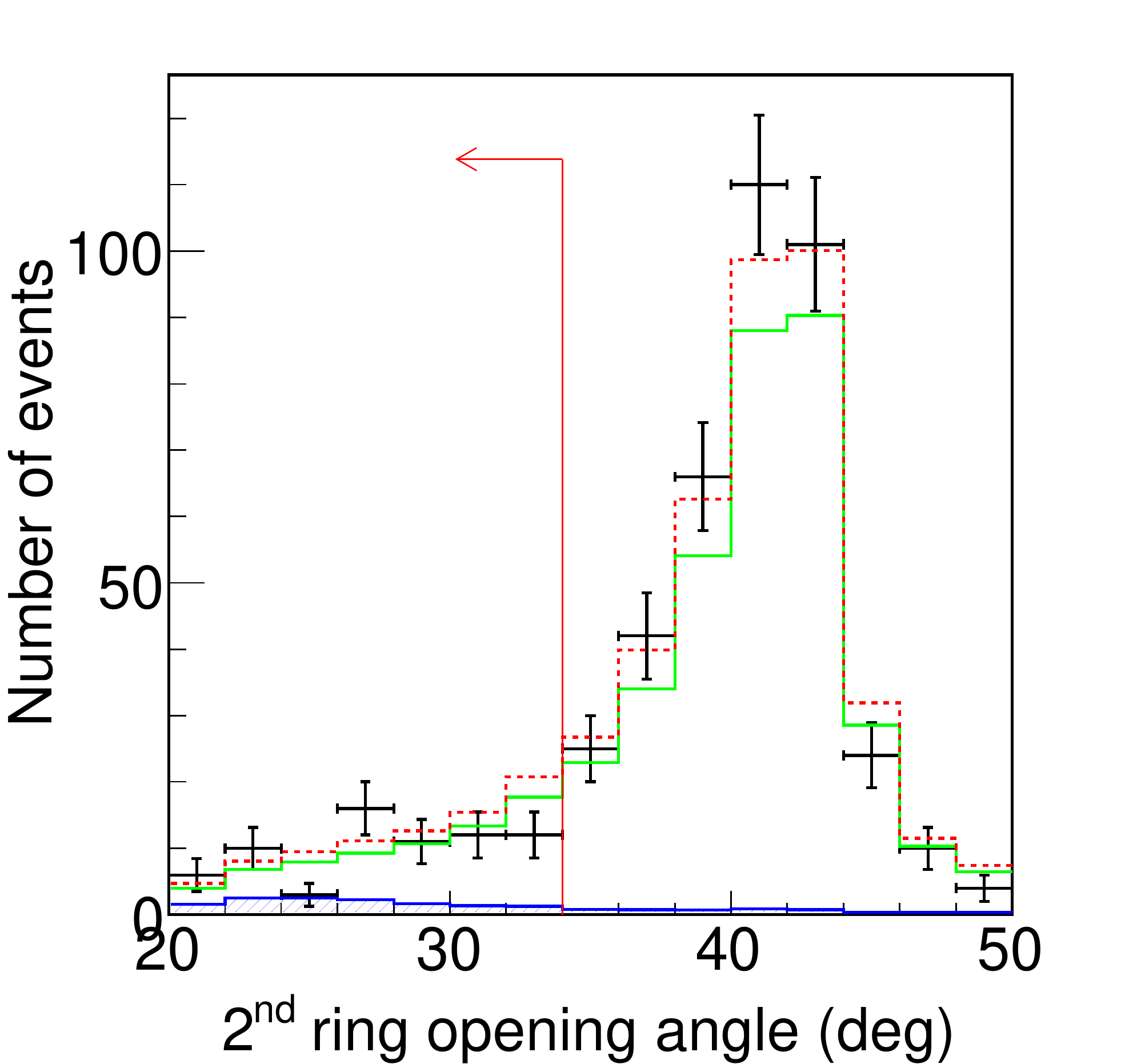}
  \includegraphics[width=2.25in]{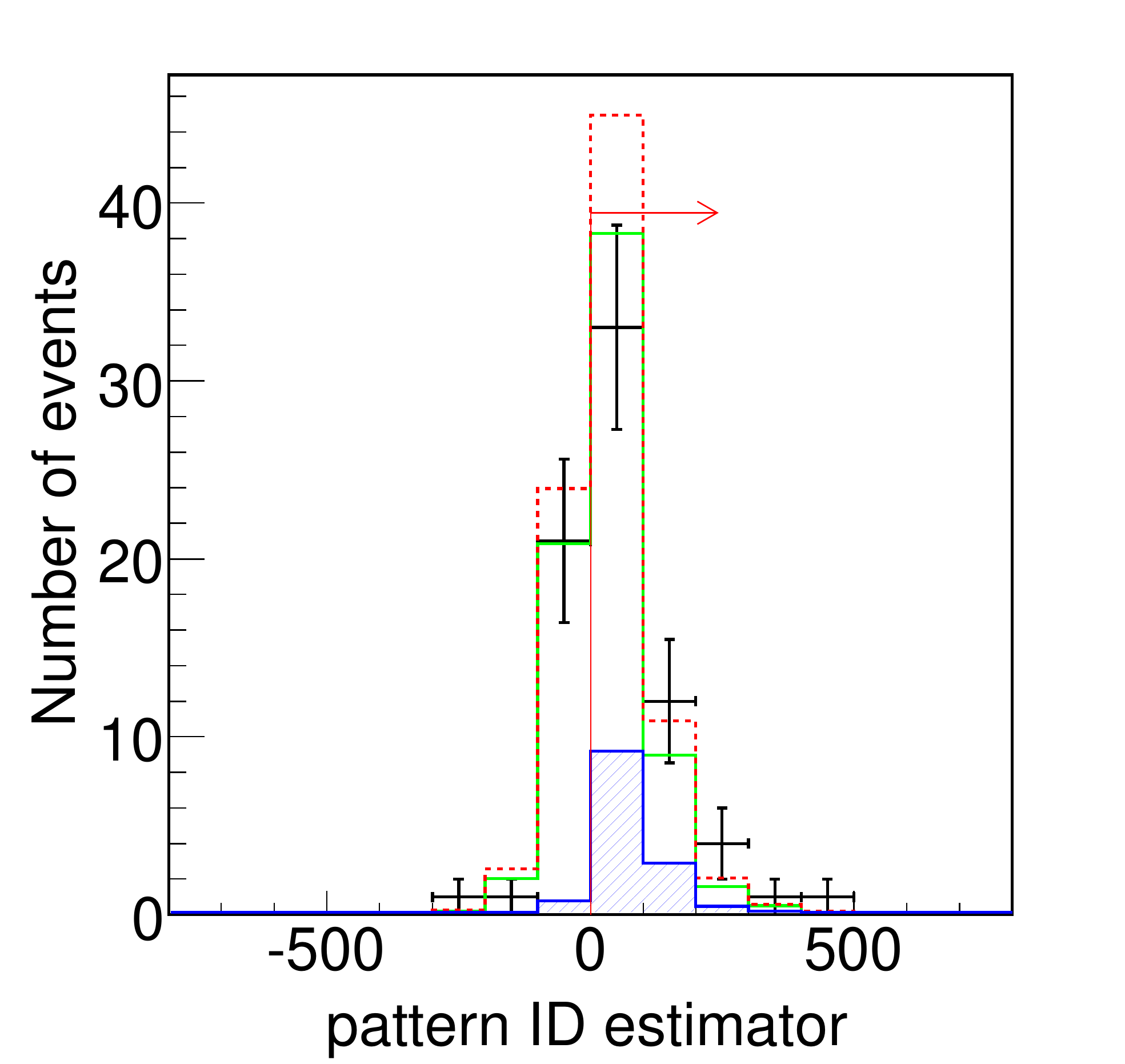}
  \includegraphics[width=2.25in]{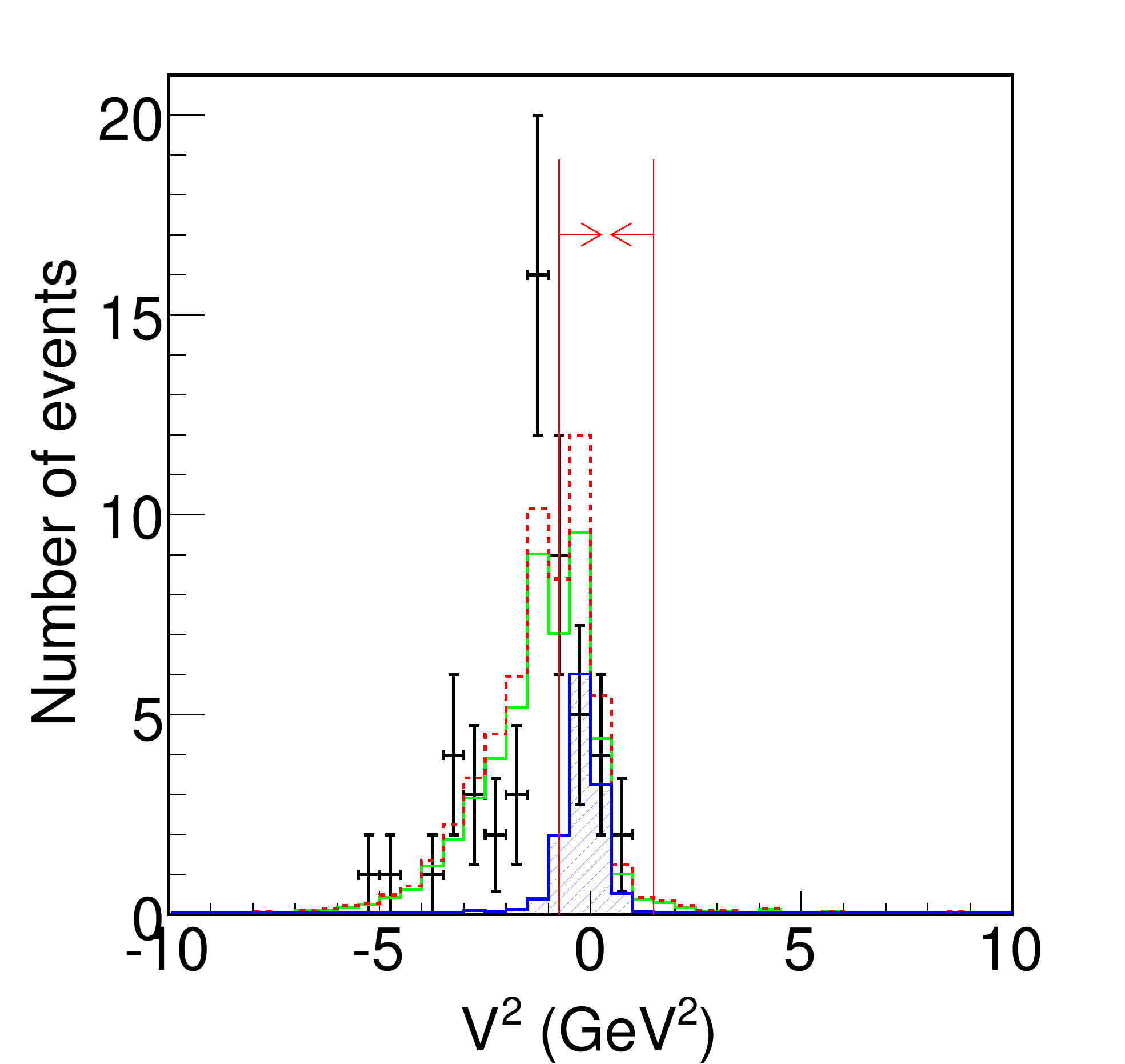}
  \end{minipage}
  \caption{Distribution of analysis variables with cuts for 2-ring
    CCQE event selection in SK-II. The top two plots are
    two-dimensional $L,P$ distributions for data and MC, shown with
    the charged pion removal cut; the arrows show the selected domain. 
    In the bottom three plots, the full
    and dashed line show Monte-Carlo expectation from NEUT and NUANCE,
    with oscillation reweighting, but no correction for the absolute
    flux normalization. The hatched region shows the contribution from
    CCQE events.}
\label{fig:ccqe2-cuts-sk2}
\end{center}
\end{figure*}

CCQE selection relies on the following cuts:
\begin{enumerate}
\item We first select only vertices inside the fiducial volume, with 2
  fully-contained rings, and $100<E_{\mathrm{vis}}<4000$ MeV (Note
  that for the CCQE search the energies are high enough that no
  spallation cut is necessary).
\item Our next cut aims at removing background coming from pion
  production, where the second ring was in fact caused by a charged
  pion. In this case, the fitted proton-like momentum of the track is
  typically about 1.6 GeV/c, because the charged pions have typical
  momenta between 200 and 300 MeV, and thus a ring opening angle of
  roughly 30\degree. However, the fitted path length is very short
  (about 20 cm) because of pion interactions.  In order to remove the
  charged pions, a linear cut on the proton-like momentum and track
  lengths $A < P < B\times L(\mathrm{m})+C$, with $A=C=1.1$ GeV/c and
  $B=7/6$ GeV/c/m.  This selection can be seen in
  Fig.~\ref{fig:ccqe2-cuts-sk1}.
\item We then require that the opening angle of the second ring be
  less than 34\degree, to select particles with low $\beta$.
\item To select proton tracks a cut on the pattern likelihood
  difference $\log(L_{\mathrm{lepton+proton}})-\log(
  L_{\mathrm{lepton+muon}})>0$ is applied.
\item At this stage the sample still contains non-CCQE events, mostly
  corresponding to mis-identified pions. To improve CCQE selection we
  use event kinematics.  Let $V$ be the
  4-vector $$V=P_p+P_l-P_n,$$ where $P_p$, $P_l$, and $P_n$ are the
  4-momenta of the proton, lepton, and target neutron.  For a true
  CCQE event, $V$ must be equal to $P_\nu$, the neutrino 4-momentum,
  and consequently the Lorentz invariant quantity $V^2$ must be
  $m_\nu^2\approx 0\ \mathrm{eV^2/c^4}$. On the other hand, for a non-CCQE
  event, $V$ is not equal to the neutrino 4-momentum and its square
  will be non-zero. In practice, nuclear effects such as Fermi motion
  of the neutron or scattering of the outgoing nucleon, and detector
  effects will smear the resolution on $V^2$, but it is still a
  powerful tool, as can be seen in Figs.~\ref{fig:ccqe2-cuts-sk1} and
  \ref{fig:ccqe2-cuts-sk2}, with CCQE events peaked around 0.  We
  apply a cut $-0.75<V^2<1.5\ \mathrm{GeV/c}^2$ for 2-ring events.
\end{enumerate}

Figures~\ref{fig:ccqe2-cuts-sk1} and \ref{fig:ccqe2-cuts-sk2} show the
distributions of these variables for data and Monte-Carlo for SK-I and
SK-II, showing good agreement. The results are shown in tables
\ref{table:ccqe2-sk1} and \ref{table:ccqe2-sk2}.  The overall signal
efficiency is 42-45\%, with an expected signal-to-background ratio of
1.4 to 1. The high signal efficiency can be explained by the fact that
those protons have high enough momenta to have been
found by the ring fitter, and are well above Cherenkov threshold.
Monte-Carlo studies of the selected events show that of the 38 (19)
events in the final SK-I (SK-II) sample, 33 (16) do have their second
ring caused by a proton (the rest mostly come from gammas and charged
pions), and 22 (11) of those are CCQE signal events.  The non-CCQE
proton rings mostly come from CC pion events, and are an irreducible
background to the CCQE search. It is worth noting that at this stage
the sample contains 93\% neutrinos and 7\% anti-neutrinos.

\begin{table*}[!htbp]
  \begin{tabular}{lccc}
    \hline
    \hline
    ~SK-I~ & ~Data~ & ~ Total MC~ & ~ Signal MC~ \\
    \hline FC, FV, two-ring,
    $0.1<\mathrm{E_{vis}}<4$ GeV                  & 1876 (100\%) & 1773 (100\%) &50.12 (100\%)
    \\ $1.1<P<7/6L(m)+1.1$ GeV/c & 829 (44.19\%)& 839.4(47.34\%)&36.12(72.1\%)
    \\ Angle of 2$^{\mathrm{nd}}$ cone $<34$\degree & 135 (7.2\%) & 124.1(7.0\%)&27.2 (54.3\%) 
    \\ Pattern ID cut & 86 (4.58\%) & 82.88 (4.67\%) & 25.26 (50.4\%)
    \\ $-0.75<V^{2}<1.5~\mathrm{GeV}^2$ & 44 (2.35\%) & 38.59 (2.18\%) & 22.71 (45.31\%)\\
    \hline \hline
  \end{tabular}
  \caption{Summary of SK-I CCQE selection for two-ring events. In this
    table the Monte-Carlo has been reweighted according to live time,
    solar wind activity for SK-I, as well as neutrino oscillations assuming
    \dms$=2.5\times 10^{-3}~\mathrm{eV}^{2}$ and \sstt $=1$.}
  \label{table:ccqe2-sk1}
\end{table*}

\begin{table*}[!htbp]
  \begin{tabular}{lccc}
    \hline
    \hline
    ~ SK-II~ & ~Data~ & ~ Total MC~ & ~ Signal MC~ \\
    \hline FC, FV, two-ring,
    $0.1<\mathrm{E_{vis}}<4$ GeV                  & 1023 (100\%) & 910.8(100\%) & 26.2 (100\%)
    \\ $1.1<P<7/6L(m)+1.1$ GeV/c & 471 (46.04\%)& 417.1(45.79\%)&18.3(68.8\%)
    \\ Angle of 2$^{\mathrm{nd}}$ cone $<34$\degree & 74 (7.23\%)& 72.73(7.99\%)& 13.7(52.3\%) 
    \\ Pattern ID cut & 51 (4.99\%) & 49.64 (5.5\%) & 12.84 (49.01\%)
    \\ $-0.75<V^{2}<1.5~\mathrm{GeV}^2$ & 12 (1.17\%) & 18.99 (2.09\%) & 11.22 (42.81\%)\\
    \hline \hline
  \end{tabular}
  \caption{Summary of SK-II CCQE selection for two-ring events. In
    this table the Monte-Carlo has been reweighted according to live
    time, solar activity for SK-II, as well as neutrino oscillations
    assuming \dms$=2.5\times 10^{-3}~\mathrm{eV}^{2}$ and \sstt $=1$.}
  \label{table:ccqe2-sk2}
\end{table*}

\subsection{Events with a single fitted ring after standard algorithms}

Monte-Carlo studies show that there are potentially $\approx 150$ CCQE
events in the SK-I and II data sets with a proton above threshold
where the proton was missed by the standard ring finder. Such events
only have a single fitted ring corresponding to the brighter lepton
track. The proton ring is weak, but visible by eye-scan.  A dedicated
ring finding algorithm to select these weak rings was developed. The
vertex and lepton track are very well known at this stage of the
reconstruction: as single ring events, the refined vertex fitter (see
section~\ref{sec:afit}) was applied, and the e-like or $\mu$-like
nature of the track is known.  The observed charge pattern is
projected into a $50\times 50$ grid in ($\cos\theta,\phi$) spherical
coordinates around the lepton track.  The expected light from the
lepton is subtracted, and the maximum amount of charge in a $25\times
10$ sliding window is found, avoiding directions too close to the
lepton track. The corresponding average direction is used as a
starting position in a precise grid fit, where a ring-shaped pattern
is fitted to the observed light pattern, varying the direction and
opening angle until the best match is found. This direction fit yields
a new ring candidate, which could be a proton or another particle, or
just a region with higher charge in a bona-fide single~ring event.

Therefore, we also calculate the distribution of charges $Q(\theta)$
as a function of opening angle $\theta$ from the candidate direction,
subtracting the expected light contribution from the lepton.  We then
calculate the ratio of the integrated charge around the maximum of
this distribution to an average of the charge at low and high values
of $\theta$.  This estimator $Q_{\mathrm{dens}}$, should be large for
a true Cherenkov ring, since a real ring should have a marked peak in
$Q(\theta)$ near the Cherenkov edge. A similar technique is also used in
standard ring counting to reject fake seeds found by the Hough
transform.

Using the lepton's fitted information and the new fitted direction,
we add a lepton and a proton's expected charge patterns.  The
likelihood of this combined pattern is maximized as already outlined
previously, yielding $\log (L_{\mathrm{lepton+proton}})$.  Another
pattern likelihood corresponding to the lepton alone $\log
(L_{\mathrm{lepton\ alone}})$ is also calculated, and the pattern ID
estimator
$$\Delta L = \log \left[\log(L_{\mathrm{lepton+proton}})
  - \log(L_{\mathrm{lepton\ alone}})\right]$$ is constructed. A high
value of this estimator corresponds to a proton-like track candidate.

\begin{figure*}[!htb]
  \begin{center}
  \begin{minipage}{7in}
  \includegraphics[width=5in]{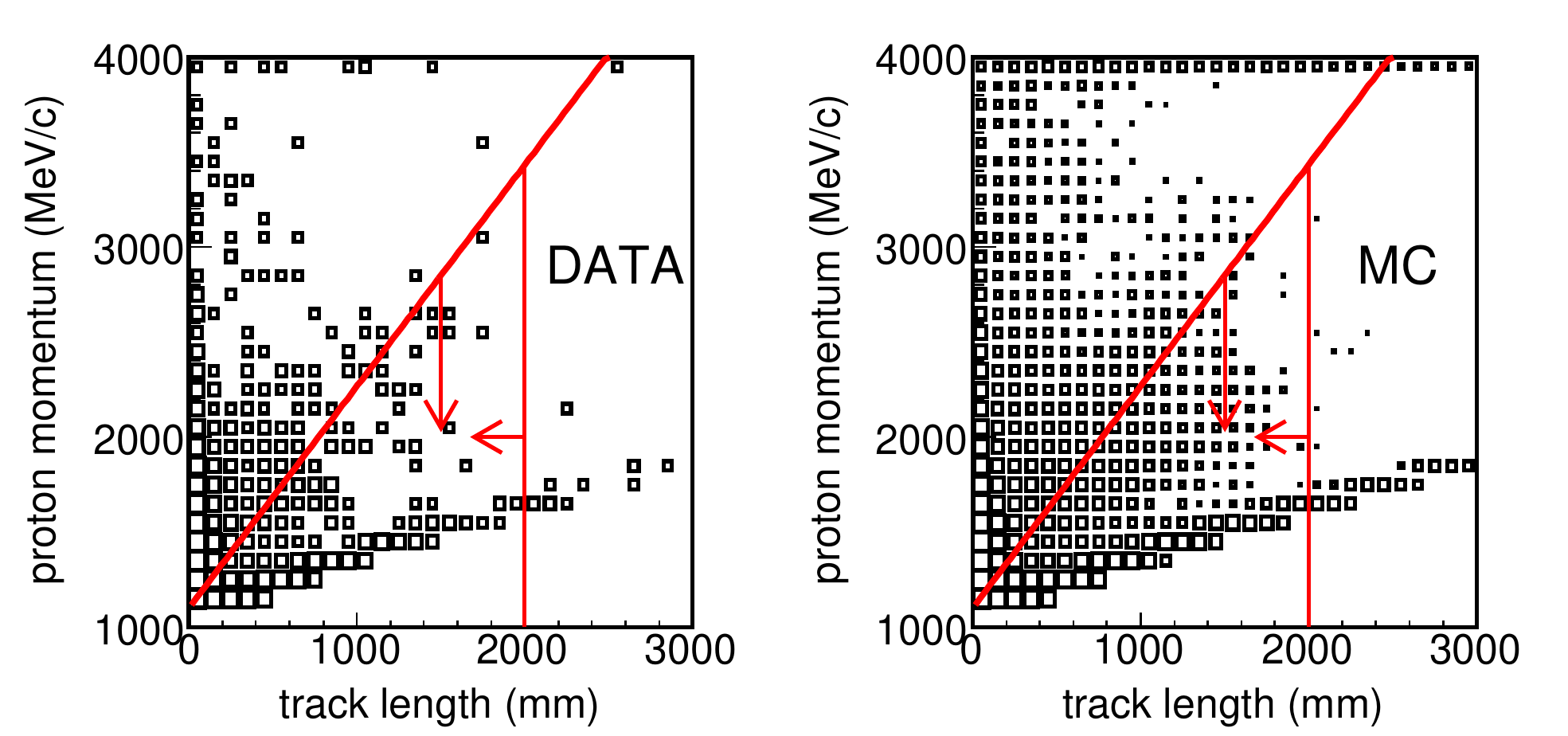}
  \end{minipage}
  \vspace{0.2in}
  \begin{minipage}{7in}
  \includegraphics[width=2.25in]{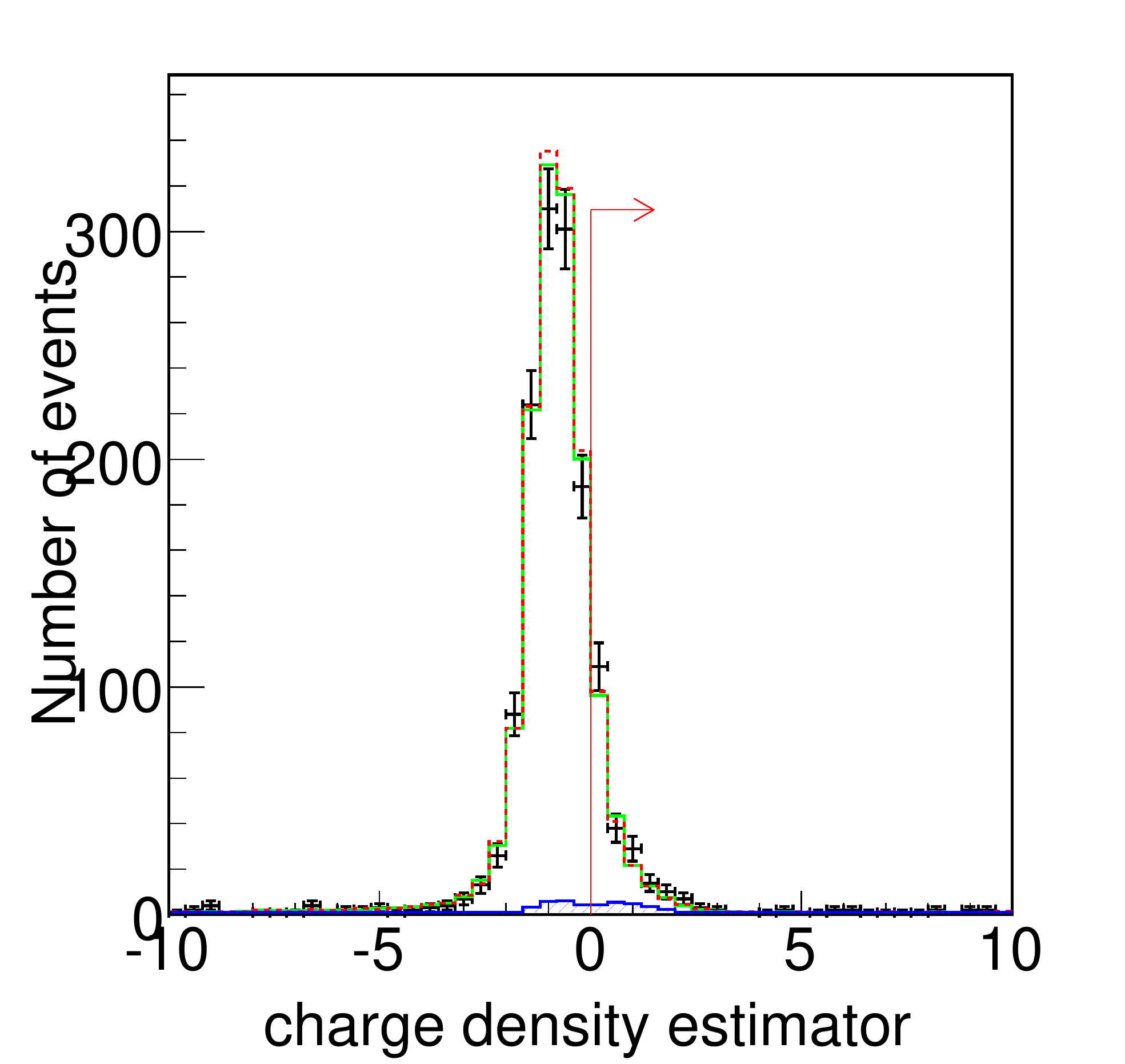}
  \includegraphics[width=2.25in]{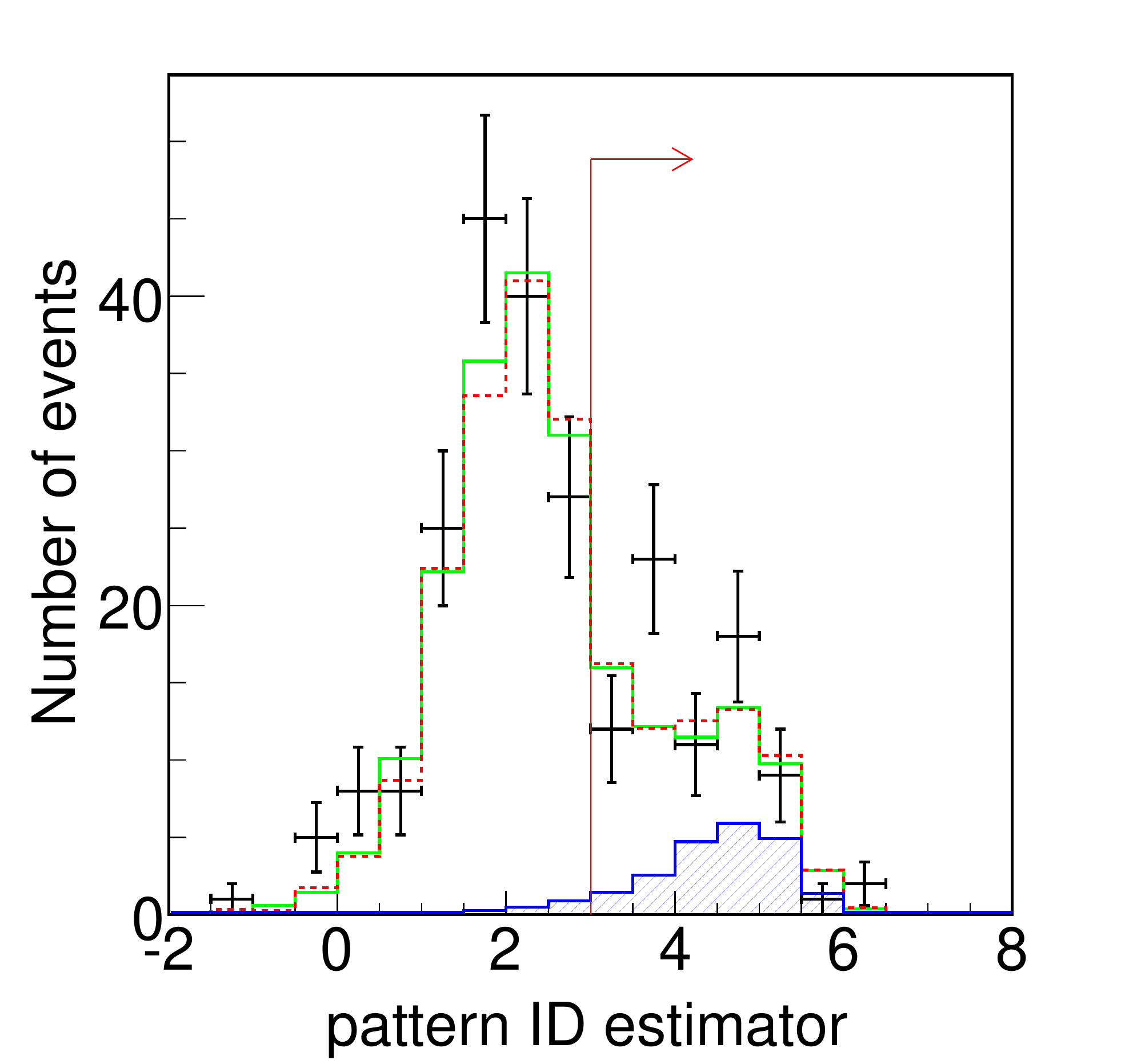}
  \includegraphics[width=2.25in]{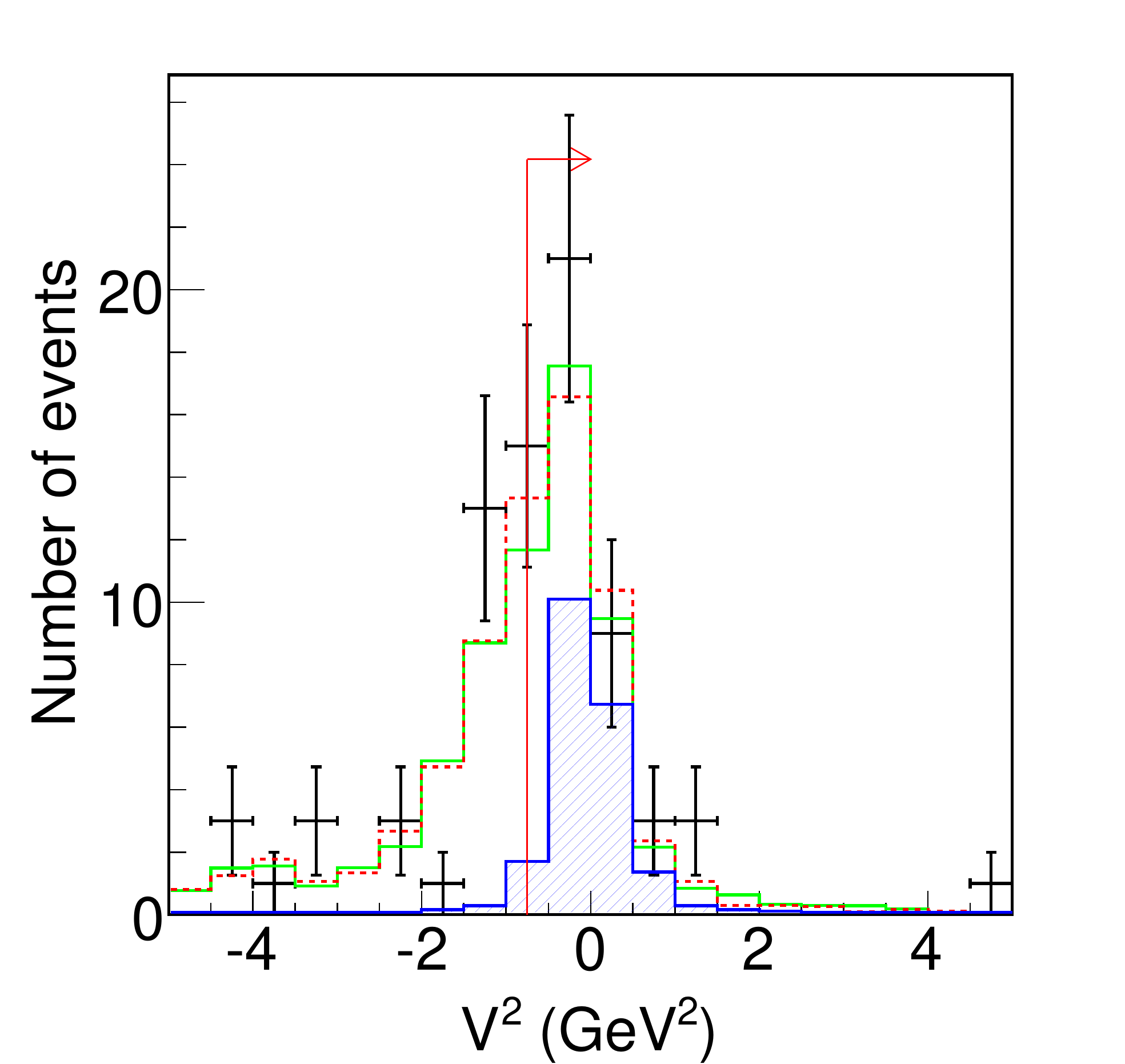}
  \end{minipage}
  \caption{Distribution of analysis variables with cuts for
    single-ring CCQE event selection in SK-I. The top two plots are
    two-dimensional $L,P$ distributions for data and MC, shown with
    the charged pion removal cut; the arrow shows the selected domain. 
    In the bottom three plots, the full
    and dashed line show Monte-Carlo expectation from NEUT and NUANCE,
    with oscillation reweighting, but no correction for the absolute
    flux normalization. The hatched region shows the contribution from
    CCQE events.}
\label{fig:ccqe1-cuts-sk1}
\end{center}
\end{figure*}

\begin{figure*}[!htb]
  \begin{center}
  \begin{minipage}{7in}
  \includegraphics[width=5in]{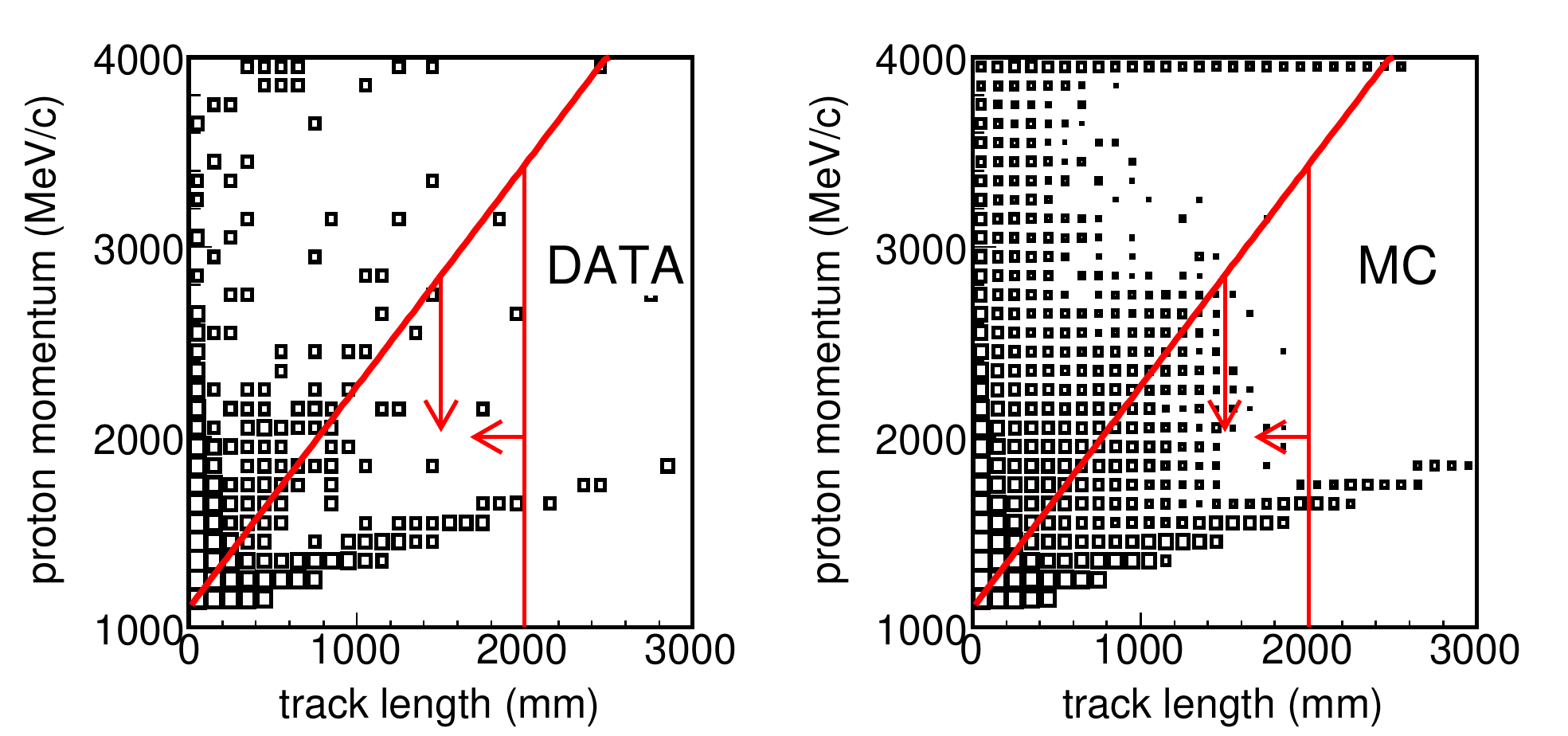}
  \end{minipage}
  \vspace{0.2in}
  \begin{minipage}{7in}
  \includegraphics[width=2.25in]{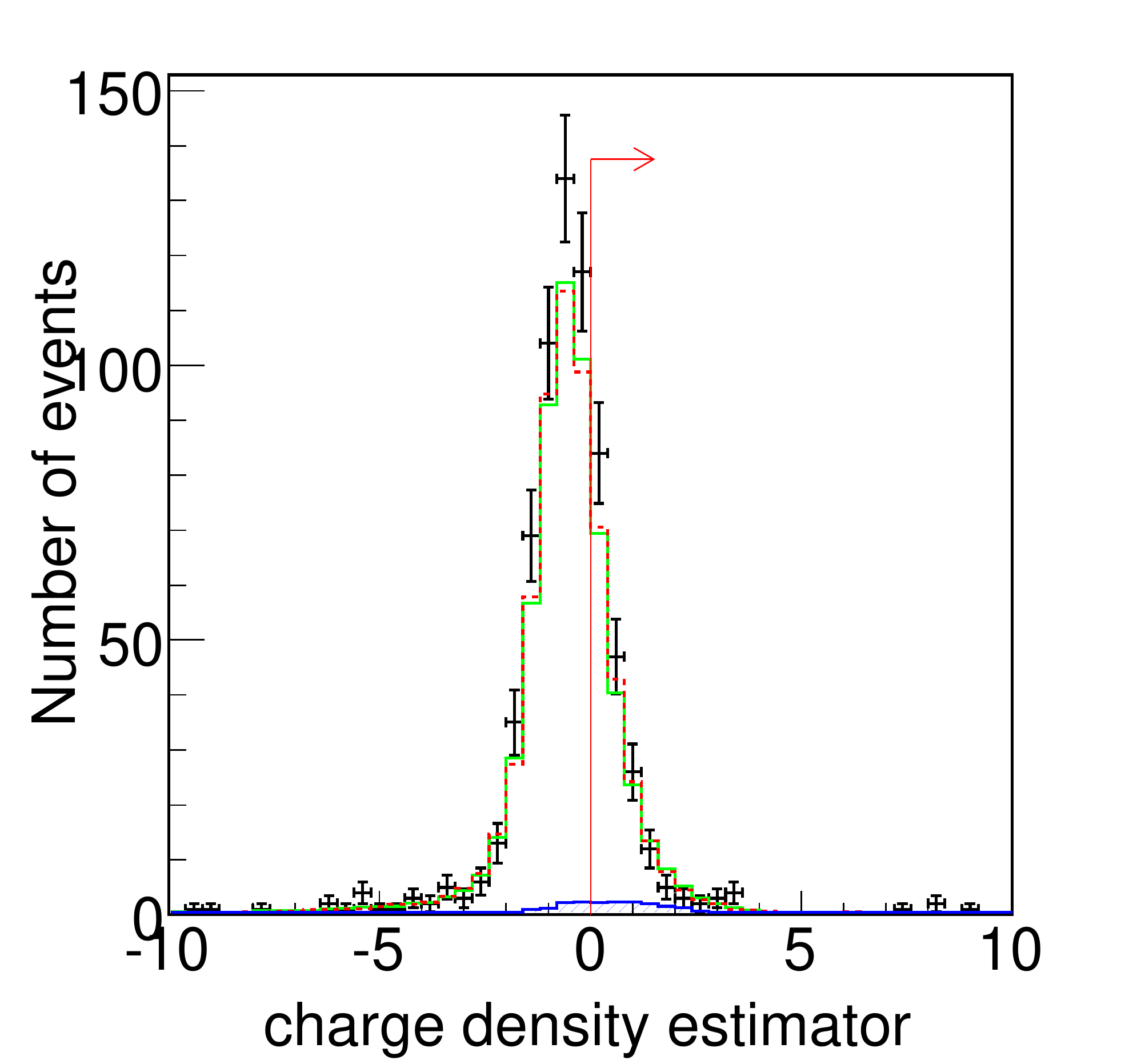}
  \includegraphics[width=2.25in]{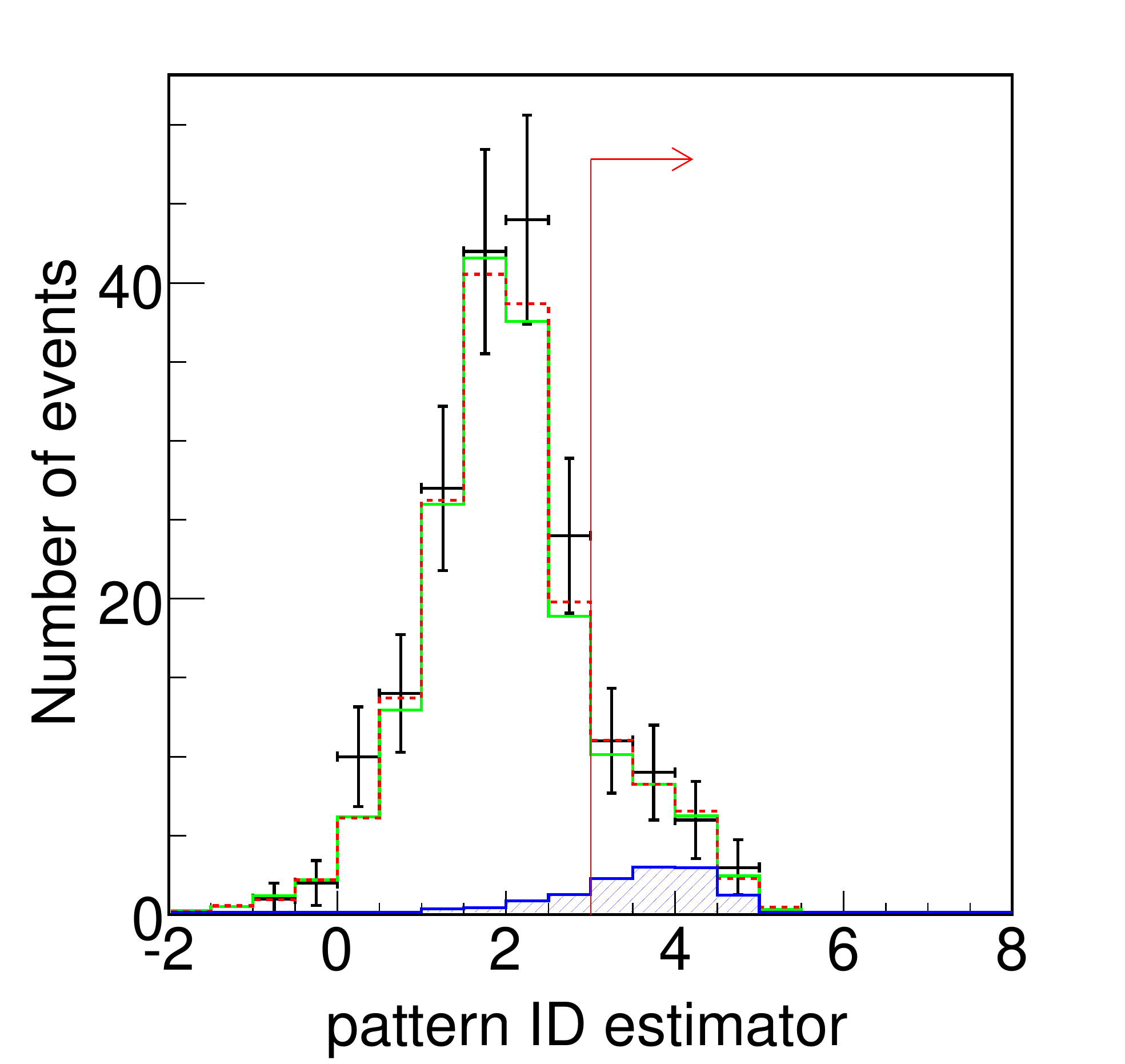}
  \includegraphics[width=2.25in]{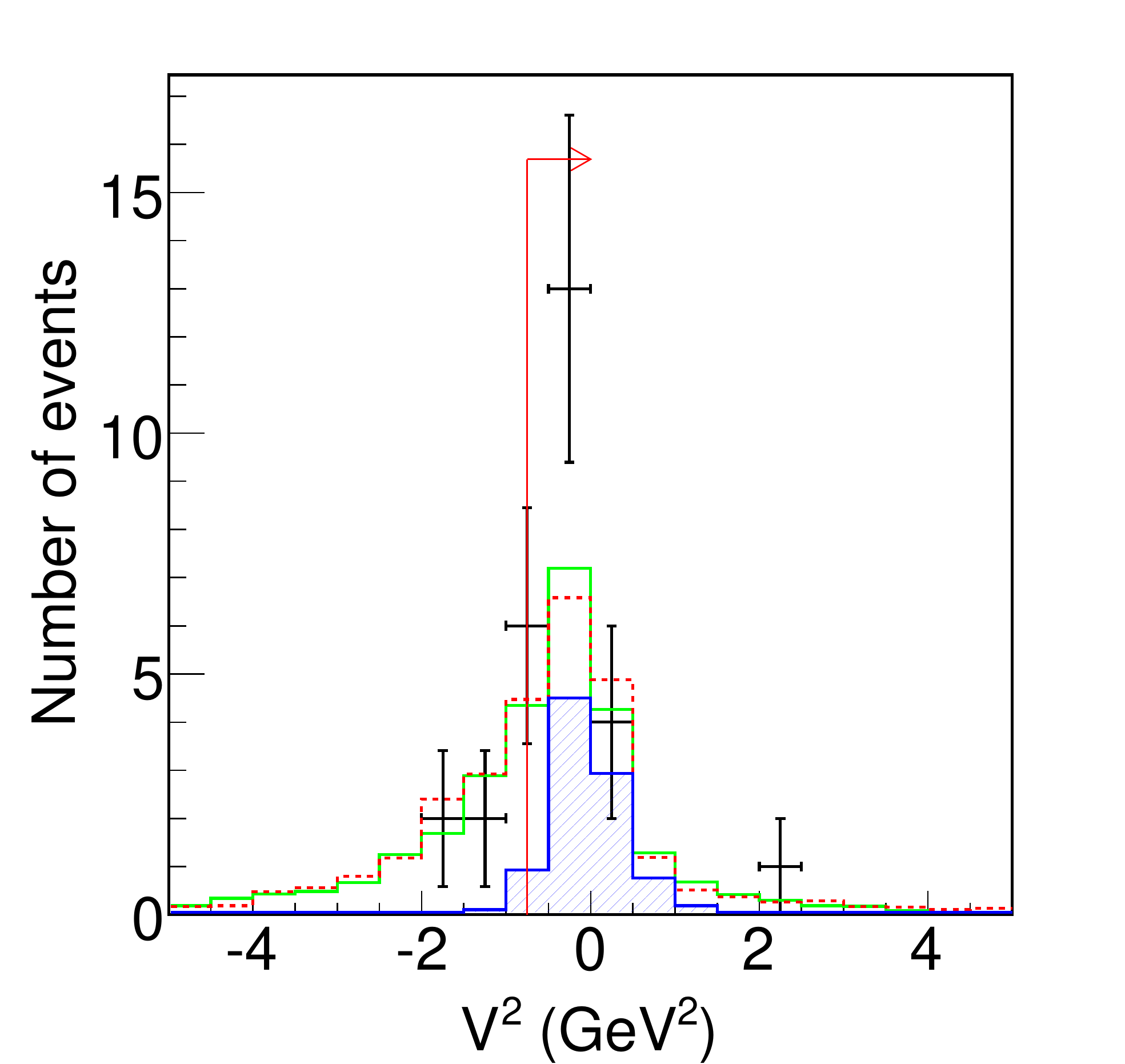}
  \end{minipage}
  \caption{Distribution of analysis variables with cuts for
    single-ring CCQE events selection in SK-II. The top two plots are
    two-dimensional $L,P$ distributions for data and MC, shown with
    the charged pion removal cut; the arrow shows the selected domain. 
    In the bottom three plots, the full
    and dashed line show Monte-Carlo expectation from NEUT and NUANCE,
    with oscillation reweighting, but no correction for the absolute
    flux normalization. The hatched region shows the contribution from
    CCQE events.}
\label{fig:ccqe1-cuts-sk2}
\end{center}
\end{figure*}

Our analysis of single-ring events closely follows the steps
outlined for the two-ring search.
\begin{enumerate}
  \item
    We select single-ring fully-contained
    events within the fiducial volume and $0.1<\mathrm{E_{vis}}<4$ GeV.
  \item Using the output of the proton pattern fitter on the new
    candidate ring, we apply the cuts $A < P < B\times
    L(\mathrm{m})+C$, with $A=C=1.1$ GeV/c and $B=7/6$ GeV/c/m.  We
    also require that $L<2$ m. These cuts aim at rejecting rings
    coming from pions as in section~\ref{sec:2ring}, 
    as well as fake rings which are usually fitted
    with a very low $P$ and $L$.
 \item We then require $\log
 Q_{dens}>0$, to remove bad ring candidates. 
 \item In order to improve proton selection we then apply the cut 
   $\Delta L >3$, where $\Delta L$ is the pattern ID estimator defined
   above.
 \item Finally to reject non-CCQE background as well as other non proton
   rings we ask that $V^2>-0.75\ \mathrm{GeV/c}^2$, using the same
   definition of $V$ as in the previous section.
\end{enumerate}
The final signal efficiency is about 21\% (18\% in SK-II), for the cases
where the proton ring is potentially visible. This is
lower than in the two-ring case because for single-ring events, 
most of those protons are just above Cherenkov threshold and 
therefore very hard to find.
Signal-to-noise after all cuts is about 1.1 to 1. We expect to
recover 20 (9) CCQE events in SK-I (SK-II), almost doubling our
statistics. The efficiencies are summarized in
tables~\ref{table:ccqe1-sk1} and \ref{table:ccqe1-sk2}, and the
distributions of the variables are shown in
Figs.~\ref{fig:ccqe1-cuts-sk1} and \ref{fig:ccqe1-cuts-sk2}.

In the final sample in SK-I (SK-II), 31 (13) events out of the final
39 (18) do have a proton track missed by the ring finder that was
recovered by this new fitter; 11 (4) of those are non-CCQE
interactions with a visible protons (mostly CC pion production) and
are an irreducible background to the CCQE sample.  After all cuts the
single-ring sample contains 90\% neutrinos and 10\% anti-neutrinos.

\begin{table*}[!htbp]
  \begin{tabular}{lccc}
    \hline
    \hline
    ~SK-I~ & ~Data~ & ~ Total MC~ & ~ Signal MC~ \\
    \hline FC, FV, single-ring,
    $0.1<\mathrm{E_{vis}}<4$ GeV & 7224 (100\%) & 6523 (100\%) & 94.4 (100\%)
    \\ $1.1<P<7/6L(m)+1.1~\mathrm{and}~L<2$ m & 1456 (20.2\%) & 1488 (22.8\%) & 44.8 (47.5\%)
    \\ Q density cut & 235 (3.25\%) & 213.1 (3.3\%) & 22.95 (24.32\%) \\ Pattern
    ID cut & 76 (1.05\%) & 66.01 (1.01\%) & 21.04 (22.29\%)
    \\ $V^{2}>-0.75~\mathrm{GeV}^2$ & 47 (0.65\%) & 38.69 (0.59\%) & 20.33
    (21.53\%) \\ \hline \hline
  \end{tabular}
  \caption{Summary of SK-I CCQE selection for single ring events. In
    this table the Monte-Carlo has been reweighted according to live
    time, solar activity for SK-I, as well as neutrino oscillations
    assuming \dms$=2.5\times 10^{-3}~\mathrm{eV}^{2}$ and \sstt $=1$.}
  \label{table:ccqe1-sk1}
\end{table*}

\begin{table*}[!htbp]
  \begin{tabular}{lccc}
    \hline
    \hline
    ~SK-II~ & ~Data~ & ~ Total MC~ & ~ Signal MC~ \\
    \hline FC, FV, single-ring,
    $0.1<\mathrm{E_{vis}}<4$ GeV & 3874 (100\%) & 3463 (100\%) & 50.71 (100\%)
    \\ $1.1<P<7/6L(m)+1.1~\mathrm{and}~L<2$ m & 702 (18.12\%) & 625.2 (18.1\%) & 20.45 (40.3\%)
    \\ Q density cut & 193 (4.98\%) & 175.2 (5.1\%) & 12.75 (25.14\%) \\ Pattern
    ID cut & 29 (0.75\%) & 27.53(0.80\%) & 9.58 (18.9\%)
    \\ $V^{2}>-0.75~\mathrm{GeV}^2$ & 22 (0.57\%) & 17.6 (0.51\%) & 9.2 (18.1\%)\\
    \hline \hline
  \end{tabular}
  \caption{Summary of SK-II CCQE selection for single ring events. In
    this table the Monte-Carlo has been reweighted according to live
    time, solar activity for SK-II, as well as neutrino oscillations
    assuming \dms$=2.5\times 10^{-3}~\mathrm{eV}^{2}$ and \sstt $=1$.}
  \label{table:ccqe1-sk2}
\end{table*}

\subsection{Kinematic reconstruction of atmospheric neutrinos}
\label{sec:kine}    
  Once we have a CCQE enriched sample, kinematic reconstruction of the
  incoming neutrino is straightforward, assuming that the target
  neutron is immobile.  Writing the total momentum as $P_{tot}$ and the
  direction as $\mathbf{d}$ the following equations hold:
  $$ P_{tot} = \sqrt{ (\mathbf{P_{\mathrm{p}}} + \mathbf{P_{\mathrm{l}}})^2 },$$ and
  $$ \mathbf{d} = \frac{1}{P_{tot}}(\mathbf{P_{\mathrm{p}}} +
  \mathbf{P_{\mathrm{l}}}),$$
  where $\mathbf{P_{\mathrm{p}}}$ and $\mathbf{P_{\mathrm{l}}}$ are the proton
  and lepton 3-momenta. Since we assume that all selected events are CCQE
  we use $E=P_{tot}$ for the neutrino energy.
  Using simulated data we have checked the
  angular resolution on neutrino track direction, as well as the energy
  resolution.  The energy reconstruction process has a tail toward
  negative values for non-CCQE events reconstructed with this technique
  and is non-Gaussian.  We have computed a 68.3\% confidence interval
  centered on the peak of the distribution. The angular and energy
  resolutions are summarized in table~\ref{table:resol}.

\begin{table*}[!htbp]
  \begin{tabular}{lccccc}
    \hline
    \hline
    & & \multicolumn{2}{c}{two-ring events} & 
    \multicolumn{2}{c}{single-ring events}\\    
    & & CCQE & TOTAL & CCQE & TOTAL \\
    \hline
    \multirow{2}{*}{SK-I} & Angular resolution (\degree) 
    & 8.0\degree & 12.3\degree & 8.1\degree & 14.9\degree \\
    & Energy resolution (\%) &$\asymerr{10.7}{4.4}$ & $\asymerr{10.6}{17.2}$
    & $\asymerr{9.9}{6.2}$ & $\asymerr{13.6}{16.7}$ \\
    \hline
    \multirow{2}{*}{SK-II} & Angular resolution (\degree) 
    & 8.6\degree & 12.0\degree & 8.1\degree & 16.2\degree \\
    & Energy resolution (\%) & $\asymerr{10.3}{5.9}$ & $\asymerr{10.8}{17.9}$ 
    & $\asymerr{9.6}{5.8}$ & $\asymerr{10.8}{22.0}$ \\
    \hline \hline
  \end{tabular}
  \caption{Energy and angular resolution of the kinematic reconstruction
  shown in the previous paragraph. Since the energy resolution distributions
  are asymmetric we report 68.3\% centered confidence intervals.}
  \label{table:resol}
\end{table*}

\subsubsection{Zenith angle distributions}

\begin{figure*}[!htbp]
  \includegraphics[width=2.25in]{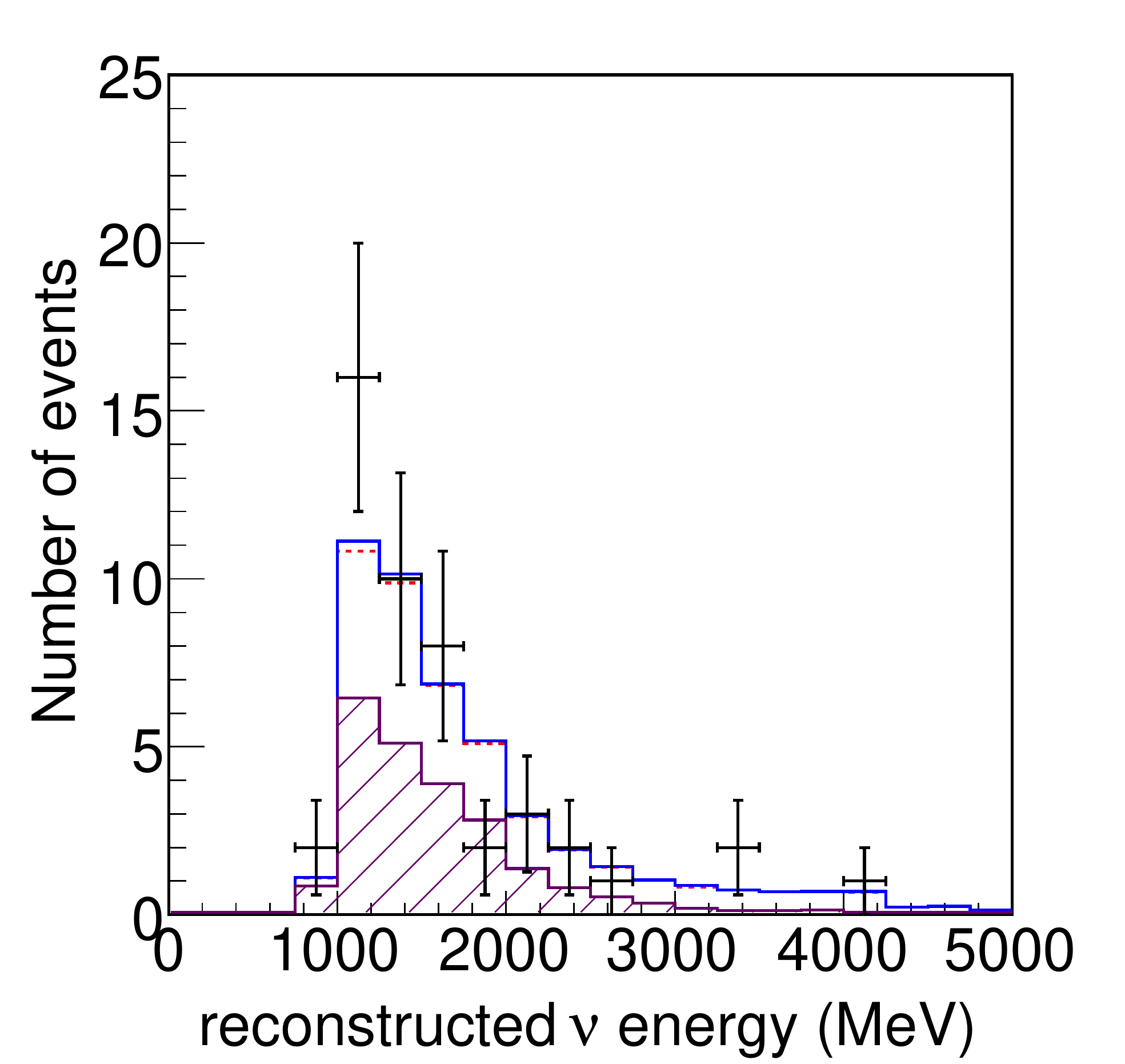}
  \includegraphics[width=2.25in]{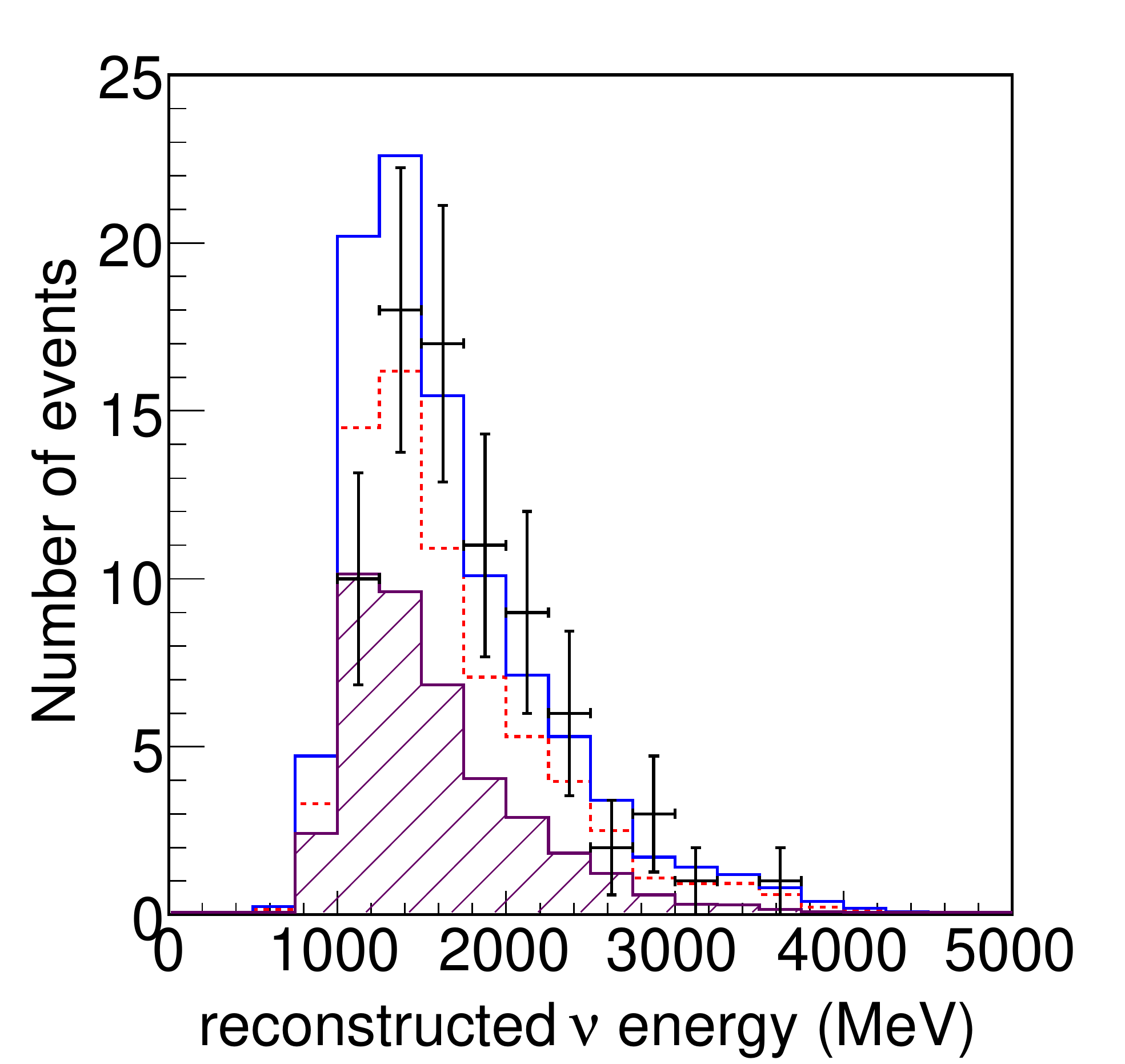}
  \caption{Reconstructed energy spectra of the CCQE enriched sample,
  e-like events are on the left and mu-like events on the right.
  The hatched region show the CCQE fraction from NEUT.
  The solid histograms correspond to our non-oscillated Monte-Carlo
  simulation, while the dashed histograms show the prediction with
  oscillations assuming \dms$=2.5\times 10^{-3}~\mathrm{eV}^{2}$ and \sstt $=1$.}
\label{fig:ccqefinalenergy}
\end{figure*}

To examine the evidence of zenith angle distortion in the enhanced
CCQE sample, we first separate it into electron and muon neutrino
interactions.  Each event is categorized according to the lepton's
e-like or mu-like identification.  After this characterization, we
kinematically reconstruct each event.  The reconstructed neutrino energy spectra
are shown in Fig.~\ref{fig:ccqefinalenergy}.
  
\begin{figure*}[!htbp]
  \includegraphics[width=2.25in]{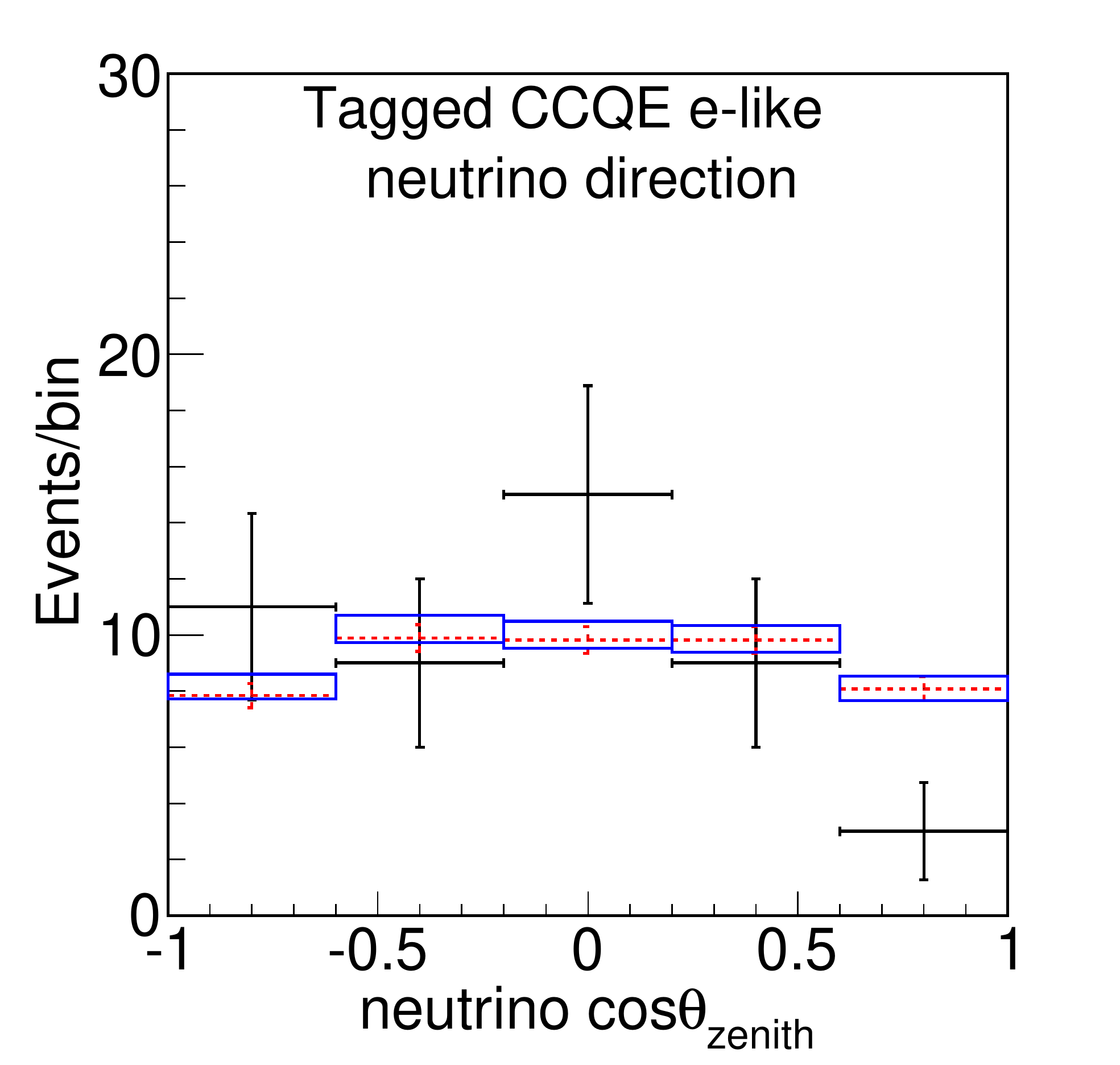}
  \includegraphics[width=2.25in]{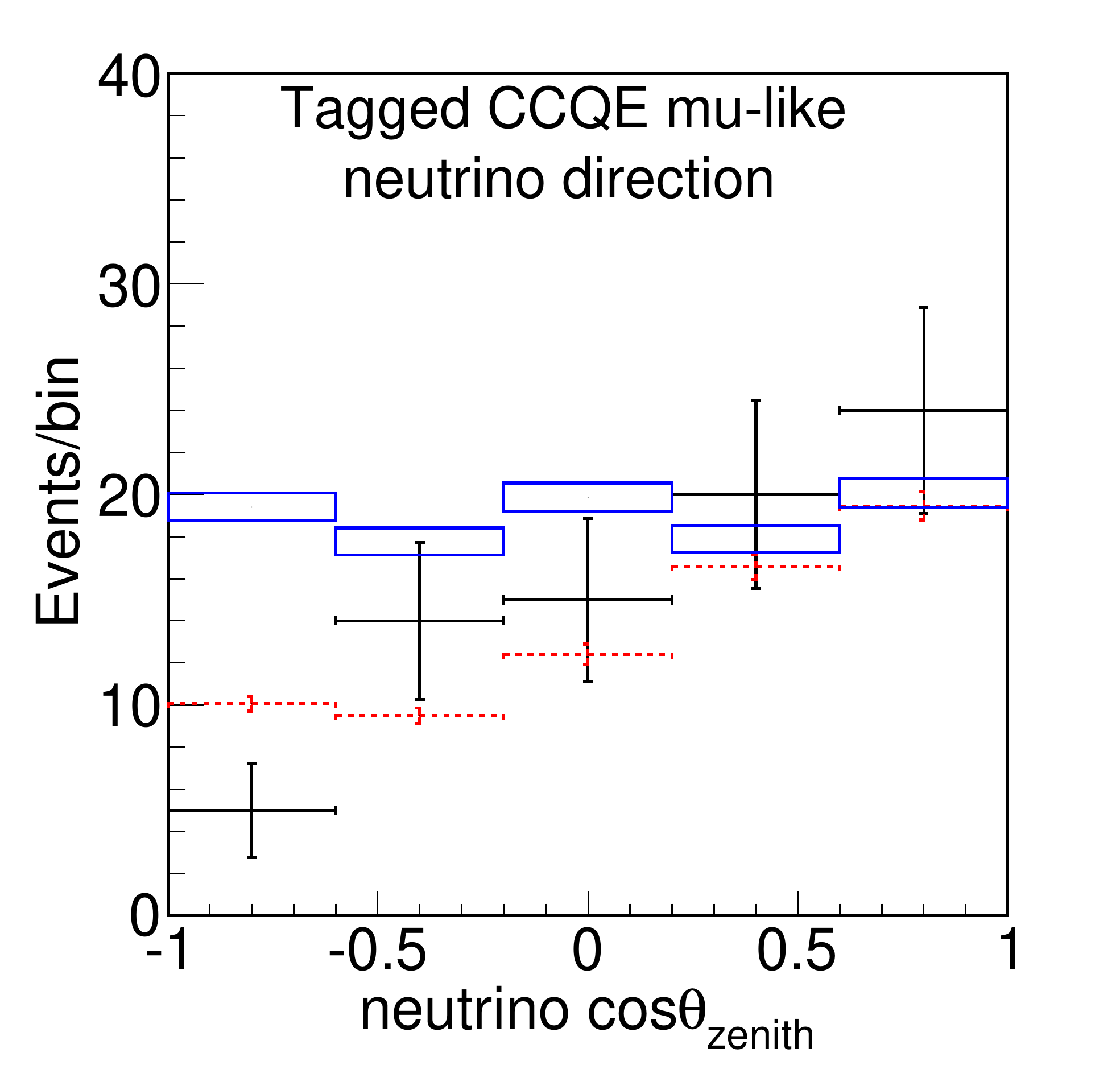}
  \caption{Reconstructed neutrino zenith angle of the CCQE enriched sample,
  e-like events are on the left and mu-like events on the right.
  The box histograms correspond to our non-oscillated Monte-Carlo
  simulation, while the dashed histograms show the prediction with
  oscillations assuming \dms$=2.5\times 10^{-3}~\mathrm{eV}^{2}$ and \sstt $=1$.}
\label{fig:ccqefinalzenith}
\end{figure*}

Figure~\ref{fig:ccqefinalzenith} shows the reconstructed neutrino
zenith angle distribution, along with the MC expectation with and
without oscillations.  Clearly, the distributions are consistent with
our the measurement of oscillation parameters presented
in~\cite{ashie:2005ik}.

For the events selected in this analysis, which are mostly in the
sub-GeV region, the lepton angular correlation to
the incoming neutrino is weak ($\sim 80$\degree resolution).  However,
with neutrino kinematic reconstruction, the angular resolution on the
neutrino track reaches 12\degree for \numu and 16\degree for \nue,
making the expected zenith angle distortion in the neutrino direction
of mu-like sample clearly visible.  The observed up-down asymmetry
using neutrino track reconstruction is $-0.52\pm0.17(\mathrm{stat})$
for $\mu$-like events and $-0.06\pm0.24$ for e-like events, compatible
with \mutau neutrino oscillations.  For comparison purposes the
asymmetries obtained using the lepton zenith angle are $-0.059\pm
0.24$ for $\mu$-like and $0\pm 0.22$ for e-like events.

\begin{figure*}[!htbp]
  \includegraphics[width=2.25in]{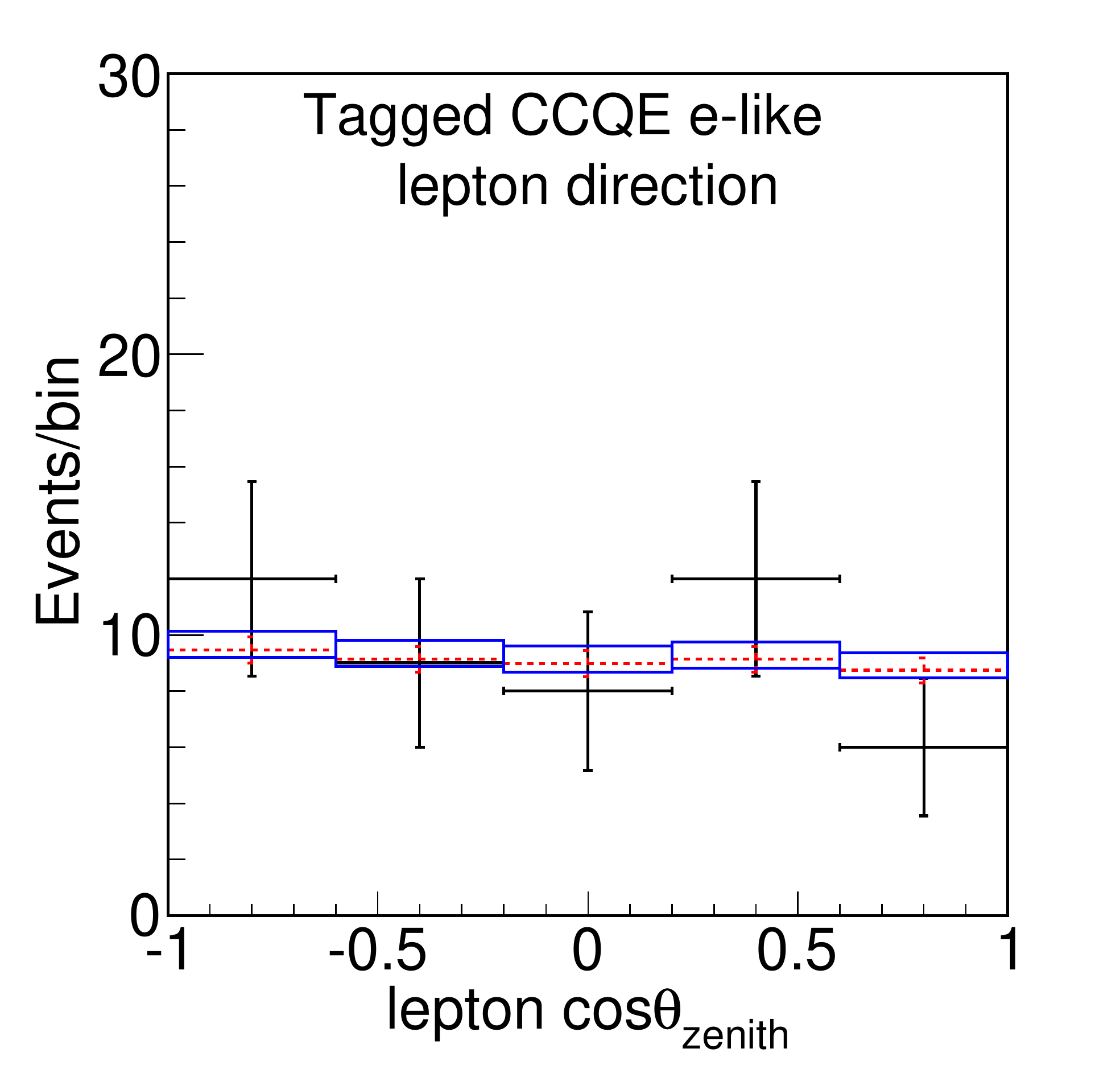}
  \includegraphics[width=2.25in]{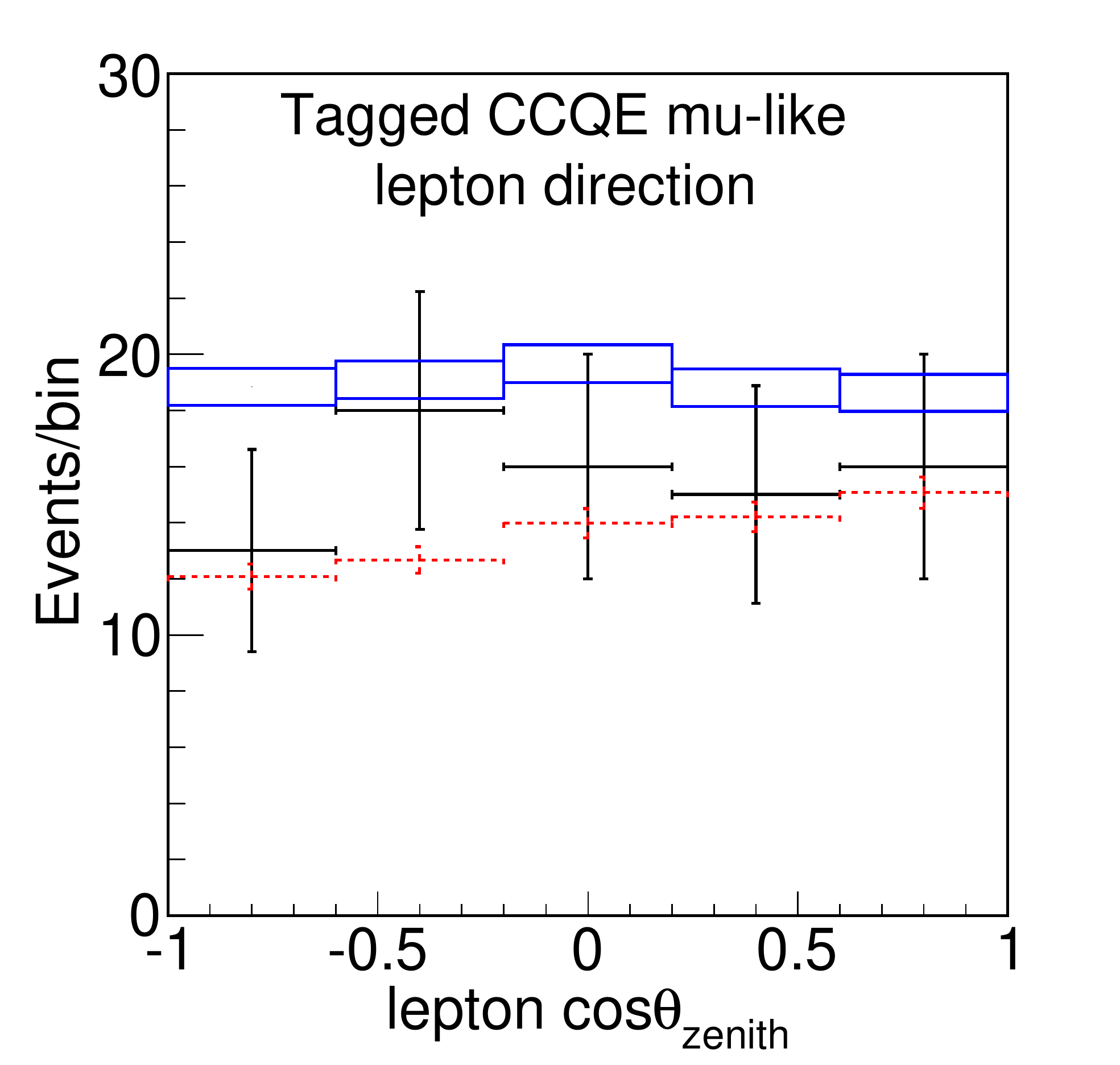}
  \caption{The lepton zenith angle of the CCQE enriched sample,
  e-like events are on the left and mu-like events on the right.
  The box histograms correspond to our non-oscillated Monte-Carlo
  simulation, while the dashed histograms show the prediction with
  oscillations assuming \dms$=2.5\times 10^{-3}~\mathrm{eV}^{2}$ and \sstt $=1$.}
\label{fig:leptonzenith}
\end{figure*}

The distributions in Fig.~\ref{fig:ccqefinalzenith} can be
contrasted with those in Fig.~\ref{fig:leptonzenith}.  In the second
set of figures, the lepton direction rather than the kinematically
reconstructed neutrino direction is shown for the CCQE enhanced
sample.  As can be seen, there is little sign of the oscillation
distortion in the lepton zenith angle distributions.  The correlation
between the incoming neutrino and outgoing lepton direction has been
lost. Most of the events in the CCQE enriched sample have a large
angle between the neutrino and lepton by virtue of the fact that a
large momentum was given to the proton in the reaction, resulting in
the proton being visible and the lepton having a large transverse
momentum to balance it.

Approximately one half of the sample is made up of two-ring events
where one of the rings has been identified as a proton.  In order to
obtain a high charged current purity, in~\cite{ashie:2005ik} only
multi-ring events with a leading muon (not an electron) were
included. For the same reason, a minimum requirement of 600~MeV/c was
imposed on the leading muon momentum for the sub-GeV multi-ring muons.
Both of those restrictions have been lifted here.  In the analysis
of~\cite{ashie:2005ik}, even applying the extra requirements used in
the mu-like multi-ring sample resulted in a CC purity of 54\% in the
e-like multi-ring sample.  The use of the proton tag in the CCQE
enhanced sample results in charged current purity of 88\% for e-like
events and 95\% for mu-like events.

In the combined single ring and two-ring enriched CCQE sample the
zenith distortion is clearly seen. Additionally, although the
statistics is too low to be usable (there are only about 20 events in
total), according to Monte Carlo, the data is expected to show a
zenith angle suppression even for events with the momentum of the
lepton less than 400~MeV.  As shown in~\cite{ashie:2005ik}, in
previous analyses, using the lepton direction for these events resulted in
the complete loss of the initial neutrino direction.

The use of the CCQE sample restores zenith angle pointing to a low
energy portion of the atmospheric sample which previously had no
observed zenith angle distortion, and adds events to the sample with
good pointing which were not previously considered.  This is an
important further cross-check of the both the analysis technique and
the oscillation hypothesis.

\subsubsection{$L/E$ distributions}

The neutrino path length $L$ is a simple function of the zenith angle
however, there are two main sources of uncertainty:
\begin{itemize}
\item Near the horizon, a small error on the zenith angle leads to a large
  error in $L$ since $dL/d\cos\theta_{zenith}$ is large. 
\item There is an additional uncertainty in the production height of
atmospheric neutrinos.  While it is negligible compared to the overall
travel length for neutrino produced on the far side of the Earth, it
is comparable to the total L for neutrinos produced directly above
\superk. The model from Honda \etal \cite{honda} is used to estimate
the production height. For a given zenith angle, $L$ is the average
path-length over 20 samplings from the production height distribution.
\end{itemize}

\begin{figure*}[!htb]
  \includegraphics[width=2.25in]{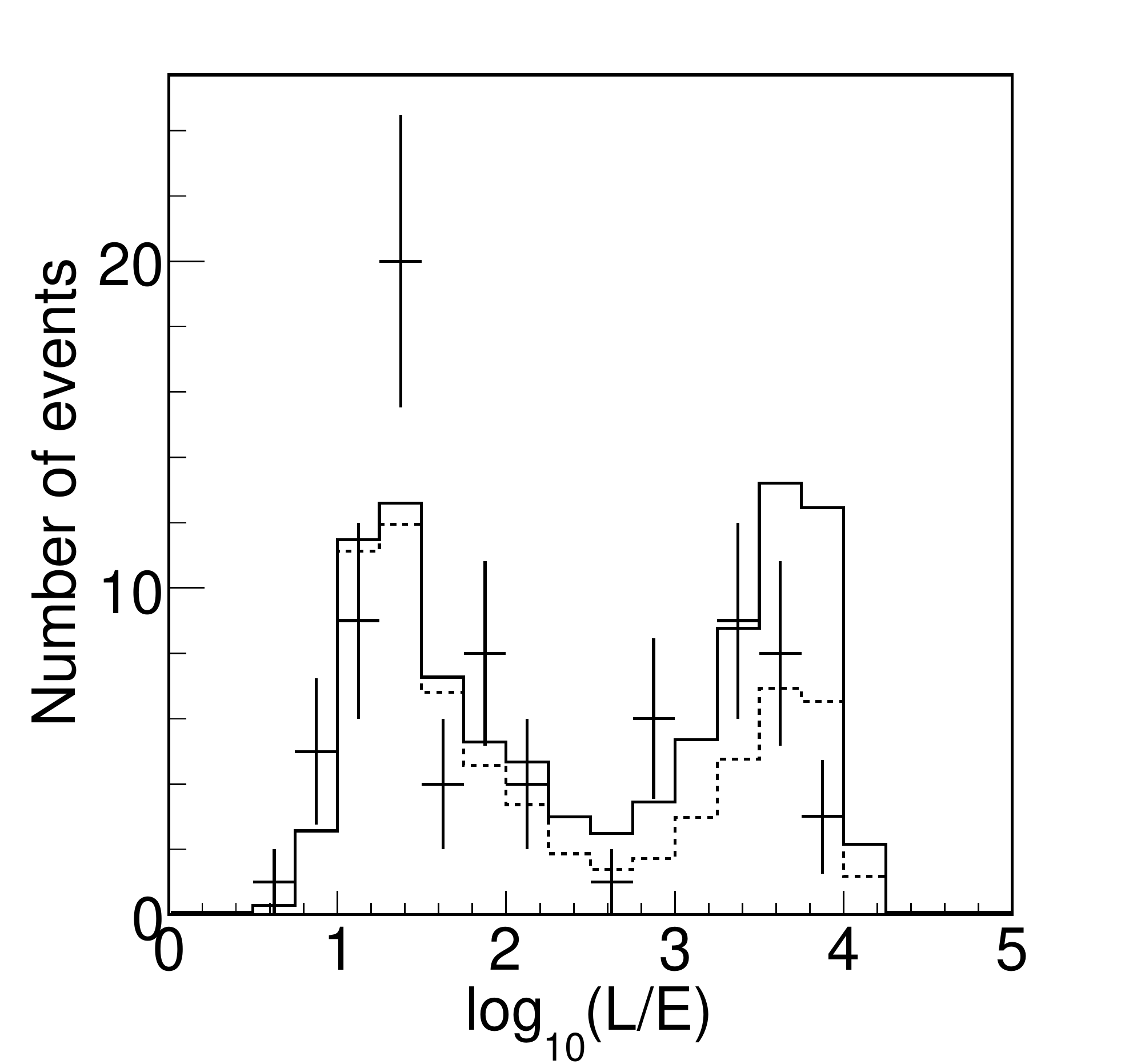}
  \includegraphics[width=2.25in]{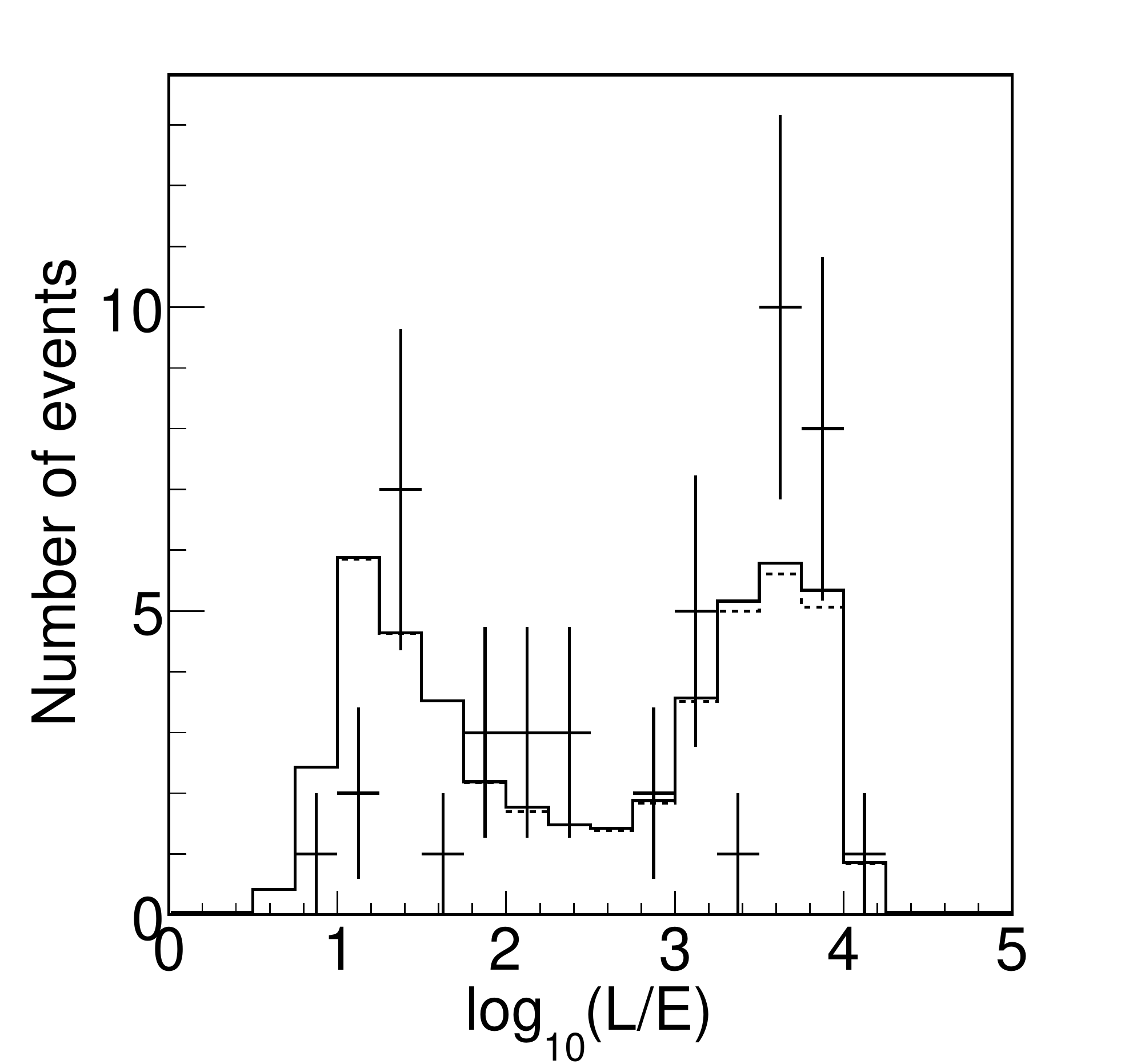}
  \caption{Distribution of L/E for the combined CCQE enriched $\mu$-like
    (left) and e-like (right) samples. The continuous lines show the
    expectation from NEUT without $\nu$ oscillations, while the
    dashed lines shows the expectation with \mutau oscillations with
    \dms$=2.5\times 10^{-3}~\mathrm{eV}^{2}$ and \sstt $=1$}
  \label{fig:lemu}
\end{figure*}

In this analysis, we use the reconstructed neutrino zenith angle to
estimate $L$; $E$ is the kinematically reconstructed $\nu$ energy
defined in section~\ref{sec:kine}. This has not been attempted so far
in \superk; the method reported in \cite{ashie:2004mr} used an energy
estimator based on the total energy of all charged particles in the
ID, and the neutrino zenith angle was obtained using the lepton zenith
angle. Figure~\ref{fig:lemu} is the distribution of L/E for events
with a primary $\mu$-like ring in the CCQE sample. SK-I and SK-II and
single- and two-ring samples were combined together to increase
statistics. The agreement with the oscillated distribution is very
good.

These distributions can be used in \mutau oscillation searches.
Binning the e-like and $\mu$-like data and MC $\log_{10}(L/E)$
distributions (5 equally spaced bins from 0 to 5 for both samples),
we use the following Poisson likelihood ratio \cite{pdg}:
\begin{widetext}
$$ \chi^2 = \sum_{i=2}^5
\left((M_{MC}-N_{DATA})+N_{DATA}\log\frac{N_{DATA}}{N_{MC}} \right)+ \sum_{j=2}^{6}
\varepsilon_j^2/\sigma_j^2.$$
\end{widetext}
$\sigma_j$ is the estimated uncertainty on
each $\varepsilon_j$ parameter; these variables describe the sources of systematic
errors and are adjusted during the fit. $N_{MC}$ is the weighted
sum of Monte-Carlo events, taking into account livetime and solar
activity. Oscillation probabilities and systematic errors are taken
into account using a technique similar to that of
\cite{fukuda:1998mi}. The weight of each Monte-Carlo event is :
\begin{widetext}
$$ w = F (1+\varepsilon_1)(E_\nu/2\mathrm{GeV})^{\varepsilon_2} P(\sin^2
2\theta,\Delta m^2,(1+\varepsilon_3)L/E_\nu) (1\pm \varepsilon_4/2) (1 \pm
\varepsilon_5/2), $$ 
\end{widetext}
where $F$ is a numerical factor accounting for livetime normalization
and solar activity, $E_\nu$ is the true neutrino energy, $L$ the neutrino path length,
and the function $P$ is the oscillation probability. 
For each event, 20 values of $L$ are obtained by sampling the production
height distribution; $P$ is the arithmetic average of the
oscillation probabilities obtained for each of these lengths.
The $\varepsilon_j$ parameters have the following meanings :
\begin{itemize}
\item $\varepsilon_1$ 
  represents the uncertainty on the absolute normalization of the sample.
  This includes flux uncertainties (8\% in this energy range \cite{honda}),
  and also the systematic uncertainty
  on our CCQE event selection, which we estimate to be 14.4\% based
  on Monte-Carlo studies.
  Since the combined error is large, it is a free parameter in the fit.
\item $\varepsilon_2$ is the uncertainty on the neutrino flux
  spectral index, with $\sigma_2=0.05$ as in \cite{fukuda:1998mi}.
\item $\varepsilon_3$ is a systematic error on the Monte-Carlo $L/E$ ratio,
  coming from uncertainties in the production height, and absolute
  neutrino energy scale. Following \cite{ashie:2005ik} its width
  is set to 10\%.
\item $\varepsilon_4$ describes the systematic error on e/$\mu$
  separation and uncertainties on the flavor content of the
  atmospheric flux: in the definition of $w$ the sign is negative for
  $\mu$-like and positive for e-like, leading to complete
  anti-correlation between the \numu and \nue samples.  Since SK-I and
  SK-II are combined, we have used a conservative estimate of 15\% for
  this error.
\item $\varepsilon_5$ accounts for the uncertainty in the non-CCQE 
  background selection efficiency, as well as for the
  CCQE/non-CCQE cross-section ratio; the sign preceding
  $\varepsilon_5$ in the definition of $w$ is positive for CCQE and
  negative for non-CCQE events. We have assumed this ratio to be known
  to 10\% in our simulation.
\item Finally one last parameter $\varepsilon_6$ is used to shift the
  reconstructed value of $\log_{10}(L/E)$ to account for biases in the
  kinematic reconstruction technique. Combining energy and direction
  reconstruction errors, we conservatively estimate the width of this
  parameter to be 10\%.
\end{itemize}

\begin{figure*}[!htb]
  \includegraphics[width=4.0in]{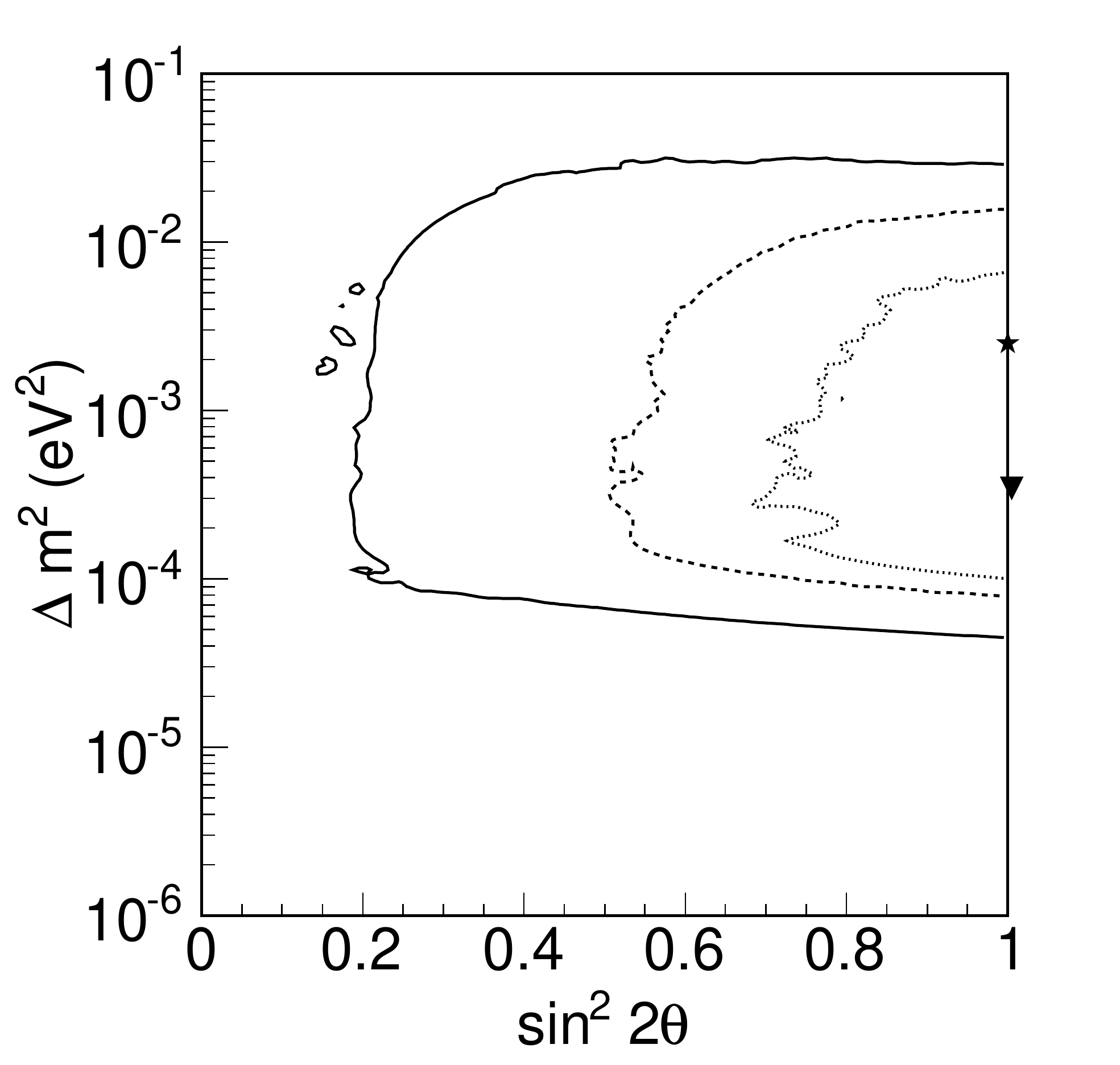}
  \caption{Allowed regions at 68.3\% (dotted), 90\% (dashed) and
    99\% (continuous) CL for \mutau oscillations from the CCQE
    kinematically reconstructed samples. The star shows the positions of
    the best fit from \cite{ashie:2005ik}, while the triangles show
    the position of the best fit for this sample in the physical region. 
    These allowed regions are consistent
    with previously published results and exclude the no-oscillation
    hypothesis at greater than 3 sigma.}
\label{fig:lecontour}
\end{figure*}

We have used a $201\times 201$ grid in $\log_{10}\Delta m^2$ and \sstt
; at each point on the grid the $\varepsilon_j$ were fitted using
MINUIT.  The best fit point is found in the unphysical region at
($\Delta m^2 = 1.3\,10^{-4}$ eV$^2$, \sstt = 1.3), with $\chi^2_{min}
= 6.70 / 8 $ dof.  The best fit in the physical region occurs at
($\Delta m^2 = 3.4\,10^{-4}$ eV$^2$, \sstt = 1.0) at
$\chi^2_{min,phys} = 7.98 / 8$ dof.  Figure~\ref{fig:lecontour} shows
the 68.3\%, 90\% and 99\% allowed regions ; these contours are located
at $\chi^2_{min,phys}+2.86$, 5.32, and 10.02 respectively, using the
same method as in \cite{ashie:2005ik}.  With the low
statistics of this sample, the allowed region is large. They
are consistent with our previous, more precise results: the best fit
point reported in our previous publications is accepted at better than
1$\sigma$. The no-oscillation point is rejected at 3$\sigma$ 
($\chi^2_{no osc} - \chi^2_{min,phys} =12.95$).
The systematic parameters at the best fit are summarized in
table~\ref{tab:syst}.
The kinematically reconstructed CCQE sample used here has roughly
half the statistics of the Soudan-2 tracking calorimeter experiment 
for which a similar analysis was performed \cite{Sanchez:2003rb}.

\begin{table*}[!htbp]
  \begin{tabular}{lccc}
    \hline
    \hline
    parameter & meaning & uncertainty & value at best fit \\
    $\varepsilon_1$ & absolute normalization  & (free) & 6.5\% \\
    $\varepsilon_2$ & spectral index  & 0.05 & -0.0006 \\
    $\varepsilon_3$ & Error on true L/E  & 10\% & -1.9\% \\
    $\varepsilon_4$ & e/$\mu$ ratio  & 15\% & -2.1\% \\
    $\varepsilon_5$ & CCQE/non-CCQE ratio  & 10\% & 0.2\%\\
    $\varepsilon_6$ & shift in reconstructed L/E  & 10\% & -5.3\% \\   
    \hline 
    \hline
  \end{tabular}
  \caption{Systematic parameters and their best fit values.}
  \label{tab:syst}
\end{table*}

It must be noted that the CCQE selection method selects a quasi-pure
sample of neutrinos ($\approx 90\%$). This could potentially have
sensitivity to the mass hierarchy and other CP-odd effects.

\section{Conclusion}

We have developed a technique for identifying proton tracks in the
\superk detector, relying on their distinctive Cherenkov ring pattern.
The proton momentum threshold for Cherenkov radiation is 1.07 GeV/c,
and above about 2 GeV/c they produce visible secondary particles
through hadronic interactions in the water, which makes identification
outside of this proton momentum range very difficult. 
Because of this we find that protons can be identified efficiently
in the momentum range between 1.25 GeV/c and 1.7 GeV/c.
The majority of atmospheric neutrino interactions produces protons
with momenta below the Cherenkov threshold, which results in low
statistics.
Using the full SK-I and SK-II data sets (2285.1 days), we have applied
this tool to identify neutral current elastic events.  The combined
data is compatible with the observation of protons ($\chi^2 = 9.3$ for
6 bins, $P=15.7\%$).

In a separate analysis, we have selected a sample of CCQE events,
reconstructing both the lepton and the proton, and thereby the
incoming neutrino kinematic parameters.  This is the first time that
kinematic reconstruction of the neutrino track is reported in a water
Cherenkov detector detector using atmospheric neutrinos. This sample
has $91.7\pm 3\%$ neutrino (as opposed to anti-neutrino) content,
potentially allowing studies where separating neutrinos and
anti-neutrinos is crucial, e.g. resolving the neutrino mass hierarchy.
The technique reported in this paper has an angular resolution of
12\degree for \numu and 16\degree for \nue on the neutrino track,
allowing the observation of a clear zenith-angle dependant asymmetry
in the neutrinos themselves. The observed up-down asymmetry using
neutrino track reconstruction is $-0.52\pm0.17\ (\mathrm{stat})$ for
$\mu$-like events and $-0.06\pm0.24$ for e-like events, compatible
with our previous results with other higher energy samples and \mutau
neutrino oscillations.

Although the CCQE selection yields low statistics in \superk and is
not competitive with previously published measurements of the
oscillation parameters, we have performed a fit of the $L/E$
distribution including relevant systematic effects. With this sample
alone, the no-oscillation hypothesis is excluded at 3 standard
deviations.  Table~\ref{tab:summary} summarizes the new data samples
selected in this paper.  We expect that this new technique will be of
particular interest to future projects involving water Cherenkov
detectors even larger than \superk, where the statistics will be much
higher.

\begin{table*}[!htb]
    \begin{tabular}{lcc}
    \hline
    \hline
    Event class  & SK-I data & SK-II data \\
    NC elastic (expected NC elastic fraction)  & 27 (64.7\%) & 11 (55.6\%) \\
    CCQE e-like (expected CCQE fraction) & 31 (53.0\%) & 16 (51.4\%)\\
    CCQE $\mu$-like (expected CCQE fraction) & 60 (62.4\%) & 18 (61.3\%) \\
    \hline 
    \hline
  \end{tabular}
  \caption{Summary of the observed data samples : single
    proton, CCQE e-like and CCQE $\mu$-like, for SK-I, SK-II.}
  \label{tab:summary}
\end{table*}

\section{Acknowledgments}

We gratefully acknowledge the cooperation of the Kamioka Mining and
Smelting Company. The Super-Kamiokande experiment was built and has
been operated with funding from the Japanese Ministry of Education,
Science, Sports and Culture, and the United States Department of
Energy.  We gratefully acknowledge individual support by the National
Science Foundation, and the Polish Committee for Scientific Research.
Some of us have been supported by funds from the Korean Research
Foundation (BK21), the Korea Science and Engineering Foundation and
the Japan Society for the Promotion of Science, and Research
Corporation.

\bibliography{bibliography}

\end{document}

%% file: authors.tex
\newcommand{\AFFicrr}{\affiliation{Kamioka Observatory, Institute for Cosmic Ray Research, University of Tokyo, Kamioka, Gifu 506-1205, Japan}}
\newcommand{\AFFkashiwa}{\affiliation{Research Center for Cosmic Neutrinos, Institute for Cosmic Ray Research, University of Tokyo, Kashiwa, Chiba 277-8582, Japan}}
\newcommand{\AFFipmu}{\affiliation{Institute for the Physics and
Mathematics of the Universe, University of Tokyo, Kashiwa, Chiba
277-8582, Japan}}
\newcommand{\AFFbu}{\affiliation{Department of Physics, Boston University, Boston, MA 02215, USA}}
\newcommand{\AFFbnl}{\affiliation{Physics Department, Brookhaven National Laboratory, Upton, NY 11973, USA}}
\newcommand{\AFFucd}{\affiliation{Department of Physics, University of California, Davis, Davis, CA 95616, USA}}
\newcommand{\AFFuci}{\affiliation{Department of Physics and Astronomy, University of California, Irvine, Irvine, CA 92697-4575, USA }}
\newcommand{\AFFcsu}{\affiliation{Department of Physics, California State University, Dominguez Hills, Carson, CA 90747, USA}}
\newcommand{\AFFcnm}{\affiliation{Department of Physics, Chonnam National University, Kwangju 500-757, Korea}}
\newcommand{\AFFduke}{\affiliation{Department of Physics, Duke University, Durham NC 27708, USA}}
\newcommand{\AFFgmu}{\affiliation{Department of Physics, George Mason University, Fairfax, VA 22030, USA }}
\newcommand{\AFFgifu}{\affiliation{Department of Physics, Gifu University, Gifu, Gifu 501-1193, Japan}}
\newcommand{\AFFuh}{\affiliation{Department of Physics and Astronomy, University of Hawaii, Honolulu, HI 96822, USA}}
\newcommand{\AFFkanagawa}{\affiliation{Physics Division, Department of Engineering, Kanagawa University, Kanagawa, Yokohama 221-8686, Japan}}
\newcommand{\AFFkek}{\affiliation{High Energy Accelerator Research Organization (KEK), Tsukuba, Ibaraki 305-0801, Japan }}
\newcommand{\AFFkobe}{\affiliation{Department of Physics, Kobe University, Kobe, Hyogo 657-8501, Japan}}
\newcommand{\AFFkyoto}{\affiliation{Department of Physics, Kyoto University, Kyoto, Kyoto 606-8502, Japan}}
\newcommand{\AFFumd}{\affiliation{Department of Physics, University of Maryland, College Park, MD 20742, USA }}
\newcommand{\AFFmit}{\affiliation{Department of Physics, Massachusetts Institute of Technology, Cambridge, MA 02139, USA}}
\newcommand{\AFFmiyagi}{\affiliation{Department of Physics, Miyagi University of Education, Sendai, Miyagi 980-0845, Japan}}
\newcommand{\AFFnagoya}{\affiliation{Solar Terrestrial Environment
Laboratory, Nagoya University, Nagoya, Aichi 464-8602, Japan}}
\newcommand{\AFFsuny}{\affiliation{Department of Physics and Astronomy, State University of New York, Stony Brook, NY 11794-3800, USA}}
\newcommand{\AFFniigata}{\affiliation{Department of Physics, Niigata University, Niigata, Niigata 950-2181, Japan }}
\newcommand{\AFFokayama}{\affiliation{Department of Physics, Okayama University, Okayama, Okayama 700-8530, Japan }}
\newcommand{\AFFosaka}{\affiliation{Department of Physics, Osaka University, Toyonaka, Osaka 560-0043, Japan}}
\newcommand{\AFFseoul}{\affiliation{Department of Physics, Seoul National University, Seoul 151-742, Korea}}
\newcommand{\AFFshizuokasc}{\affiliation{Department of Informatics in
Social Welfare, Shizuoka University of Welfare, Yaizu, Shizuoka, 425-8611, Japan}}
\newcommand{\AFFshizuoka}{\affiliation{Department of Systems Engineering, Shizuoka University, Hamamatsu, Shizuoka 432-8561, Japan}}
\newcommand{\AFFskk}{\affiliation{Department of Physics, Sungkyunkwan University, Suwon 440-746, Korea}}
\newcommand{\AFFtohoku}{\affiliation{Research Center for Neutrino Science, Tohoku University, Sendai, Miyagi 980-8578, Japan}}
\newcommand{\AFFtokyo}{\affiliation{The University of Tokyo, Bunkyo, Tokyo 113-0033, Japan }}
\newcommand{\AFFtokai}{\affiliation{Department of Physics, Tokai University, Hiratsuka, Kanagawa 259-1292, Japan}}
\newcommand{\AFFtit}{\affiliation{Department of Physics, Tokyo Institute
for Technology, Meguro, Tokyo 152-8551, Japan }}
\newcommand{\AFFtsinghua}{\affiliation{Department of Engineering Physics, Tsinghua University, Beijing, 100084, China}}
\newcommand{\AFFwarsaw}{\affiliation{Institute of Experimental Physics, Warsaw University, 00-681 Warsaw, Poland }}
\newcommand{\AFFuw}{\affiliation{Department of Physics, University of Washington, Seattle, WA 98195-1560, USA}}

\AFFicrr
\AFFkashiwa
\AFFipmu
\AFFbu
\AFFbnl
\AFFucd
\AFFuci
\AFFcsu
\AFFcnm
\AFFduke
\AFFgifu
\AFFuh
\AFFkanagawa
\AFFkek
\AFFkobe
\AFFkyoto
\AFFmiyagi
\AFFnagoya
\AFFsuny
\AFFniigata
\AFFokayama
\AFFosaka
\AFFseoul
\AFFshizuoka
\AFFshizuokasc
\AFFskk
\AFFtokai
\AFFtokyo
\AFFtsinghua
\AFFwarsaw
\AFFuw

\author{M.~Fechner}
\altaffiliation{Present address: CEA, Irfu, SPP, Centre de Saclay, F-91191, Gif-sur-Yvette, France}
\AFFduke

\author{K.~Abe}
\AFFicrr
\author{Y.~Hayato}
\AFFicrr
\AFFipmu
\author{T.~Iida}
\author{M.~Ikeda}
\author{J.~Kameda}
\author{K.~Kobayashi}
\author{Y.~Koshio}
\author{M.~Miura} 
\AFFicrr
\author{S.~Moriyama} 
\author{M.~Nakahata} 
\AFFicrr
\AFFipmu
% ...

\collaboration{The Super-Kamiokande Collaboration}
\noaffiliation